\documentclass[12pt]{article}

\usepackage{graphicx}
\usepackage{amssymb}
\usepackage{epstopdf}
\usepackage{epsfig,rotating}
\usepackage{subfigure}
\def\nomQ2{0.030}

\newcommand{\Qweakproton}{$Q_W^{p}$}
\newcommand{\Qweak}{$Q_{weak}\;$}

\newcommand{\qw}{$Q_W^{p}\;$}

\newcommand{\gev}{\,{\rm GeV}}

\newcommand{\be}{\begin{equation}}
\newcommand{\ee}{\end{equation}}

\catcode`@=11
\def\@citex[#1]#2{\if@filesw\immediate\write\@auxout{\string\citation{#2}}\fi
  \def\@citea{}\@cite{\@for\@citeb:=#2\do
    {\@citea\def\@citea{,\penalty\@m}\@ifundefined
      {b@\@citeb}{{\bf ?}\@warning
       {Citation `\@citeb' on page \thepage \space undefined}}%
\hbox{\csname b@\@citeb\endcsname}}}{#1}}

\def\citer{\@ifnextchar
[{\@tempswatrue\@citexr}{\@tempswafalse\@citexr[]}}
\def\@citexr[#1]#2{\if@filesw\immediate\write\@auxout{\string\citation{#2}}\fi
  \def\@citea{}\@cite{\@for\@citeb:=#2\do
    {\@citea\def\@citea{--\penalty\@m}\@ifundefined
       {b@\@citeb}{{\bf ?}\@warning
       {Citation `\@citeb' on page \thepage \space undefined}}%
\hbox{\csname b@\@citeb\endcsname}}}{#1}}
\catcode`@=12

\def\Cerenkov{\v{C}erenkov }
\def\LH2{LH$_2$}

\newcommand{\AmS}{{\protect\the\textfont2
  A\kern-.1667em\lower.5ex\hbox{M}\kern-.125emS}}

\def\deg{$^{\circ}$\ }

\begin{document}
\date{December 10, 2007}
\title{The \Qweak{} Experiment:\\
``A Search for
New Physics at the TeV Scale \\
via a Measurement of the Proton's Weak Charge''}

\maketitle

\begin{center}
PAC Proposal Update:   JLab E02-020
\end{center}
\vspace*{0.2cm}

\noindent{\bf The Collaboration}

\noindent {D.S. Armstrong$^{1}$}, {A. Asaturyan$^{13}$}, {T. Averett$^{1}$}, {J. Benesch$^{9}$}, {J. Birchall$^{7}$},
{P. Bosted$^{9}$}, {A. Bruell$^{9}$}, {C. L. Capuano$^{1}$}, {G. Cates$^{21}$}, {C. Carrigee$^{14}$},
{\underline{R. D. Carlini}$^{9}$ (Principal Investigator)}, {S. Chattopadhyay$^{16}$}, {S. Covrig$^{12}$},
{C.A. Davis$^{11}$, {K. Dow$^{8}$}, {J. Dunne$^{14}$}, {D. Dutta$^{14}$}, {R. Ent$^{9}$},
{J.Erler$^{3}$}, {W. Falk$^{7}$}, \\{H. Fenker$^{9}$}, {\underline{J.M. Finn}$^{1}$}, {T.A. Forest$^{24}$}, {W. Franklin$^{8}$},
{D. Gaskell$^{9}$}, {M. Gericke$^{7}$}, {J. Grames$^{9}$}, \\{ K. Grimm$^{9}$}, {F.W. Hersman$^{12}$},
{D. Higinbotham$^{9}$}, {M. Holtrop$^{12}$}, {J.R. Hoskins$^{1}$}, {K. Johnston$^{6}$}, \\{E. Ihloff$^{8}$}, {M. Jones$^{9}$}, {R. Jones$^{2}$}, 
{K. Joo$^{2}$}, {J. Kelsey$^{8}$}, {C. Keppel$^{18}$},
{M. Khol$^{8}$}, {P. King$^{17}$}, \\{E. Korkmaz$^{15}$}, {\underline{S. Kowalski}$^{8}$}, {J. Leacock$^{10}$}, {J.P. Leckey$^{1}$}, {L. Lee$^{7}$},
{A. Lung$^{9}$}, {D. Mack$^{9}$}, {S. Majewski$^{9}$}, {J. Mammei$^{10}$}, 
{J. Martin$^{20}$}, {D. Meekins$^{9}$}, {A. Micherdzinska$^{20}$}, {A. Mkrtchyan$^{13}$}, {H. Mkrtchyan$^{13}$}, {N. Morgan$^{10}$},
{K. E. Myers$^{22}$}, {A. Narayan$^{14}$}, {A. K. Opper$^{22}$}, {J. Pan$^{7}$}, {\underline{S.A. Page}$^{7}$}, {K. Paschke$^{21}$},
\\{M. Pitt$^{10}$}, {M. Poelker$^{9}$}, {Y. Prok$^{25}$}, {W. D. Ramsay$^{7}$},
{M. Ramsey-Musolf$^{4}$}, {J. Roche$^{17}$}, \\{N. Simicevic$^{6}$}, {G. Smith$^{9}$ (Project Manager)},
{T. Smith$^{19}$}, {P. Souder$^{23}$}, {D. Spayde$^{5}$}, {B. E. Stokes$^{22}$}, {R. Suleiman$^{9}$}, {V. Tadevosyan$^{13}$}, {E. Tsentalovich$^8$}, 
{W.T.H. van Oers$^{7}$}, {W. Vulcan$^{9}$}, {P. Wang$^{7}$}, \\{S. Wells$^{6}$}, {S.A. Wood$^{9}$}, {S. Yang$^{1}$}, 
{R. Young$^{9}$}, {H. Zhu$^{12}$}, {C. Zorn$^{9}$}

\vspace{.3cm}

\noindent {\bf The Institutions}

\noindent $^{1}${\em College of William and Mary}, $^{2}${\em University of Connecticut,}, 
$^{3}${\em Instituto de Fisica, Universidad Nacional Autonoma de Mexico},
$^{4}${\em University of Wisconsin}, $^{5}${\em Hendrix College}, $^{6}${\em Louisiana Tech University}, 
$^{7}${\em University of Manitoba}, $^{8}${\em Massachusetts Institute
of Technology}, $^{9}${\em Thomas Jefferson National Accelerator facility}, $^{10}${\em Virginia Polytechnic
Institute}, $^{11}${\em TRIUMF}, $^{12}${\em University of New Hampshire}, $^{13}${\em Yerevan Physics Institute}, 
$^{14}${\em Mississippi State University}, $^{15}${\em University of Northern British Columbia}, 
$^{26}${\em Cockroft Institute of Accelerator Science and Technology}. $^{17}${\em Ohio University}, 
$^{18}${\em Hampton University}, $^{19}${\em Dartmouth College}, $^{20}${\em University of Winnipeg}, $^{21}${\em University of
Virginia}, $^{22}${\em George Washington University}, $^{23}${\em Syracuse University}, $^{24}${\em Idaho State
University}, $^{25}${\em Christopher Newport University}

\tableofcontents \pagebreak

            %
\section{Introduction} \label{intro}

The Standard Model (SM) has been extremely successful at describing a comprehensive range of phenomena in
nuclear and particle physics. After three decades of rigorous experimental testing, the only indication of a
shortcoming of the SM lies in the discovery of neutrino oscillations~\cite{Ahn:2002up}. That discovery has
renewed interest in identifying other places where physics beyond the Standard Model might
be observed. There
are two principal strategies in the search for new physics, and ultimately a more fundamental description of
nature. The first is to build increasingly energetic colliders, such as the Large Hadron Collider (LHC) at CERN,
which aim to excite matter into a new form. The second, more subtle approach is to perform precision
measurements at moderate energies, where any observed discrepancy with the Standard Model will reveal the
signature of these new forms of matter \cite{Erler:2004cx,RamseyMusolf:2006vr}. Results from the \Qweak
measurement at Jefferson Laboratory, in conjunction with existing measurements of parity-violating electron
scattering, will constrain the possibility of relevant physics beyond the Standard Model to the multi-TeV energy
scale and beyond.

\begin{figure} [hh!]
\begin{center}
\rotatebox{0.}{\resizebox{5.5in}{4.0in}{\includegraphics{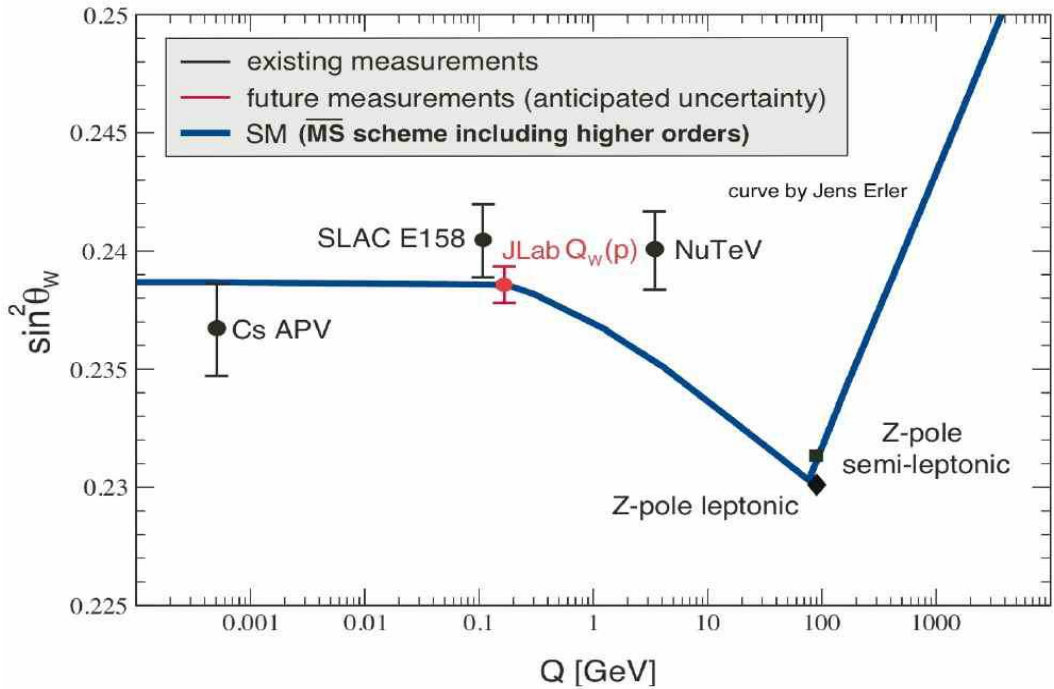}}}
\end{center} \vspace*{-0.5cm}
\caption{\em Calculated running of the weak mixing angle in the
  Standard Model, as defined in the modified minimal subtraction
  scheme\cite{Erler:2004in}. The uncertainty in the predicted running
  corresponds to the thickness of the blue curve. The black error bars show the current situation, while the
  red error bar (with arbitrarily chosen vertical location) refers to
  the proposed 4\% \Qweak{} measurement. The existing measurements are
  from atomic parity violation (APV)~\cite{Bennett:1999pd}, SLAC E-158~\cite{Anthony:2005pm}, deep inelastic
  neutrino-nucleus scattering (NuTeV)~\cite{Zeller:2001hh}, and from $Z^{0}$ pole
  asymmetries (LEP+SLC)~\cite{Yao:2006px}.} \label{RUNNINGTHETA}
\end{figure}

The \Qweak collaboration proposes\footnote{This proposal and other documents are available
  at the home page of the \Qweak Collaboration:
  ``http://www.jlab.org/Qweak/''.} to carry out the first precision measurement of
  the proton's weak charge:
  \begin{equation}
Q^p_w =1-4\sin^{2}\theta_{W} \label{eq:qweak}
\end{equation}
at JLab, building on technical advances that have been made in the laboratory's world-leading parity-violation
program and using the results of earlier experiments to constrain hadronic corrections. The experiment is a high
precision measurement of the parity-violating asymmetry in elastic $ep$ scattering at $Q^{2} = 0.026\,{\mathrm
GeV}^2$ employing approximately 180 $\mu A$ of 85\% polarized beam on a 35 cm liquid hydrogen target. It will
determine the proton's weak charge with about $4\%$ combined statistical and systematic errors.

In the absence of physics beyond the Standard Model, our experiment will provide a $\simeq$0.3\% measurement of
$\sin^{2}\theta_{W}$, making this the most precise stand alone measurement of the weak mixing angle at low
$Q^{2}$,  and in combination with other parity measurements, a high precision determination of the weak charges
of the up and down quarks. Our proposed measurement of \qw will be performed with significantly smaller
statistical and systematic errors than existing low $Q^2$ data.  Any significant deviation from the Standard
Model prediction at low $Q^{2}$ would be a signal of new physics, whereas agreement would place new and
significant constraints on possible Standard Model extensions.

The Standard Model makes a firm prediction for $Q^p_w$, based on the running of the weak mixing angle,
$\sin^{2}\theta_{W}$, from the $Z^{0}$ pole down to low energies, as shown in Figure~\ref{RUNNINGTHETA}. The
precise measurements near the $Z^{0}$ pole anchor the curve at one particular energy scale. The shape of the
curve away from this point is a prediction of the Standard Model, and to test this prediction one needs precise
off-peak measurements. Currently there are several precise off-peak determinations of $\sin^{2}\theta_{W}$: one
from atomic parity violation (APV)~\cite{Bennett:1999pd}; and another from E-158 at SLAC which measured
$\sin^{2}\theta_{W}$ from parity-violating $\vec{e}e$ (M\o ller) scattering at low $Q^2$~\cite{Anthony:2005pm}.
The result from deep inelastic neutrino-nucleus scattering~\cite{Zeller:2001hh} is less clearly interpretable.

It is worth noting that radiative corrections affect the proton and electron weak charges rather differently; in
addition to the effect from the running of $\sin^{2}{\hat\theta}_{W}(\mu^2)$, there is a relatively large WW box
graph contribution to the proton weak charge that does not appear in the case of the electron.  This
contribution compensates numerically for nearly all of the effect of the running of the weak mixing angle, so
that the final Standard Model result for the proton's weak charge is close to what it would be at tree level,
which is not so for the electron.

The \Qweak experiment (E02-020) was initially approved at the 21st meeting of the Jefferson Laboratory Program
Advisory Committee in January, 2002, and was awarded an ``A'' scientific rating, which was reconfirmed in
January, 2005.  Major equipment construction activities are underway at collaborating institutions and
commercial vendors. A schedule has been adopted for the experiment, with the aim of initial installation in
JLab's Hall C in 2009.  This document is a review of the current status of the experiment, with emphasis on
critical systems requirements and performance, schedule, and a beam time request to complete the measurements to
the proposed accuracy in 2010-2012 at JLab.

\subsection{Extracting \qw from Experimental Data}

Electroweak theory can rigorously derive a low-energy effective interaction between the electron and the quarks
that can be used to predict low-energy electroweak observables. Assuming that conventional, Standard Model
effects arising from non-perturbative QCD or many-body interactions are under sufficient theoretical control,
any deviation from the predictions of that effective interaction is then an unambiguous signal of physics beyond
the Standard Model. The recent measurements of parity-violating electron scattering (PVES) on nuclear targets
have made possible a dramatic improvement in the accuracy with which we probe the weak neutral-current sector of
the Standard Model at low energy. The existence of this set of high-precision, internally-consistent PVES
measurements provides the critical key to the interpretation of the asymmetry to be measured by the proposed
\Qweak experiment. Specifically, they provide direct experimental determination of the contribution of hadronic
form factors to our very low $Q^2$ asymmetry measurement.

For the purpose of this measurement, the relevant piece of the weak
force which characterizes the virtual-exchange of a $Z^0$-boson
between an electron and an up or down quark can be parameterized by
the constants, $C_{1u(d)}$, that are defined through the effective
four-point interaction by \cite{Yao:2006px}
\begin{equation}
{\cal L}_{\rm NC}^{eq}=-\frac{G_F}{\sqrt{2}}\bar{e}\gamma_\mu\gamma_5e \sum_q C_{1q}\bar{q}\gamma^\mu q\,.
\label{eq:LSM}
\end{equation}
These effective couplings are known to high-precision within the Standard Model, from precision measurements at
the $Z$-pole \cite{ZPOLE:2005em} and evolution to the relevant low-energy scale
\cite{Marciano:1983ss,Erler:2003yk,Erler:2004in}.  There are also parity-violating contributions arising from
the lepton vector-current coupling to the quark axial-vector-current, with couplings, $C_{2q}$, defined in a
similar manner. Although the PVES asymmetries are also dependent on the $C_{2q}$'s, they cannot be extracted
from these measurements without input from nonperturbative QCD computations.

As currently summarized by the Particle Data Group (PDG)~\cite{Yao:2006px}, existing data, particularly the
determination of the Cesium weak charge using atomic parity violation~\cite{Bennett:1999pd},  constrain the
combination of the up and down quark ``charges'', $ZC_{1u} + NC_{1d}$. Since $Z=55$ and $N=78$ for Cesium,  its
weak charge has comparable sensitivities to $C_{1u}$ and $C_{1d}$. The proton weak charge, in contrast, is more
strongly-dependent on $C_{1u}$: $Q_W^P=-2(2C_{1u}+C_{1d})$. Thus, knowldge of the two weak charges can permit a
separate determination of $C_{1u}$ and $C_{1d}$. As illustrated in Figure \ref{fig:C1qNEW}, combining the \qw
measurement with  with the previous experimental results will lead to a significant improvement in the allowed
range of values for $C_{1u}$ and $C_{1d}$. This constraint will be determined within the experimental
uncertainties of the electroweak structure of the proton.  Assuming the Standard Model holds, the resulting new
limits on the values allowed for these fundamental constants will severely constrain relevant new physics to a
mass scale for new weakly coupled physics of $\sim$2--6~TeV.

During the past 15 years much of the experimental interest in precision PVES measurements on nuclear targets has
been focussed on the strange-quark content of the nucleon.  Progress in revealing the strangeness form factors
has seen a dramatic improvement with experimental results being reported by SAMPLE at
MIT-Bates~\cite{Ito:2003mr,Spayde:2003nr}, PVA4 at Mainz~\cite{Maas:2004ta,Maas:2004dh} and the HAPPEX
\cite{Aniol:2005zf,Aniol:2005zg} and G0 \cite{Armstrong:2005hs} Collaborations at Jefferson Lab. Depending on
the target and kinematic configuration, these measurements are sensitive to different linear combinations of the
strangeness form factors, $G_E^s$ and $G_M^s$, and the  effective axial form factor $G_A^e$  that receives
${\cal O}(\alpha)$ contributions from the nucleon anapole  form factor
\cite{Haxton:1989ap,Musolf:1990ts,Zhu:2000gn}.

A global analysis \cite{Young:2006jc} of the present PVES data yields a determination of the strange-quark form
factors, namely  $G_E^s=0.002\pm0.018$ and $G_M^s=-0.01\pm0.25$ at $Q^2$ =0.1 GeV$^2$ (correlation coefficient
$-0.96$). This fit does not include the value of the neutral current axial form factor determined from neutron
$\beta$-decay, Standard Model electroweak corrections to vector electron-axial vector quark couplings, $C_{2q}$,
and theoretical estimates of the anapole contribution obtained using chiral perturbation theory. Should one
further adopt the value of $G_A^e$ obtained from these inputs\cite{Zhu:2000gn}, these values shift by less than
one standard deviation (with $G_E^s=-0.011\pm0.016$ and $G_M^s=0.22\pm0.20$). Nevertheless, even with the fits
constrained by data alone, one can now ascertain that, at the 95\% confidence level (CL), strange quarks
contribute less than 5\% of the mean-square charge radius and less than 6\% of the magnetic moment of the
proton. This determination of the strangeness form factors intimately relies on the accurate knowledge of the
low-energy electroweak parameters of Eq.~\ref{eq:LSM}. Therefore, this potential uncertainty as it relates to
our \qw measurement has turned out to be of minimal significance and is absorbed into the general fitting
procedure to separate the hadronic background from the weak charge, as described below.

A global analysis of the PVES measurements can fit the world data with a systematic expansion of the relevant
form factors in powers of $Q^2$. In this way one can make the greatest use of the entire data set, including the
extensive study of the dependence on momentum transfer between $0.1$ and $0.3\gev^2$ by the G0 experiment
\cite{Armstrong:2005hs}. By including the existing world PVES data and the anticipated results from the \qw
measurement, the two coupling constants, $C_{1u}$ and $C_{1d}$, and the hadronic background term can  be
determined by the data. Most of the existing PVES data have been acquired with hydrogen targets. For small momentum transfer, in the
forward-scattering limit, the parity-violating asymmetry can be written as
\begin{equation}
\label{eq:alrq}
A_{LR}^p \simeq A_0 \left[ Q_{\rm weak}^p Q^2 + B_4 Q^4 +\ldots \right]\,,
\end{equation}
where the overall normalization is given by $A_0=-G_\mu/(4\pi\alpha\sqrt{2})$. The leading term in this
expansion directly probes the weak charge of the proton, related to the quark weak charges by $Q_{\rm
weak}^p=G_E^{Zp}(0)=-2(2C_{1u}+C_{1d})$.  The next-to-leading order term, $B_4$, is the first place that
hadronic structure enters, with the dominant source of uncertainty coming from the neutral-weak, mean-square
electric radius and magnetic moment. Under the assumption of charge symmetry, this uncertainty translates to the
knowledge of the strangeness mean-square electric radius and magnetic moment. By considering different
phenomenological parameterizations of the elastic form factors, it has been confirmed that the potential
uncertainties from this source will have a small impact on our final result from the \qw measurement. Indeed,
the existing PVES data alone, taken over the range $0.1<Q^2<0.3\gev^2$, allow a reliable extrapolation in $Q^2$
to extract the $B_4$ $Q^4$ term contribution to the measured asymmetry.  Thus, we are confident that when the
\qw data become available, a clean extraction of the proton's weak charge will be possible.

\begin{figure}[hh!]
\begin{center}
\includegraphics[width=14cm,angle=0]{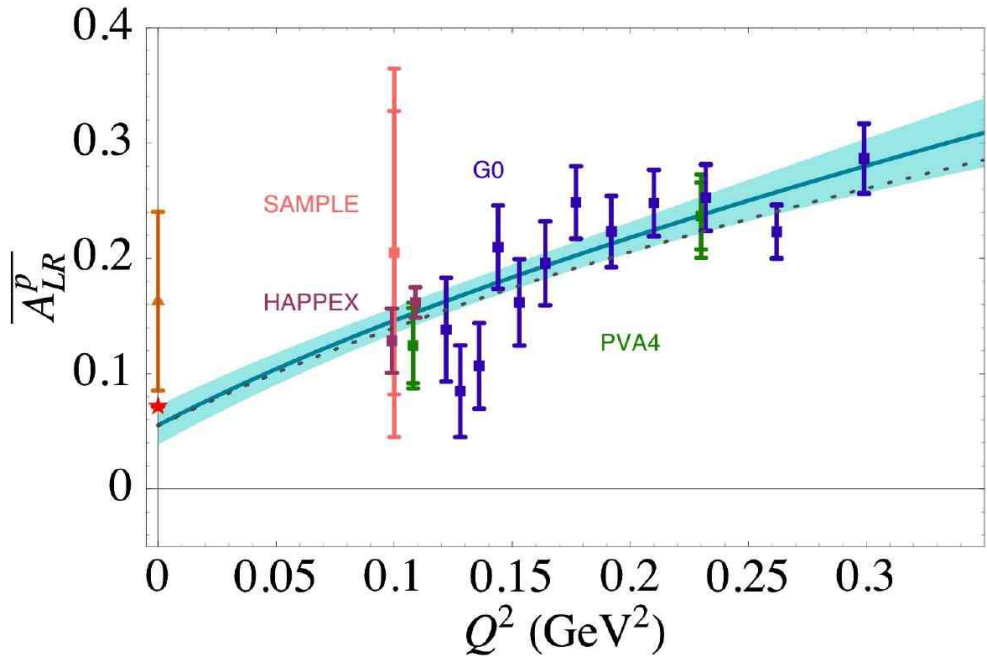}
\caption{{\em Normalized $ep$ parity-violating asymmetry measurements, extrapolated to the forward-angle limit
using
    all current world data~\cite{Young:2007zs}. The extrapolation to $Q^2=0$
    illustrates the methodology we plan to use to measure the proton's
    weak charge after the JLab \qw results are obtained. The previous
    experimental limit on \qw (within uncertainties on the neutron weak
    charge) is shown by the triangular data point, and the Standard
    Model prediction is indicated by the star. The solid curve and shaded region
    indicate, respectively, the best fit and 1-$\sigma$ bound, based
    upon a global fit to all electroweak data. The dotted curve shows
    the resulting fit if one incorporates the theoretical
    value of $G_A^e$, the effective axial vector form factor of the nucleon~\cite{Zhu:2000gn}.
    With the inclusion of the anticipated data from the \Qweak experiment at
    $Q^2$ = 0.026 $GeV^2$, a new global analysis will be able extract
    the weak charges separated from hadronic form factor
    contributions.} } \label{fig:extrap}
\end{center}
\end{figure}

Figure~\ref{fig:extrap} shows the various existing $ep$ asymmetry measurements, extrapolated to zero degrees as
explained below. The data are normalized as $\overline{A_{LR}^p} \equiv A_{LR}^p/(A_0 Q^2)$, such that the
intercept at $Q^2=0$ has the value \qw. The fitted curve and uncertainty band are the result of the full global
fits, where helium, deuterium and all earlier relevant neutral-weak current measurements
\cite{Yao:2006px,Erler:2004cx} are also incorporated.

Because the existing PVES measurements have been performed at different scattering angles, the data points
displayed in Fig.~\ref{fig:extrap} have been rotated to the forward-angle limit using the global fit of this
analysis, with the outer error bar on the data points indicating the uncertainty arising from the $\theta\to 0$
extrapolation. The dominant source of uncertainty in this extrapolation lies in the determination of the
contribution of the effective axial vector form factor $G_A^e$. The experimentally-constrained uncertainty on
$G_A^e$ is relatively large compared to computations obtained using the value of $g_A$ from neutron
$\beta$-decay plus isospin symmetry, Standard Model electroweak radiative corrections to the $C_{2q}$ couplings,
and a chiral perturbation theory computation of the anapole contribution  supplemented with a vector meson
dominance model estimate of the corresponding low-energy constants\cite{Zhu:2000gn} . Further constraining our
fits to this theoretical value for $G_A^e$ yields the dotted curve in Fig.~\ref{fig:extrap}, where the
difference with the experimentally determined (less precise) fit is always less than one standard deviation;
this effect will have a small impact on the final weak charge extraction.

The resulting measurement of the proton's weak charge by the \Qweak experiment provides an independent
constraint to combine with the precise atomic parity-violation measurement on Cesium
\cite{Bennett:1999pd,Ginges:2003qt}, which primarily constrains the isoscalar combination of the weak quark
charges.  The preliminary combined analysis using only existing data is shown in Fig.~\ref{fig:C1qNEW}. This
analysis involves the simultaneous fitting of both the hadronic structure (strangeness and anapole) and
electroweak parameters ($C_{1u,d}$, $C_{2u,d}$) and demonstrates excellent agreement with the data.
\begin{figure}[hb!]
\begin{center}
\vspace*{0.2cm}
\includegraphics[width=12cm,angle=0]{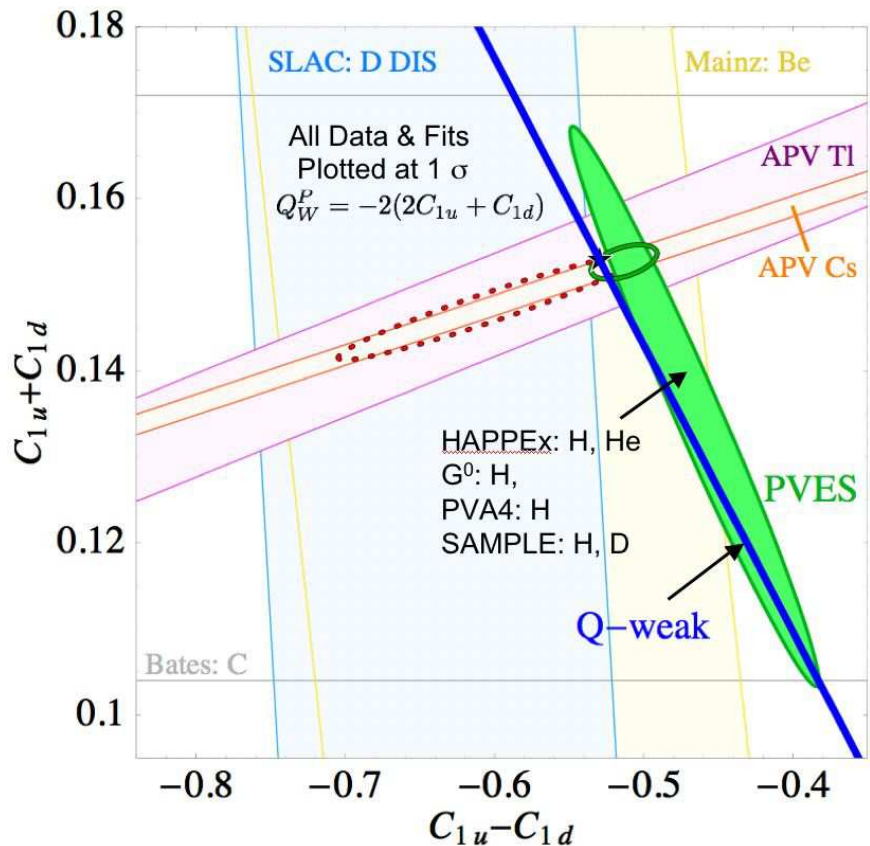}
\caption{{\em Constraints on the neutral weak effective couplings of
    Eq.~(\ref{eq:LSM}).  The dotted contour displays the experimental
    limits (95\% CL) reported in the PDG~\cite{Yao:2006px} together
    with the prediction of the Standard Model (black star). The filled
    ellipse denotes the current constraint provided by recent PVES
    scattering measurements on hydrogen, deuterium and helium targets
    (at 1 standard deviation), while the smaller solid contour (95\%
    CL) indicates the full constraint obtained by combining all
    existing results.  The solid blue line indicates the anticipated
    constraint from the planned \qw measurement, assuming the
    SM value.  All other experimental limits are
    at 1 $\sigma$.}} \label{fig:C1qNEW}
\end{center}
\end{figure}
\newpage
Whatever the dynamical origin, new physics can be expressed in terms
of an effective contact interaction \cite{Erler:2003yk},
\begin{equation}
\label{eq:lnew}
{\cal L}_{\rm NP}^{eq}=\frac{g^2}{\Lambda^2}\bar{e}\gamma_\mu\gamma_5 e \sum_q h_V^q \bar{q}\gamma^\mu q\,.
\end{equation}
With the characteristic energy scale, $\Lambda$, and coupling
strength, $g$, the values of the effective couplings $h_V^q$ will vary depending on the particular new physics scenario leading to Eq.~(\ref{eq:lnew}) \cite{RamseyMusolf:1999qk}. In the case of a low-energy E$_6$ $Z'$ boson, for example, one could expect a   non-zero value of $h_V^d$ (depending on the pattern of symmetry breaking) and a vanishing coupling $h_V^u$, whereas a right-handed $Z'$ boson would induce non-zero $h_V^u$ and $h_V^d$. More generally, in any given scenario, the values of the $h_V^q$ will determine the sensitivity of \qw to the mass-to-coupling ratio, $\Lambda/g$, which can be as large as a several TeV in some cases. The reach of the \Qweak experiment for different illustrative models is given in Table \ref{tab:newphysicsscale}.

\begin{table}[hh!]
\vspace*{-0.1cm} \caption{{\em The sensitivity of  various current and future low energy precision measurements
to the new physics scale $\Lambda$ in different models.  Also shown are the direct search limits from the
current colliders (LEP, CDF and Hera) and the indirect search limits from the current electroweak precision fit.
The various new physics scales presented here are the mass of $Z^\prime$ associated with an extra ${\rm
U}(1)_\chi$ group arising in $E_6$ models  [$m({Z_\chi})$] or in left-right symmetric models [$m(Z_{LR})$];  the
mass of a leptoquark in the up quark sector [$m_{LQ}$(up)], or the down quark sector [$m_{LQ}$(down)]; the
compositeness scale for the $e-q$ or the $e-e$ compositeness interaction. Entries with ``--'' either do not
exist or do not apply.  This Table is adapted from Ref.~ \cite{RamseyMusolf:2006vr}.}}
\label{tab:newphysicsscale} \vspace*{0.4cm}
\begin{tabular}{|c|cc|cc|cc|}
\hline
&\multicolumn{2}{c|}{$Z^\prime$ models}&
\multicolumn{2}{c|}{leptoquark}&\multicolumn{2}{c}{compositeness}\\
&$m(Z_\chi)$&$m(Z_{LR})$&$m_{LQ}$(up)&$m_{LQ}$(down)&$e-q$&$e-e$\\ \hline
Current direct search limits&0.69&0.63&0.3&0.3&--&--\\
Current electroweak fit&0.78&0.86&1.5&1.5&11$-$26&8$-$10\\
0.6\% $Q_W({\rm Cs})$&1.2&1.3&5.1&5.4&28&-- \\
13.1\% $Q_W(e)$&0.66&0.34&--&--&--&13\\
4\% $Q_W(p)$&0.95&0.45&6.5&4.6&28&--\\
\hline \hline
\end{tabular}

\end{table}
\vspace*{0.1cm}

The proposed analysis technique for the anticipated \Qweak experiment's data in conjunction with the world's
existing data on PVES demonstrates that the effect of the hadronic form factors can be separated from the
low-energy, effective weak charges $C_{1u}$ and $C_{1d}$~\cite{Young:2007zs}.  Combining the resulting constraint with that obtained
from the study of atomic parity violation data will result in an extremely tight range of allowed values for
{\bf both} $C_{1u}$ and $C_{1d}$, as illustrated in Fig.~\ref{fig:C1qNEW}. Even if the results of the \qw
measurement are in agreement with the predictions of the Standard Model, the reduction in the range of allowed
values of $C_{1u}$ and $C_{1d}$ is such that it will severely limit the possibilities of relevant new physics
below a mass scale $\sim$1--6 TeV for weakly coupled theories (see Table \ref{tab:newphysicsscale}).
Of course, it is also possible that \Qweak could discover a deviation from the Standard Model which would
constrain both the mass--coupling ratio and flavor dependence of the relevant new physics, such as a $Z^\prime$
or leptoquark. In the event of a discovery at the LHC, then experiments such as \Qweak will play a key role in
determining the characteristics of the new interaction.

\subsection{\qw:  the Standard Model Prediction and Beyond}

The prospect of the \Qweak experiment has stimulated considerable theoretical activity related to both the
interpretability of the measurement and its prospective implications for new physics. As indicated above, the
interpretability of the experiments depends on both the precision with which \qw can be extracted from the
measured asymmetry as well as the degree of theoretical confidence in the Standard Model prediction for the weak
charge. In the case of the \qw extraction, the issue is illustrated by Eq.~(\ref{eq:alrq}), indicating that one
must determine the hadronic \lq\lq B" term that describes the subleading $Q^2$-dependence of the asymmetry with
sufficient precision. The \lq\lq B"-term is constrained by the existing world data set of PV electron scattering
measurements performed at MIT-Bates, Jefferson Lab, and Mainz. Recently the authors of Ref.~\cite{Young:2007zs}
analyzed the implications of the world PVES data set in the range $0.1\ (\textrm{GeV}/c)^2\leq Q^2\leq 0.3\
(\textrm{GeV}/c)^2 $ and extrapolated the results to $Q^2=0$ to obtain the current PVES value for \be
Q_W^p=0.055\pm 0.017 \ee (present world average).  Inclusion of this planned low-$Q^2$, high-statistics
measurement by the \Qweak collaboration will reduce this extracted uncertainty to $0.003$. This error includes
both the impact of the experimental uncertainty in the hadronic``B" term, as well as the anticipated uncertainty
in $A_{PV}$.

\begin{figure}[b]
\begin{center}
\epsfig{figure=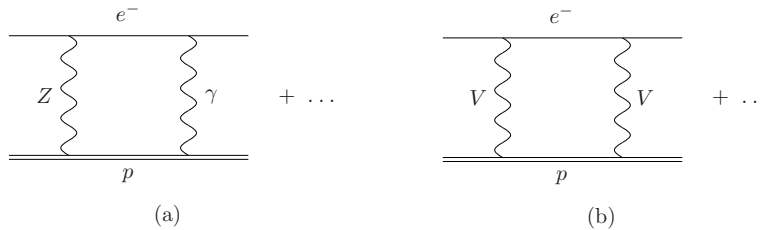,width=4.in} \caption{{\em Standard Model box graph contributions to the weak
charge of
  the proton. Panel (a) gives the $Z\gamma$ corrections while panel
  (b) gives the WW and ZZ box contributions. }}
\label{fig:box}
\end{center}
\end{figure}

The impact of a four percent determination of $Q_W^p$ depends on both the precision with which the Standard
Model value for the weak charge can be computed, as well as on its sensitivity to various possible sources of
new physics. Writing \be Q_W^p=Q_W^P(\textrm{SM})+\Delta Q_W^p(\textrm{new}), \ee the present theoretical
prediction in the SM gives\cite{Erler:2004in,Erler:2003yk}
\be
Q_W^p(\textrm{SM})=0.0713\pm 0.0008
\ee
where the uncertainty (1.1\%) is determined by combining several sources of theoretical uncertainty in
quadrature. The largest uncertainty arises from the value of the MS-bar weak mixing angle at the $Z^0$-pole:
$\Delta\sin{\hat\theta}_W(M_Z)$, followed by uncertainties in hadronic contributions to the $Z\gamma$ box graph
corrections  [see Fig. \ref{fig:box}a], hadronic contributions to the \lq\lq running" of $\sin{\hat\theta}_W(Q)$
between $Q=M_Z$ and $Q\approx 0$ , and higher order perturbative QCD contributions to the WW and ZZ box graphs
[see Fig. \ref{fig:box}b]. Charge symmetry violations rigorously vanish in the $Q^2=0$ limit and their effects
at non-vanishing $Q^2$ can be absorbed into the hadronic \lq\lq B" term that is experimentally constrained. Note
that the theoretical, hadronic physics uncertainties in $Q_W^p$ have been substantially reduced since the time
of the original \Qweak proposal. These theoretical errors are summarized in Table \ref{tab:qwerror}.

The precision with which the \qw measurement can probe the effects of new physics in $\Delta
Q_W^p(\textrm{new})$ depend on the combined experimental error in $Q_W^p$ and the theoretical uncertainty in
$\Delta Q_W^p(\textrm{SM})$. Since the anticipated experimental error $\pm 4.1\%$ is much larger than the
theoretical uncertainty in $\Delta Q_W^p(\textrm{SM})$, the sensitivity to new physics is set by the \Qweak
experimental precision. A comprehensive study of contributions from various scenarios for new physics has been
outlined in Refs.~\cite{RamseyMusolf:1999qk,Erler:2003yk,RamseyMusolf:2006vr,Erler:2004cx,Kurylov:2003zh}. The
results of those studies indicate that the \qw measurement is highly complementary as a probe of new physics
when compared to other electroweak precision measurements as well as studies at the LHC.

As a semileptonic process, PV $ep$ scattering is a particularly unique probe of leptoquark (LQ) interactions or
their supersymmetric analogs, R-parity violating interactions of supersymmetric particles with leptons and
quarks. Given the present constraints from the global set of direct and indirect searches, many LQ models could
lead to 10\% or larger shifts in \qw from its SM value, with larger corrections possible in some cases. LQ
interactions are particularly interesting in the context of grand unified theories that evade constraints from
searches for proton decay and that generate neutrino mass through the see-saw mechanism (see, {\em e.g.},
Ref.~\cite{Perez:2006hj} and references therein). Similarly, if TeV-scale R-parity violating (RPV) interactions
are present in supersymmetry, they would imply that neutrinos are Majorana particles and could generate
contributions to neutrinoless double beta-decay ($0\nu\beta\beta$) at an observable level in the next generation
of $0\nu\beta\beta$ searches. Given the present constraints on RPV interactions derived from both low- and
high-energy precision measurements, effects of up to $\sim 15\%$ in $Q_W^p$ could be generated by such
interactions\cite{RamseyMusolf:2006vr}.

\begin{table}[hb]
\vspace*{0.2cm} \caption{{\em Contributions to the uncertainty in $Q_W^p(\textrm{SM})$ \cite{Erler:2004in}.}}
\begin{center}
\begin{tabular}{|c|c|}
\hline Source & uncertainty\\ \hline
$\Delta\sin{\hat\theta}_W(M_Z)$ & $\pm 0.0006$\\
$Z\gamma$ box & $\pm 0.0005$\\
$\Delta\sin{\hat\theta}_W(Q)_\mathrm{hadronic}$ & $\pm 0.0003$ \\
$WW$, $ZZ$ box - pQCD & $\pm 0.0001$ \\
Charge sym & 0 \\
\hline
Total & $\pm 0.0008$ \\
\hline
\end{tabular}
\end{center}

\label{tab:qwerror}
\end{table}
\section{Overview of the Experiment}
\label{overview}

The \Qweak  collaboration will carry out the first precision measurement of the proton's weak charge, $Q^p_w
=1-4\sin^{2}\theta_{W}$, at JLab, building on technical advances that have been made in the laboratory's
world-leading parity violation program and using the results of earlier PVES experiments to experimentally
constrain hadronic corrections.   The experiment is a high precision measurement of the parity violating
asymmetry in elastic $ep$ scattering at $Q^{2} = 0.026 \;\;GeV^2$;  the results will determine the proton's weak
charge with $4\%$ combined statistical and systematic errors.

\begin{figure}[ht]
\vspace*{0.2cm}\begin{center}
\includegraphics[width=14cm,angle=0]{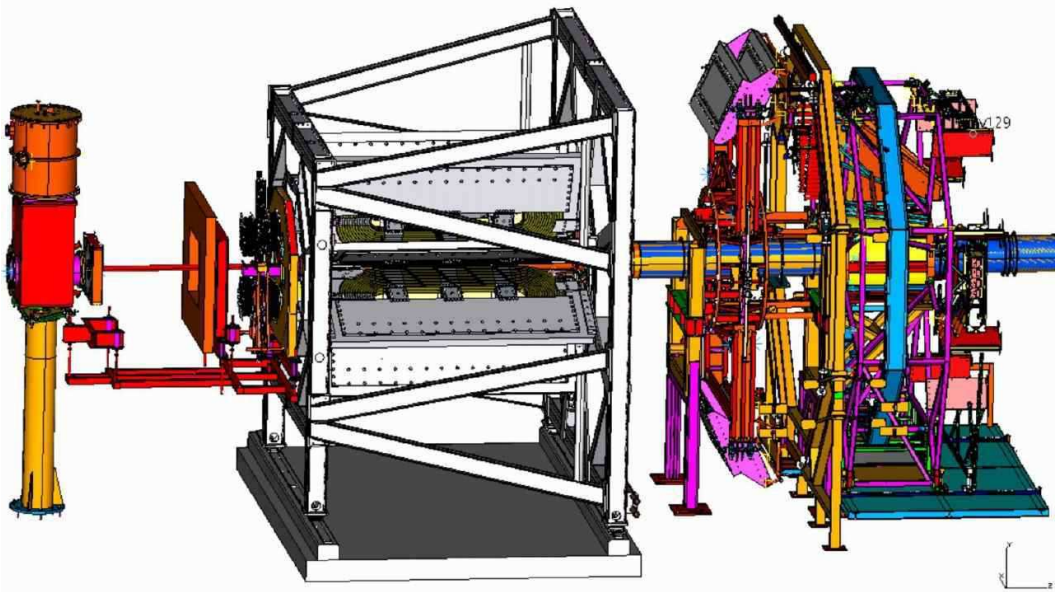}
\caption{\em CAD layout of the \Qweak apparatus. The beam and scattered electrons travel from left to right,
through the target, the first collimator,  the Region 1 GEM detectors, the two-stage second precision collimator
which surrounds the region 2 drift chambers, the toroidal magnet, the shielding wall, the region 3 drift
chambers, the  trigger scintillators and finally through the quartz \Cerenkov detectors. The tracking system
chambers and trigger scintillators will be retracted during high current running when \Qweak asymmetry data are
acquired. Luminosity monitors, (not shown) will monitor target fluctuations and provide a sensitive null
asymmetry test.  \label{fig:New_Layout}}
\end{center}
\end{figure}

A sketch showing the layout of the experiment is shown in Figure~\ref{fig:New_Layout}. The major systems of the
experiment include: A 2.5 kW $LH_2$ cryo-target system, a series of Pb collimators which define the $Q^2$
acceptance, an 8 segment toroidal magnet, 8 \Cerenkov detectors plus electronics, beamline instrumentation and
the rapid helicity reversing polarized source.  The toroidal magnetic field will focus elastically scattered
electrons onto the main \Cerenkov detectors, while bending inelastically scattered electrons out of the detector
acceptance. The experiment nominally requires 180 $\mu A$ of 1.2 GeV/c primary electron beam current with 85\%
average longitudinal polarization.

The experimental technique relies on hardware focusing of the e-p elastic electron peak onto 8 radially
symmetric synthetic quartz \Cerenkov detectors which will be read out in current mode via low gain
photo-multiplier tubes which drive custom low-noise high-gain current-to-voltage converters. Each voltage signal
is input into a custom 18-bit ADC and then read out phase locked with the reversal of the polarized beam
helicity. The asymmetry is then computed by calculating the beam helicity correlated normalized scattering rate.
To suppress noise and random variations in beam properties, a very rapid helicity reversal rate is employed.
Additional instrumentation monitors the various critical real time properties of the beam.

Basic parameters of the experiment are summarized in Table~\ref{EXTB_11}.  The production time shown includes
the time required for a separate initial 8\% measurement as well as the final 4\% result. Significant additional
beam time is required for systematics and calibration studies, as detailed later in the beam time request.

\begin{table}[htb]
\caption{\em Basic parameters of the $Q_{weak}^{p}$ experiment.} \label{EXTB_11}
\begin{center}
\begin{tabular}{lc}
\multicolumn{1}{c}{Parameter}&
\multicolumn{1}{c}{Value}\\
\hline\hline
  Incident Beam Energy          &       1.165 GeV   \\
  Beam Polarization             &       85\%                \\
  Beam Current                  &       180 $\mu$A  \\
  Target Thickness      &       35 cm (0.04$X_{0}$)     \\
  Full Current Production Running        &   2544 hours  \\
  Nominal Scattering Angle      &       7.9$^{\circ}$ \\
  Scattering Angle Acceptance   &       $\pm$3$^{\circ}$        \\
  $\phi$ Acceptance             &   49\% of 2$\pi$            \\
  Solid Angle               &       $\Delta\Omega$ = 37  msr       \\
  Acceptance Averaged $Q^{2}$   &   $<Q^{2}>$= 0.026 $(GeV/c)^{2}$ \\
  Acceptance Averaged Physics Asymmetry &   $<A>$ = -0.234 ppm           \\
  Acceptance Averaged Expt'l Asymmetry &    $<A>$ = -0.200 ppm           \\
  Integrated Cross Section      &       4.0 $\mu$b          \\
  Integrated Rate (all sectors) &       6.5 GHz (or .81 GHz per sector) \\
\hline\hline
\end{tabular}
\end{center}
\end{table}

The main technical challenges result from the small expected asymmetry of approximately -0.3 ppm; we will
measure this asymmetry to $\pm 2.1$\% statistical  and $\pm1.3$\% systematic errors. The optimum kinematics
corresponds to an incident beam energy of E$_0$ = 1.165 GeV and nominal scattered electron angle $\theta_e = 7.9
$ degrees.   Fixing $Q^2$ = 0.026 (GeV/c)$^2 $ limits nucleon structure contributions which increase with $Q^2$
and avoids very small asymmetries where corrections from helicity correlated beam parameters begin to dominate
the measurement uncertainty. With these constraints applied, the figure-of-merit becomes relatively insensitive
to the primary beam energy;  using a higher beam energy will result in a longer measuring time with stronger
magnetic field requirements, smaller scattering angles, and the possibility of opening new secondary production
channels that might contribute to backgrounds.

The high statistical precision required implies high beam current (180 $\mu$A), a long liquid hydrogen target
(35 cm) and a large-acceptance detector operated in current mode. The polarized source now routinely delivers
reasonably high beam currents at 85\% polarization;  developments for \Qweak are focusing on more reliable
operation at higher current, control of helicity correlated properties and rapid helicity reversal rates up to
500 Hz (1 ms). Radiation hardness, insensitivity to backgrounds, uniformity of response, and low intrinsic noise
are criteria that are optimized by the choice of quartz \Cerenkov bars for the main detectors. The combined beam
current and target length requirements lead to a cooling requirement of approximately 2.5 kW, considerably over
the present capacity of the JLab End Station Refrigerator (ESR). This will require us to draw additional
refrigeration capacity from the central helium liquefier (CHL), providing a cost effective solution for the
required target cooling power. We note that the combination of high beam current and a long target flask will
make the \Qweak target the highest power cryotarget in the world by a factor of several.

It is essential to maximize the fraction of the detector signal (total \Cerenkov light output in current mode)
arising from the electrons of interest, and to measure this fraction experimentally. In addition, the asymmetry
due to background must be corrected for, and we must measure both the detector-signal-weighted $<Q^2>$ and
$<Q^4>$ -- the latter in order to subtract the appropriate hadronic form factor contribution -- in order to be
able to extract a precise value for \qw from the measured asymmetry.

The $Q^2$ definition will be optimized by ensuring that the entrance aperture of the main collimator will define
the acceptance for elastically scattered events. Careful construction and precise surveying of the collimator
geometry together with optics and GEANT Monte Carlo studies are essential to understand the $Q^2$ acceptance of
the system.  This information will be extracted from ancillary measurements at low beam current, in which the
quartz \Cerenkov detectors are read out in pulse mode and individual particles are tracked through the
spectrometer system.  The \Cerenkov detector front end electronics are designed to operate in both current mode
and pulse mode for compatibility with both the parity measurements and the ancillary $<Q^2>$ calibration runs.
The tracking  system will be capable of mapping the $<Q^2>$ acceptance to $\pm 1\%$ in two opposing octants
simultaneously; the tracking chambers will be mounted on a rotating wheel assembly as shown in Figure
\ref{fig:New_Layout}  so that the entire system can be mapped in 4 sequential measurements.  The front chambers
are based on the CERN `GEM' design, chosen for their fast time response and good position resolution. The
chambers plus trigger scintillator system will be retracted during normal \Qweak data taking at high current.

The experimental asymmetry must be corrected for inelastic and room background contributions as well as hadronic
form factor effects. Simulations indicate that the former will be small, the main contribution coming from
target walls, which can be measured and subtracted. The quadrature sum of systematic error contributions to \qw,
including the hadronic form factor uncertainty, is expected to be 2.6\%.  Experimental systematic errors are
minimized by construction of a symmetric apparatus,  optimization of the target design and shielding,
utilization of feedback loops in the electron source to null out helicity correlated beam excursions and careful
attention to beam polarimetry.  We will carry out a program of ancillary measurements to determine the system
response to helicity correlated beam properties and background terms.

The electron beam polarization must be measured with an absolute uncertainty at the 1\% level.  At present, this
can be achieved in Hall C using an existing M$\o$ller polarimeter, which can only be operated at currents below
8 $\mu$A.  Work is progressing to upgrade the M$\o$ller for higher beam current operation. A major effort to
build a Compton polarimeter in Hall C at Jefferson Lab is also underway;  the Compton polarimeter will provide a
continuous on-line measurement of the beam polarization at full current (180 $\mu$A) which would otherwise not
be achievable. During the commissioning period, the new Compton will become an absolute measurement device by
calibrating it using the proven Hall C high precision M$\o$ller Polarimeter and cross checking against its
sister Compton polarimeter in Hall A.

The \Qweak apparatus also include two luminosity monitors consisting of an array of \Cerenkov detectors located
on the upstream face of the primary collimator and located downstream of the \Qweak experiment at a very small
scattering angle. The detectors will be instrumented with photomultiplier tubes operated at unity gain and
read out in current mode; the high rate of forward scattered electrons and the resulting small statistical error
in the luminosity monitor signals will enable us to use this device for removing our sensitivity to target
density fluctuations. In addition, the luminosity monitor will provide a valuable null asymmetry test,  since it
is expected to have a negligible physics asymmetry as compared to the main detector.  We will apply the same
corrections procedure for helicity correlated beam properties to both the main detectors and to the luminosity
monitor - if the systematic error sensitivities are well understood, we should be able to correct the luminosity
monitor to zero asymmetry within errors, which gives an independent validation of the corrections procedure used
to analyze the main detector data.

\begin{table}[h]
\centering \caption{\em \label{errorbudget}Total error estimate for the \Qweak experiment. The contributions to
both the physics asymmetry and the extracted \Qweakproton{} are given. In most cases, the error magnification
due to the 33\% hadronic dilution is a factor of 1.49. The enhancement for the $Q^2$ term is somewhat larger.
 }

\small
\begin{tabular}{ccc}
&&\\
{\bf Source of }&{\bf Contribution to}&{\bf  Contribution to}\\
{\bf error}&{\bf $\Delta A_{phys}/A_{phys}$  }&{\bf $\Delta$\Qweakproton /\Qweakproton }  \\ \hline
Counting Statistics& 2.1\% & 3.2\% \\
Hadronic structure  & --- & 1.5 \% \\
Beam polarimetry &  1.0 \% & 1.5\% \\
Absolute $Q^{2}$ &  0.5\% & 1.0\%  \\
Backgrounds      &  0.5\% & 0.7\%    \\
Helicity-correlated  &  &  \\
beam properties      &  0.5\%  & 0.7\%  \\ \hline
TOTAL:          & 2.5\%     & 4.1\%   \\
\end{tabular}
\end{table}

Table ~\ref{errorbudget} summarizes the statistical and systematic error contributions to the \qw  measurement
that are anticipated for the experiment. Note that the hadronic and statistical uncertainties were determined by
assuming the Standard Model asymmetry at the reference design $Q^2$=0.026 $GeV^2$. The actual asymmetry
precision and hadronic uncertainty will be affected slightly by the $Q^2$ (incident beam energy) the final world
PVES data set used (additional results are anticipated from G0 ``backwards" and PVA4) and the degrees of freedom
allowed in the global fit.
\section{Qweak Magnetic Spectrometer}

A key component of the \Qweak apparatus is the magnetic spectrometer `QTOR', whose toroidal field will focus
elastically scattered electrons onto a set of eight V-shaped, rectangular in cross section synthetic quartz
\v{C}erenkov detectors. The axially symmetric acceptance in this geometry is very important because it reduces
the sensitivity to a number of systematic error contributions.  A resistive toroidal spectrometer magnet with
water-cooled coils was selected for \Qweak{} because of the low cost and inherent reliability relative to a
superconducting solution.

\begin{figure}[h!]
\begin{center}
\vspace*{0.2cm}
\includegraphics[width=150mm]{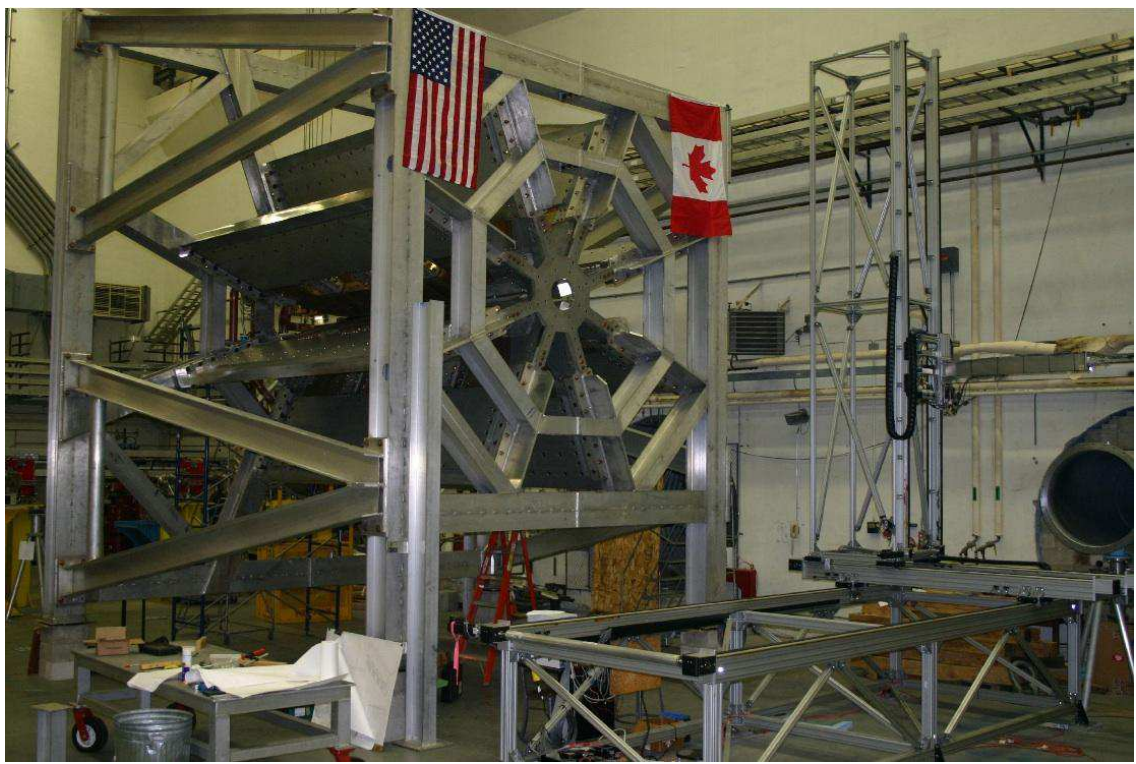}
\caption{{\em \Qweak spectrometer magnet and G0/TRIUMF field mapper at MIT-Bates. }} \label{fig:magnetpic}
\end{center}
\end{figure}

The coil geometry was optimized in a series of simulation studies using GEANT plus numerical integration over
the conductor's current distributions to determine the magnetic field.  The simplest and least expensive QTOR
coil design that  meets the needs of the \Qweak{} experiment is a simple racetrack structure.  Each coil package
consists of a double pancake structure, with each layer consisting of two, 2.20 m long straight sections, and
two semicircular curved sections with inner radius 0.235 m and outer radius 0.75 m. The copper conductor has a
cross section of 2.3 in by 1.5 in with a center hole of 0.8 in in diameter. The total DC current under operating
conditions will be 8650 A at 146 V.  A GEANT Monte Carlo simulation was used to study the effects of coil
misalignments on the $Q^2$  distribution at the focal plane as well as on the symmetry of the 8-octant system as
required for systematic error reduction. The simulation results have been used to set coil alignment and field
uniformity requirements for the assembly of the spectrometer.

The \Qweak magnetic spectrometer and support structure were assembled at MIT-Bates in the spring and summer of
2007, as shown in Figure \ref{fig:magnetpic}.   After assembly of pairs of pancakes in their coil holders, all
eight coils were placed in the main magnet frame and aligned using the QTOR survey monument system installed in
the assembly hall at MIT-Bates.  As the coils were nearly touching along the radial inside edges, the decision
was made to move the coils radially outward by 0.50 inches with respect to the original design. This has some
negative effect on the focal plane image, but simulations have shown that the good focus can be maintained by
slightly increasing the magnetic field strength, within the overhead of the power supply specifications.  An
updated field map for QTOR was generated using a custom Biot-Savart numerical code in the spring of 2007. This
new field map incorporates as-built dimensions of each individual coil and also the modified radial coil
positions noted above.  The field map covers a very large volume which extends well beyond the physical limits
of the magnet. The full field map has been incorporated in the \Qweak GEANT simulation package.

The next steps in commissioning the \Qweak spectrometer are to test the magnet at high power and obtain a
precise experimental field map. Once the power supply has been delivered, which is anticipated in December 2007,
it will be connected to the AC power feed, QTOR, and the cooling water. Magnetic field measurements and fine
alignments of the coil positions will then follow. MIT has negotiated a new rate structure with the local power
utility company which will allow us to test the QTOR power supply to the full design current while the power
supply is still under warranty, and will allow the coils to settle into their fully energized positions before
precision field mapping and adjustments to the coil alignments take place. Proceeding in this manner will reduce
the uncertainties during installation in Hall C at JLab with its inherent pressure for time.

\subsection{QTOR Magnetic Verification}

A magnetic field mapping apparatus, built by the Canadian group for the G0 experiment, will be employed to map
the QTOR spectrometer field.   Two types of measurements will be made to assess the QTOR magnetic field.
Initially, the spatial current distribution in the eight coil windings will be ascertained by using the
zero-crossing technique developed for G0, as described below, and adjustments will be made to the individual
coil positions as necessary.   Subsequently, absolute field strengths will be determined at selected points
along the central electron trajectories to verify that the associated $\int \vec{B}.d\vec{\ell}$ is matched to
the required 0.4\% for all sectors.

For the G0 experiment, an automated field measuring apparatus was used to determine the locations of a set of
the zero-crossing locations of specific field components at selected points of symmetry of the magnet.
Determination of these zero-crossing points then allowed the determinations of the actual coil locations and
hence, in principle, the complete specification of the magnetic field. The system is capable of providing an
absolute position determination of $\pm$0.2 mm, and a field determination of $\pm$0.2 Gauss, in order to resolve
a zero-crossing position to within $\pm$0.3 mm. The field mapping system consists of a programmable gantry with
full 3D motion within a (4 x 4  x 2) m$^3$ volume, and a set of high precision Hall probes, thermocouples and
clinometers  mounted on the end of a probe boom on the gantry.

The objective of the zero-crossing measurements for Q$_{weak}$ is to determine all coil positions to the
required tolerances of $\pm 1.5$ mm and coil angles to $\pm 0.1^\circ$.  The analysis program originally
developed for G0 to extract the coil positions from zero crossing measurement data has recently been reworked.
For G0, the analysis procedure used to extract the coil positions from the measured zero-crossing points was
tested against computer simulations, where known coil displacements were used to generate simulated data. Not
only were the `displaced' coil positions correctly extracted, but the relative orientations and positions of the
Hall probes themselves could also be extracted. Analysis of the experimental data resulted in a full
determination of the residual coil displacements for all 8 coils. Initial tests on the code as modified for
Q$_{weak}$ with simulated zero-crossing displacements provided excellent reproduction of the actual coil
displacements imposed in software.
Results are illustrated in Figure \ref{zx}, lending confidence in the technique for \Qweak. \\

\begin{figure}[h]
\centerline{\psfig{figure=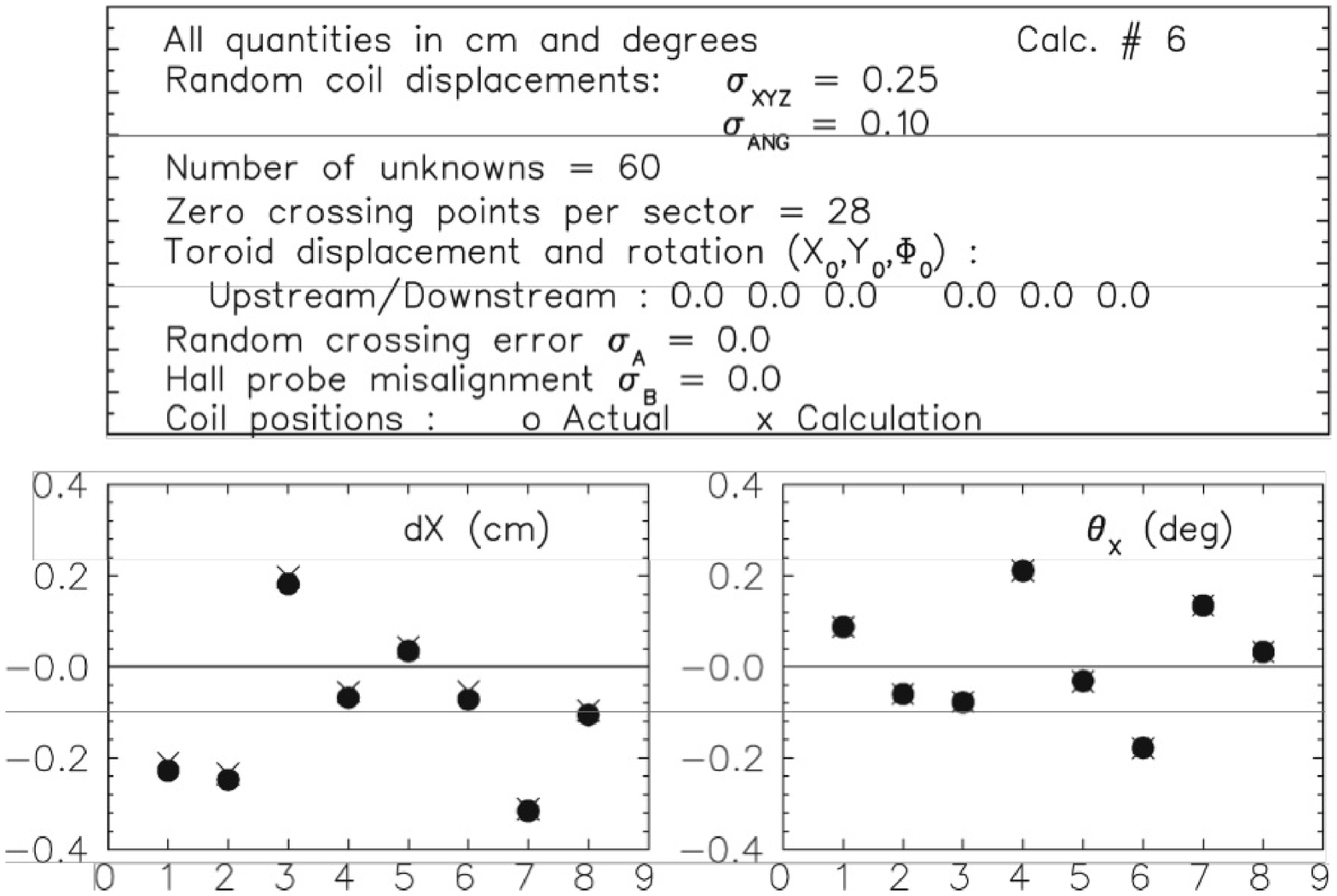,height=5.8cm} \hspace*{0.02cm} \psfig{figure=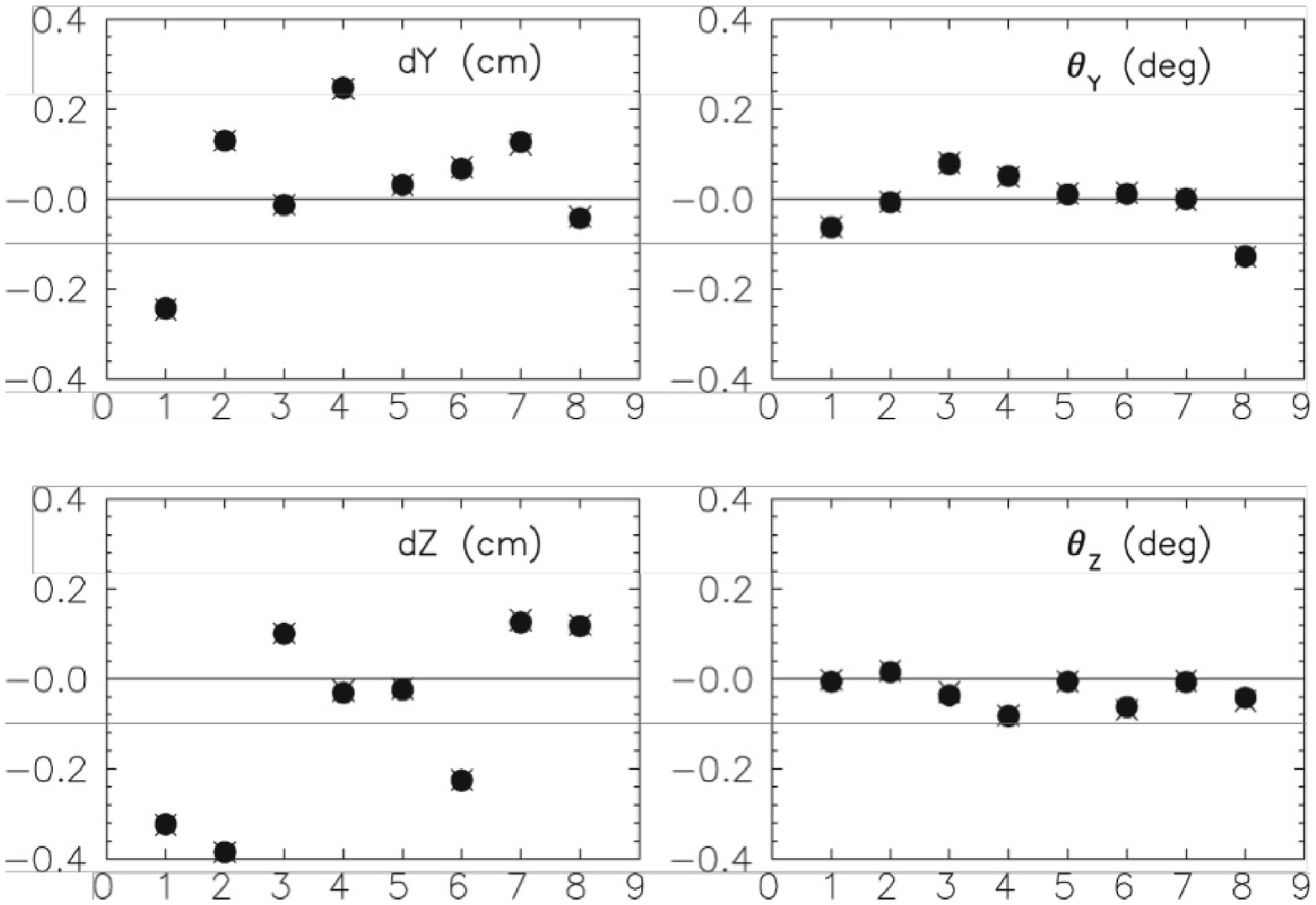,height=5.8cm} }
\caption{\em  Illustration of the zero-crossing technique, tested in software, for extracting coil positions
from the QTOR field mapping data.  Simulated input coil displacements (solid circles) and positions fitted to
the zero-crossing data (crosses) are shown here for all 8 coils.  } \label{zx}
\end{figure}

The G0 field mapper was shipped from UIUC, and it arrived at MIT-Bates in early August, 2007. Following this,
TRIUMF and U.Manitoba personnel travelled to MIT-Bates to reassemble and recommission the system. The gantry
motion was tested and appeared to be moving smoothly, the magnetic field sensors were reading out correctly, and
the control software appeared to be working as expected.  Updated collision-avoidance software was installed,
but has not been fully tested yet against the proposed zero-crossing points.

In mid-October 2007, TRIUMF and U.Manitoba personnel again travelled to MIT-Bates to tune the gantry electronics
and to recalibrate the gantry motion using a laser-tracker. At this time, the field mapper has essentially been
recommissioned.  Depending on the arrival date and installation of the QTOR power supply, the full magnetic
verification measurements are expected to begin in the spring of 2008.
\section{Detector System and Low Noise Electronics}

The \Qweak main detector collects \Cerenkov light produced by electrons passing through thin, fused silica
(synthetic quartz) radiators. After many bounces, photons reach the ends of the rectangular bars by total
internal reflection, and are then collected by 5'' PMT's with UV-transmitting windows. The distribution of
elastic and inelastic tracks at the focal plane is shown in Figure \ref{MDenvelope}. While the signal during
parity violation measurements is the DC anode current, for background and systematic checks at very low
luminosities we will increase the PMT gains for pulsed-mode data taking.

\begin{figure}[h]
\begin{center}
\rotatebox{0.}{\resizebox{6.0in}{2.5in}{\includegraphics{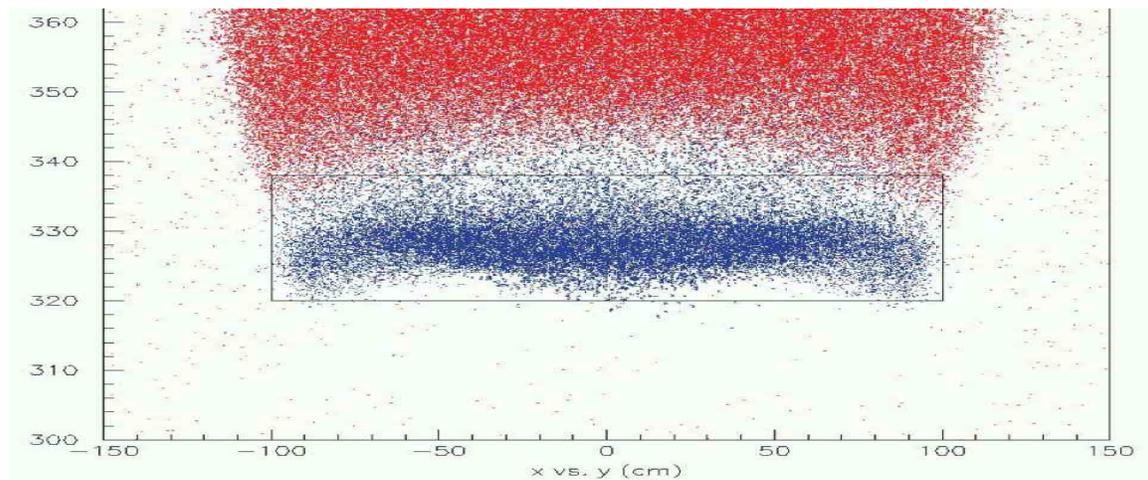}}}
 \end{center}
\caption{{\em Dotplot showing the approximate distributions of the elastic (blue) and inelastic (red) events
with respect to a radiator bar. The properly weighted fraction of inelastic to elastic tracks on a bar is
0.04\%.}} \label{MDenvelope}
\end{figure}

In the following sections, we summarize progress since our last Jeopardy proposal on the design and construction
of the optical assemblies, the PMT and voltage dividers, and our understanding of the detector's expected
performance.

\subsection{Progress on the Optical Assembly}

\paragraph*{Final Radiator Specifications:}

We limited the radiator length to 200 cm to allow all the detectors to lie in a single plane without
interference between adjacent octants. But because a single 200 cm long bar would have cost 4 times as much as a
single 100 cm long bar, we ordered pairs of 100 cm long bars which will have to be glued.

The quartz bar thickness was carefully optimized to minimize excess noise. Excess noise is a scale factor which
multiplies the statistical error of the experiment calculated assuming that all electrons are detected with the
same weight. If this factor is 1.01, for example, then the experiment would have to run 2\% longer (or more
efficiently) to achieve the same statistical error as if there were no excess noise. Simulations showed that
radiators that are too thin have poor resolution due to low average photoelectron yields, while radiators that
are too thick have poor resolution due to shower fluctuations. (Low light production from electrons traversing
edges of the bars was also taken into account.) We ordered bars of 1.25 cm thickness, half the thickness of our
prototype bars, because this was near the minimum excess noise of 3.8\% \cite{Gericke}. The width in the bend
direction was also changed from 16 cm to 18 cm to
 capture more of the elastic beam envelope.

The net effect of these changes in radiator geometry allowed us to reduce the material costs by 50\% while
keeping the error bar on \qw constant. The final dimensions for one active radiator are 200 cm x 18 cm x 1.25
cm. To keep the PMT's away from the scattered beam envelope, UV-transmitting lightguides of dimensions 18 cm x
18 cm x 1.25 cm are attached to each end. A complete optical assembly for one octant of the spectrometer
therefore has dimensions of 236 cm x 18 cm x 1.25 cm. Parameters of the optical assemblies are summarized in
Table \ref{BARspecs}.

\begin{table}[h]
\centering \caption{\label{BARspecs}{\em Updated parameters for the optical assembly. The bar tilt angle is
measured with respect to the vertical.}} \small \vspace*{0.2cm}
\begin{tabular}{c|c}
  Parameter             &    Value                              \\ \hline
  shape                 & rectangular solid                                    \\
 radiator size          & 200 cm (L) x 18 cm (W) x 1.25 cm (T)                 \\
 optical assembly size  & 236 cm (L) x 18 cm (W) x 1.25 cm (T)                 \\
 radiator material      & fused silica: Spectrosil 2000                        \\
 lightguide material    & fused silica: JGS1-UV                                \\
 glue                   & Shin-Etsu Silicones SES406                           \\
 expected excess noise  &  3.8\%                                                \\
 detector position      & Z = 570 cm downstream of the magnet center            \\
                        & R = 319 cm from the beam axis (inner edge)             \\
 bar tilt angle         & 0 degrees              \\ \hline \hline
\end{tabular}
\end{table}
\normalsize

\paragraph*{Procurement and Quality Control:}

      The radiator bars were procured from St.\ Gobain Quartz. Delivery was completed in fall,
2006. The bars are made of Spectrosil 2000, which is an artificial fused silica with low fluorescence and
excellent radiation hardness due to the very low concentration of impurities such as iron. The ingot foundry for
Spectrosil 2000, as well as the polishing subcontractor, are in England.
About 2/3 of the \$300K cost was for labor for the optical grade polish needed to ensure a total internal
reflection coefficient of 0.996. Overall dimensions, flatness, and polish quality appear to be within
specifications\cite{Elliott}. The bevels on some bars were occasionally wider than even our relaxed
specification of 1 mm $\pm$ 0.5 mm, but simulations showed that the resulting loss of photoelectrons would be
modest.

      The lightguides were procured from Scionix and delivery was completed early 2007.
They are made of a Chinese brand of artificial fused silica termed JGS1-UV. According to BaBar DIRC group
tests, this brand is equal or superior in quality to Spectrosil 2000 for the small pieces that were
tested. The delivered lightguides have tiny scattered pits, but these occupy too small a fraction of the surface
area to cause observable deterioration in performance.

Scintillation in the radiator bars could potentially cause a dilution of the elastic $e-$ light yield from the
absorption of x-ray backgrounds. To verify that our batch of Spectrosil 2000 had the expected low scintillation
coefficient, we had to expand upon a technique previously used at SLAC. The basic idea\cite{SLAC}
 was to use a 300
$\mu$Ci $^{55}Fe$ source which produces a 6 keV x-ray. Because this x-ray energy is far too low to produce
\Cerenkov radiation via Compton scattering, any prompt light production by these x-rays must be due to
scintillation. Unfortunately, source-in minus source-out rate measurements did not produce results which were
precise enough for our needs, due  to  instability in the dark rate of the PMT. We dramatically improved our
sensitivity by chopping the x-rays and detecting the rate modulation in a spectrum analyzer. After normalizing
the result, we found a non-zero scintillation coefficient of order 0.01 photons/MeV\cite{Mack}. Combining this
coefficient with the simulated x-ray background spectrum, the dilution of our PV signal from scintillation
should be negligible even if the background were several orders of magnitude larger. However, we are not reliant
on simulations for the background spectrum, as we will measure it directly during the experiment.

\begin{figure}[h]
\begin{center}
\rotatebox{0.}{\resizebox{2.5in}{3.75in}{\includegraphics{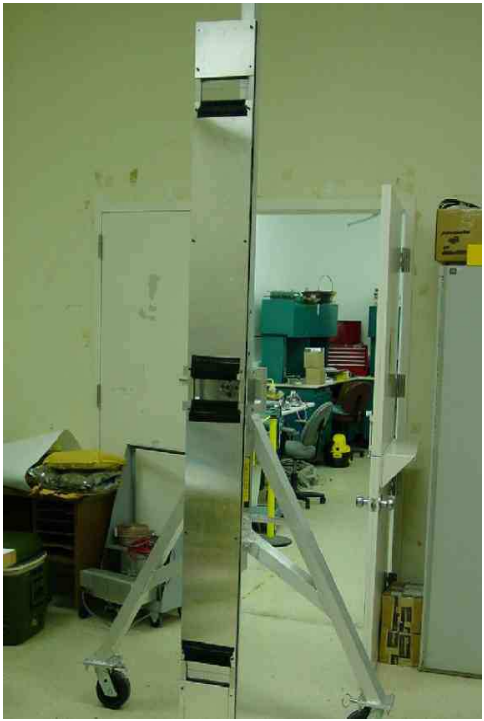}}}
 \end{center}
\caption{{\em Gluing jig in the vertical position. The height is approximately 2.5 meters.}} \label{gluejig}
\end{figure}


\paragraph*{Gluing:}

    Our central glue joint has to be reasonably strong and UV transparent even after a dose
of 100 kRad. After considering several glues, we found that Shin-Etsu Silicones SES-406 cured into a tough
material which adhered extremely well to even our ultra-smooth quartz surfaces\cite{Patrick}. Spectrophotometer
measurements were then made of the transmission through two glued slides of Spectrosil 2000, over the wavelength
range 250-500 nm.
 The glue joint was found to absorb less than 2\% of the light below 350 nm, and was completely
transparent above 350 nm. Since the light only crosses 2-3 glue joints in the optical assembly (the PMTs will
also be glued to the  lightguides), this degree of transmission is very satisfactory.

We then tested whether the light transmission would suffer radiation damage during the experiment. The glued
slides and several experimental controls were sent to Nuclear Services at NC State for irradiation to 100 kRad.
On return of the samples, we found no additional loss in transmission. We then increased the total dose to 1.1
MRad, to make sure the glue would still pose no problem if a pre-radiator were used. Again, no deterioration in
transmission was seen at the level of $\pm$0.1\%. These tests were completed in August 2007, and are partially
summarized in Reference \cite{Katie}.

Because the main detector must transmit photons down to 250 nm in wavelength, we could not use convenient,
quick-hardening UV-catalyzed glues. We therefore designed and built a gluing jig to hold the long quartz pieces
in alignment during a 24 hour curing period, as illustrated in Figure \ref{gluejig}. Our first attempts used
full-scale plastic models of the quartz pieces. After modifying our procedures and our equipment to minimize
potential damage to the expensive fused silica elements, we finally began gluing quartz in mid-October, 2007.
When our Yerevan collaborators return to JLab, we should be able to finish one complete optical assembly every
few days. Since we are only manufacturing a total of 9 assemblies (8 for production data-taking and one hot
spare), completion of all optical assemblies will take less than one month.


\subsection{Progress on PMT's}

Updated parameters for the PMT signal chain are given in Table \ref{PMTspecs}. The few changes from the last
proposal are due to our production bars being half as thick as the prototype bars, hence dropping the
photoelectron yield by a factor of 2. The nominal gains were increased by the same factor to keep the signal
magnitudes into the ADC approximately the same.

\begin{table}[h]
\centering \caption{\label{PMTspecs}{\em Updated parameters for the PMT signal chain. }} \small
\begin{tabular}{c|c}
                        &                                                      \\
  {\bf Parameter }      &  {\bf  Value}                                        \\  \hline
{\bf current mode:} &                                                       \\
   $I_{cathode}$        &  3 nA                                                 \\
    gain                &  2000                                           \\
  $I_{anode}$           &  6 $\mu$A                                       \\
  non-linearity (achieved) &  5$\times10^{-3}$                                                  \\
                        &                                                       \\
{\bf pulsed mode:} &                                                       \\
   $I_{cathode}$        &  3.2 pA at 1 MHz                                  \\
    gain                &  2$\times 10^6$                                           \\
  $I_{anode}$           &  6.4 $\mu$A                                       \\
  $V_{signal}$ (with x10 amp.)
                        &  16 mV for 1 pe; \hspace{.25cm} 320mV for 20 pe       \\
  non-linearity (goal)  & $<$ $10^{-2}$                                                   \\ \hline \hline
\end{tabular}
\end{table}
\normalsize

\paragraph*{Tube Procurement and Quality Control:}

The Electron Tubes D753WKB is a short, low-cost PMT with a 5'' diameter UVT glass window which allows detection
of UV \Cerenkov photons down to an effective low wavelength cutoff of 250 nm. These tubes have SbCs dynodes for
improved dynode stability, and custom S20 cathodes to minimize potential nonlinearities at our high cathode
current of 3 nA. These high cathode currents limit the maximum PMT gain to $\cal O$(1000). The 6 $\mu$A signals
are converted to 6 V  signals using low-noise preamplifiers before being sent on cables to upstairs ADC's. After
a painless procurement, all 28 PMT's had arrived by summer 2005\cite{Mack2} and were all checked for basic
functionality by January 2006\cite{Gericke2}.

\paragraph*{Voltage Dividers:}

At this time, the design of our current mode divider is complete and ready for procurement. We have a promising
pulsed-mode divider design but need to do one more test before it can be finalized. Details are discussed below.

\underline{Current-Mode Divider}

Our current-mode divider has to provide the nominal gain of 2000 with low noise, high linearity, and good
operational flexibility. After some initial modelling using test-ticket parameters, we began using the 5th
dynode as an anode in an attempt to optimize the linearity at this unusually low gain. Techniques for measuring
gain and linearity were developed, several prototypes were studied\cite{Mitchell}, and a satisfactory 7-stage
design was finally selected at the end of summer 2007.

The left panel of Figure \ref{PMTperformance} shows that although the nominal gain is 2000, we can vary the gain
from 500 to 16,000. This huge dynamic range will allow considerable freedom in remotely adjusting the gain.
The dark current for the 7-stage design is shown on the right panel of Figure \ref{PMTperformance}. The
corresponding signal dilution is at most 0.05\%, and possibly negligible if the dark current is stable enough to
be treated as part of the ADC pedestal.
Preliminary measurements show the nonlinearity to be 0.5\% at the nominal 6 $\mu$A operating load. Although this
is higher than our goal of 0.1\%\cite{Mack3}, it is acceptable since the most important corrections (distortion
of the charge and physics asymmetries) would still be relatively small.

\begin{figure}[h]
\begin{center}
\subfigure[]{\includegraphics[width=0.48\textwidth]{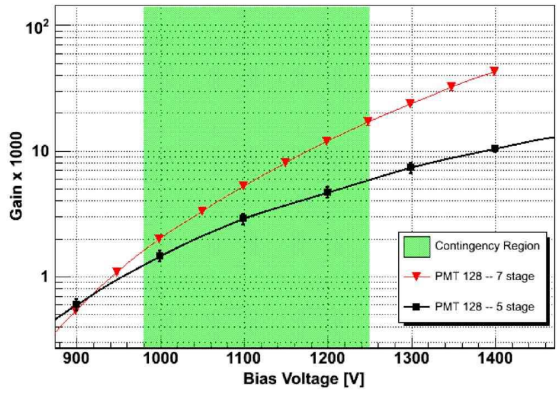}}
\subfigure[]{\includegraphics[width=0.48\textwidth]{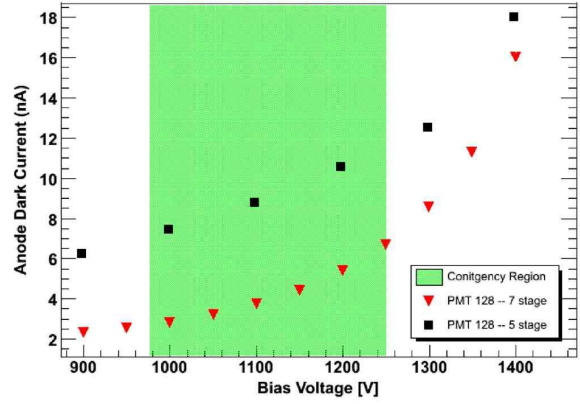}}
 \end{center}
\caption{{\em Left: Gain vs Voltage for 5- and 7-stage prototypes. Right: Dark Current vs Voltage for 5- and
7-stage prototypes. (In both cases, the green marks the operating range for the chosen 7-stage design. The PMT
selected for these measurements has a representative performance.)} }\label{PMTperformance}
\end{figure}

\underline{Pulsed Mode Divider}

    Our high-gain dividers will only be used for event-mode tests at very low luminosities.
One important application will be to provide discriminated radiator signals during tracking-based acceptance
studies. Another important application will be in bias-free background studies in which we use flash ADC's to
acquire 1 second long buffers of 100\% live radiator signal history. Because some backgrounds will take the form
of single photoelectron pulses, it is important that we be able to resolve single photoelectrons in our
event-mode ADC's. The gain of our 10-stage PMT's is too low for this purpose (2$\times$10${^6}$), so the anode
signals will be amplified with external PS777 units owned by Hall C. A batch of noisy zener diodes slowed down
our prototyping efforts at the beginning of summer 2007, but by the end of the summer we had both good cosmic
ray pulses and a quiet baseline. Once we have proven that we can resolve single photoelectrons in the upstairs
counting house, we will begin procurement of these dividers as well.


\subsection{New Detector Performance Studies}

{\bf Optimizing the Radiator Tilt Angle}

While both PMT's detect some light from every track, the average number of photoelectrons and the uniformity of
the sum of the two ends is sensitive to the tilt angle as defined in the left panel of Figure \ref{PEvsTilt}. To
optimize the tilt angle, we used a GEANT model which was previously benchmarked using cosmic and in-beam test
data with half-size prototypes\cite{Carlini}.

\begin{figure}[h!]
\begin{center}
\rotatebox{0.}{\resizebox{6.in}{2.5in}{\includegraphics{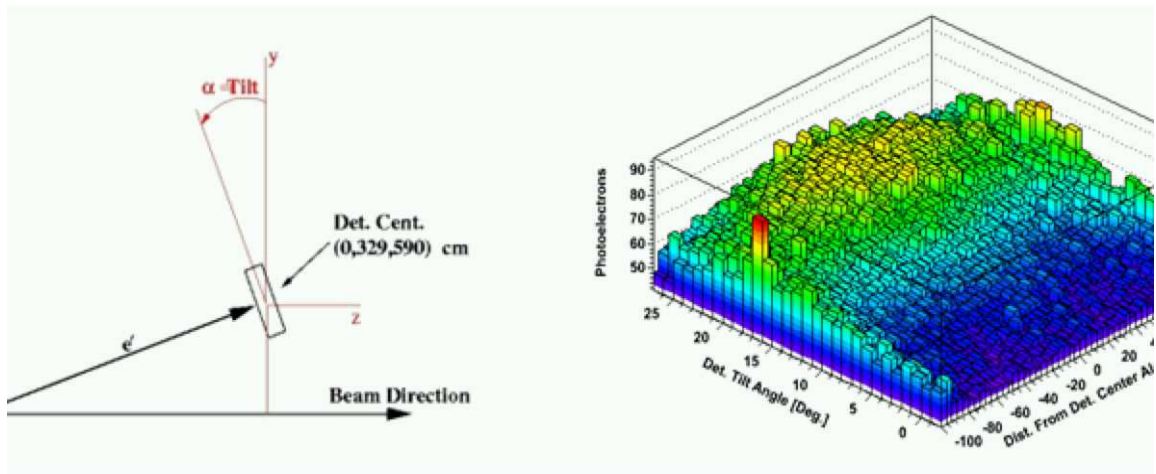}}}
 \end{center}
\caption{{\em Left: Definition of the detector tilt angle. Right: Simulation of photoelectron number versus bar
longitudinal coordinate for different tilt angles.}} \label{PEvsTilt}
\end{figure}

Results of the simulation are shown in the right panel of Figure \ref{PEvsTilt}, in which average photoelectron
number is plotted versus the tilt angle and a coordinate along the length of the bar. When the electrons are
close to normal incidence on the bar (about 23 degrees in this coordinate system), one obtains the highest
average photoelectron number. This position also minimizes the excess noise, but the uniformity of the
collection is not good. The best uniformity, with an adequate average photoelectron number and only 1.5\% higher
excess noise, is found when the radiator is facing nearly 
vertically (0 degrees). We chose this more uniform
configuration to control the magnitude of a systematic correction discussed immediately below.

\paragraph*{Detector Bias:}
This section deals with the somewhat subtle issue of detector biases in a precision, integrating experiment that
determines the PV asymmetry from an average over the $Q^2$ acceptance of the apparatus, weighted by the light
yield in the \Cerenkov detectors.  From the left panel of Figure \ref{Q2bias}, one sees that lower $Q^2$ events
are focused more toward the central half of the detector bar. Because the parity violating asymmetry is
approximately proportional to $Q^2$, this means that any nonuniformity in the detector response along the bar
will bias the asymmetry from the average value a naive Monte Carlo would predict if it were weighted by the
number of scattering events.  We will correct this by measuring the detector response during the experiment
using the Region III chambers. Here we estimate the magnitude of the effect.

\begin{figure}[h!]
\begin{center}
\subfigure[]{\includegraphics[width=0.53\textwidth]{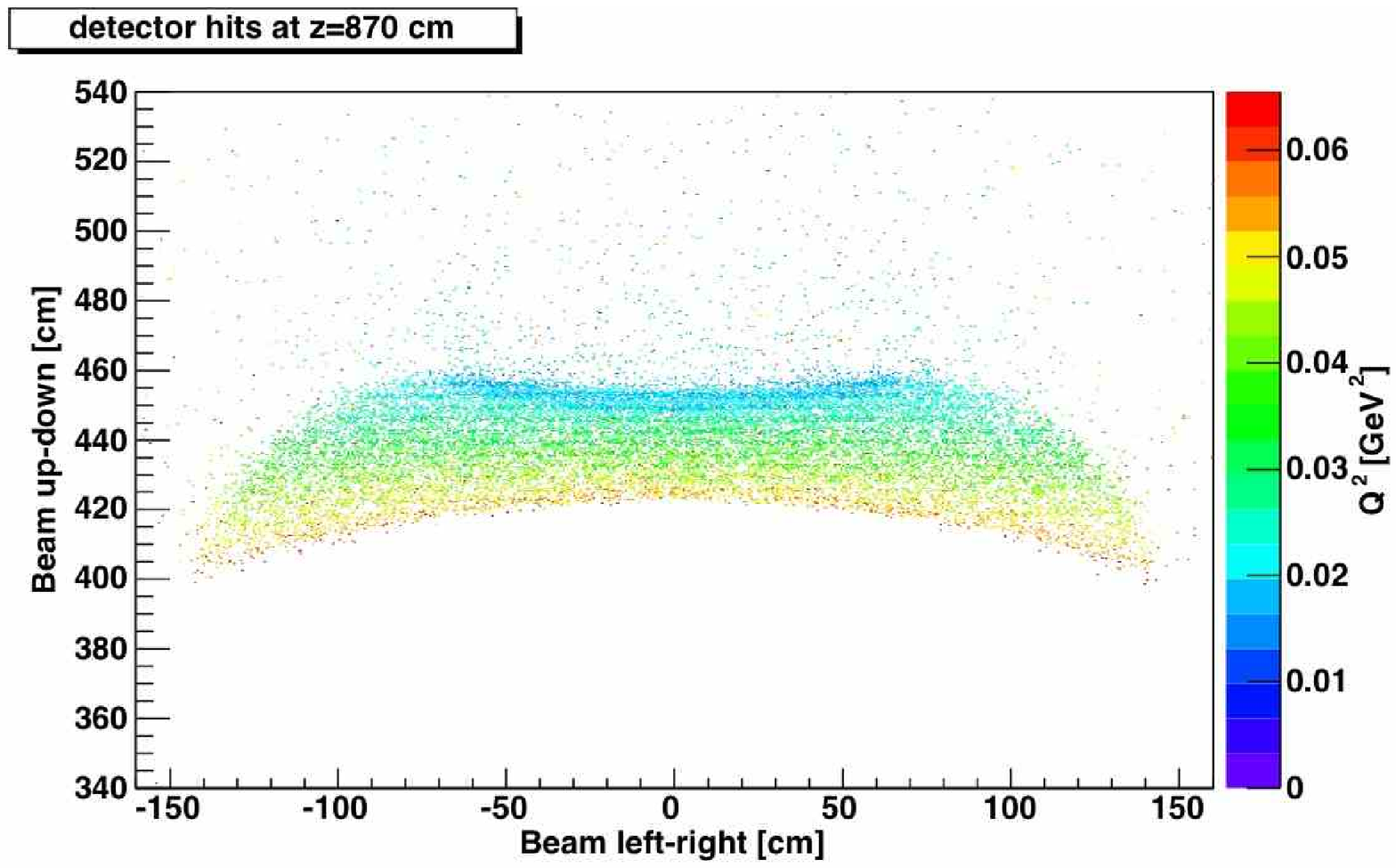}}
\subfigure[]{\includegraphics[width=0.44\textwidth]{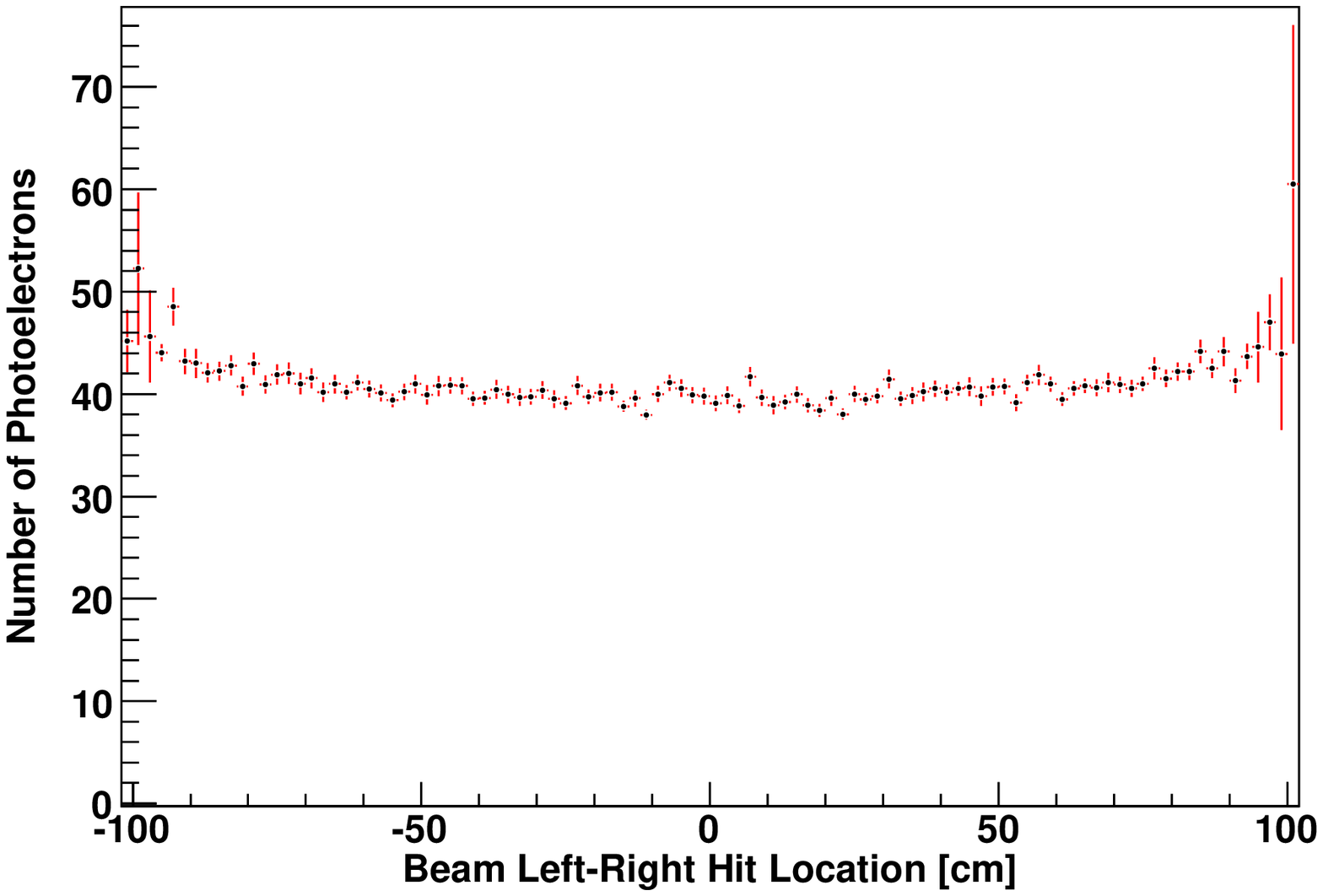}}
\end{center}
\caption{{\em Left panel: At a point 3 meters downstream from the $Q^2$ focus on the main detector, it is easy
to see that the lower $Q^2$ events are more focused toward the center half of the radiator. Right panel:
Simulation of photoelectron number versus bar longitudinal coordinate for the nominal tilt  angle of zero
degrees. Tracks near the ends of the bars receive about 5\% higher weight than those near the center of the
bars. }} \label{Q2bias}
\end{figure}

For the nominal tilt angle of 0 degrees, the predicted average photoelectron number versus position along the
bar is given in the right panel of Figure \ref{Q2bias}. The response is uniform to within 5\%, with slightly
less weight being given to the lower $Q^2$ events which are concentrated in the center of the bars. When we
include this bias in our simulation, we find the detector would measure an asymmetry (or average $Q^2$) which is
2.5\% higher than if the detector bias were not taken into account. However, the contribution to our final error
bar will be negligible since we will accurately map out the detector
 response with the Region III chambers. The fact that our detectors are radiation-hard will
greatly simplify this effort.

\paragraph*{Costs vs Benefits in Using a Pre-radiator:}

The decision of whether to use a pre-radiator involves a trade-off between statistical and systematic errors.
For a thin electron detector as we have chosen for \Qweak, soft backgrounds can be reduced by using a
pre-radiator. Not only would the pre-radiator amplify the electron signal by showering (Figure \ref{planB} left
panel), but it would also attenuate soft background. However, while the Signal/Background ratio would definitely
improve by at least an order of magnitude, shower fluctuations would lead to an increase in the statistical
error of the experiment. Here we estimate the cost of using a pre-radiator, expressed in units of lost beam
hours.

\begin{figure}[h]
\begin{center}
\rotatebox{0.}{\resizebox{6.in}{2.5in}{\includegraphics{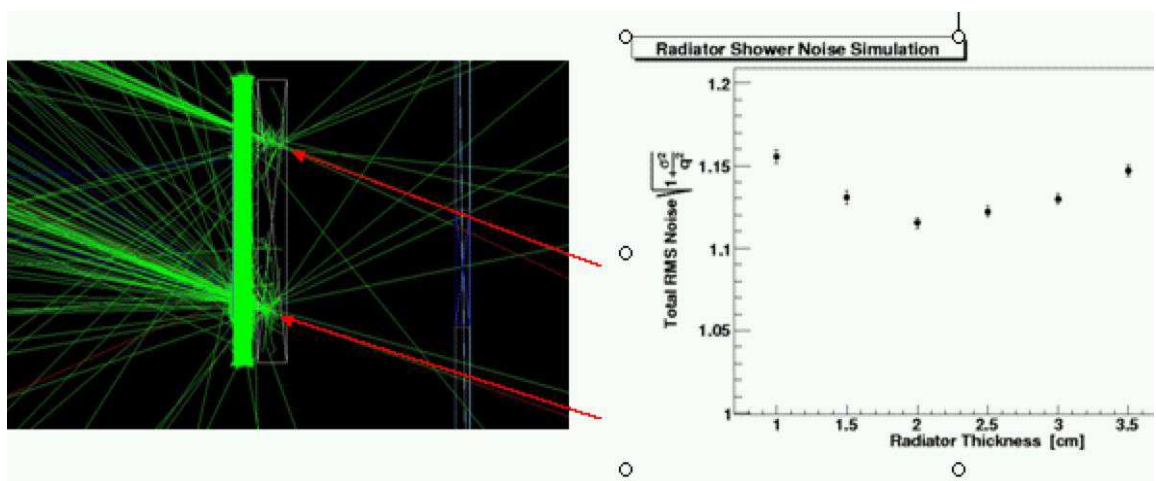}}}
 \end{center}
\caption{{\em Left panel: Simulation showing shower production by a lead pre-radiator located before the fused
silica \Cerenkov radiator. Right panel: Excess noise versus thickness of the pre-radiator, showing a minimum of
12\% for an optimal ``shower-max'' pre-radiator consisting of 2 cm of lead. }} \label{planB}
\end{figure}

We simulated the excess noise as a function of pre-radiator thickness as shown in the right panel of Figure
\ref{planB}. The fluctuations are minimized for a 2 cm thickness of lead corresponding to shower-max at our beam
energy. The minimum excess noise with an optimal pre-radiator is 12\%, which would represent a loss of about 350
beam hours when compared to our nominal detector with no pre-radiator.\footnote{The excess noise would be
smaller at significantly higher beam energies.}

In conclusion, since our expected soft backgrounds are only a few times 0.1\%, and the cost of using a
pre-radiator in beam hours is significant, we do not plan to use a pre-radiator. However, the detector housing
will contain mounting brackets for the heavy lead panels in case they are needed.
\subsection{The Q$_{weak}$ Current Mode Electronics}

\paragraph*{Preamplifiers:}

All the TRIUMF current to voltage preamplifiers have now been made and tested. Two versions have been prepared;
we have 14 ``Main-style'' at JLab with a gain selection of $V_{out}/I_{in}$ = 0.5, 1, 2, or 4 M$\Omega$, and 14
``Lumi-style'' at Virgina Tech with a gain selection of $V_{out}/I_{in}$ = 0.5, 1, 25, or 50 M$\Omega$.

The preamplifiers were tested at JLab for radiation hardness. No changes in the gain, noise, or DC level were
noticed after 18 krad integrated dose. This easily meets the experimental specification of no deterioration
after 1 krad.  The amplifiers were also used during a test run with the G0 lumi detectors in March, 2007.

\paragraph*{Digital Integrators:}

The TRIUMF current mode electronics consists of the low noise current-to-voltage preamplifiers followed by
digital integrators. The integrators are triggered at the start of each spin state and integrate for a precise
pre-set spin duration. The system clock of all the digital integrators will be slaved to the same 20 MHz clock
used to generate the spin sequence at the electron source. Figure \ref{fig:VME-int} shows the layout of an
8-channel digital integrator. When triggered, the device integrates all the input signals for the preset time.
The integration time and many other parameters can be set through the VME bus.

\begin{figure}[ht]
\begin{center}
\includegraphics[width=110mm]{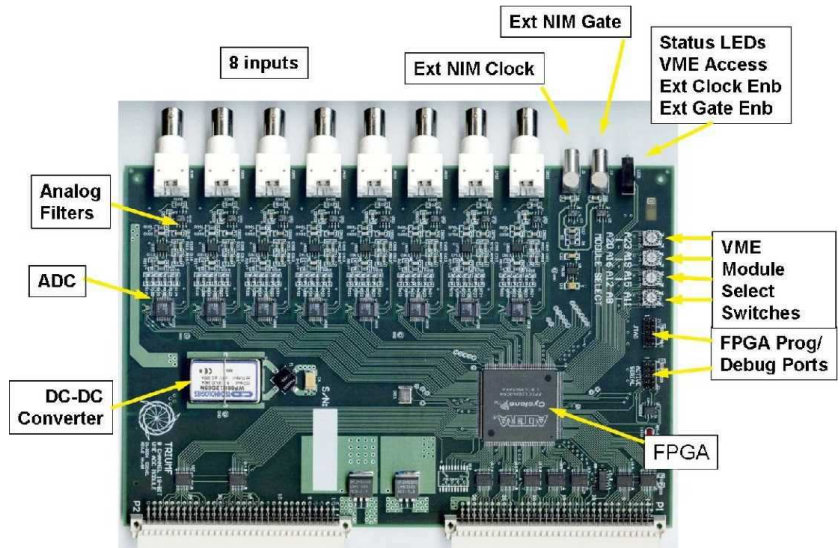}\\[2mm]
\caption{{\em Layout of the VME digital integrator. The analog signals from the 8 inputs first pass through
sharp cutoff 50 kHz anti-aliasing filters then are digitized by 18-bit ADCs operating at up to 500 ksps. The
Field Programmable Gate Array (FPGA) calculates the sums over the selected interval and delivers the results to
the VME bus. The outputs are 32 bit words, allowing integrals as long as 30 ms at 500 kcps. }}
\end{center}
\label{fig:VME-int}
\end{figure}

Internally, the analog signals to be integrated first pass through sharp cutoff 50 kHz anti-aliasing filters
then are digitized by 18-bit ADCs operating at up to 500 kilosamples per second (ksps).  The Field Programmable
Gate Array (FPGA) calculates the sums over the selected interval and delivers the results to the VME bus. Since
the 18 bit ADC digitizes each sample as one of $2^{18}$ possible codes, a 1 ms integral at 500 ksps has almost
$2^{27}$ possible values, and quantization noise on the integral is negligible compared to other sources of
noise.

Four prototypes of the VME digital integrator are now ready. One has has been tested at TRIUMF and is now at
Ohio University undergoing further tests with a realistic Q$_{weak}$ DAQ system.  The other three are at TRIUMF
and will be delivered to JLab and Ohio for further testing. We hope to have these tests complete by the end of
2007, at which time TRIUMF will proceed with building of the remaining integrators.

\paragraph*{Short Spin States:}

The heat load on the \Qweak target will be over 2000 Watts. Great care has been taken in the target design to
suppress boiling at high current.  These efforts will be complemented by a data acquisition strategy designed to
minimize the effect of target density fluctuations on the asymmetry widths.  Since the noise from target boiling
falls off at higher frequencies, the experiment now plans to use very short spin states, perhaps as short as 1
ms per spin state. In such cases it will be important that the time taken to settle on a new spin state be very
short; the JLab injector group has indicated that less that 0.1 ms can likely be achieved.

In the past, we planned to read out beamline instrumentation such as beam position monitors (BPMs) and beam
current monitors (BCMs) with the existing voltage-to-frequency converters. In the event of very short spin
states, however, the least count error on the VFCs would be excessive. For this reason we now plan to replace
the VFCs with TRIUMF digital integrators. In addition to the 14 on order for the main experiment, 16 modules
have been ordered for Hall-C instrumentation and 6 modules for the injector.

\paragraph*{Noise:}

Tests of the 1 M$\Omega$ preamplifiers with 200 pf capacitance input cable and a 50 kHz sharp-cutoff filter on
the output showed noise of 70 $\mu$V$_{rms}$ to 80 $\mu$V$_{rms}$, corresponding to a density of less than 0.4
$\mu$V$/\sqrt{Hz}$. Tests of a prototype VME integrator with the inputs terminated were made at TRIUMF. The
noise on a 2 ms integral was 11 $\mu$V$_{rms}$, implying an effective noise of 0.7 $\mu$V$/\sqrt{Hz}$ at the
integrator input.

Figure \ref{fig:signals}  shows the nature of the Qweak current-mode signals. During primary data-taking, the
6.4 $\mu$A current from the photomultiplier tube anode is made up of rather large charge quanta of 50,000$e$
that set the shot noise. Table \ref{noisetable} compares the shot noise under various running conditions to the
purely electronic noise. The beam-ON case assumes a count rate of 800 MHz, with 20 photoelectrons per event and
a noiseless PMT gain of 2500, giving an anode current of 6.4 $\mu$A. For the ``LED'' and ``lowest possible''
cases, it is assumed that the current into the preamplifier is still 6.4 $\mu$A, and that it is delivered to an
I to V preamplifier with $V_{out}/I_{in}$ = 1 M$\Omega$. The purely electronic noise is much smaller than it
needs to be during running conditions, but low noise is valuable in that it permits zero-asymmetry control
measurements using current sources to be made at the part per billion level in a relatively short time ($<$1
day).

\begin{figure}[h]
\begin{center}
\includegraphics[width=110mm]{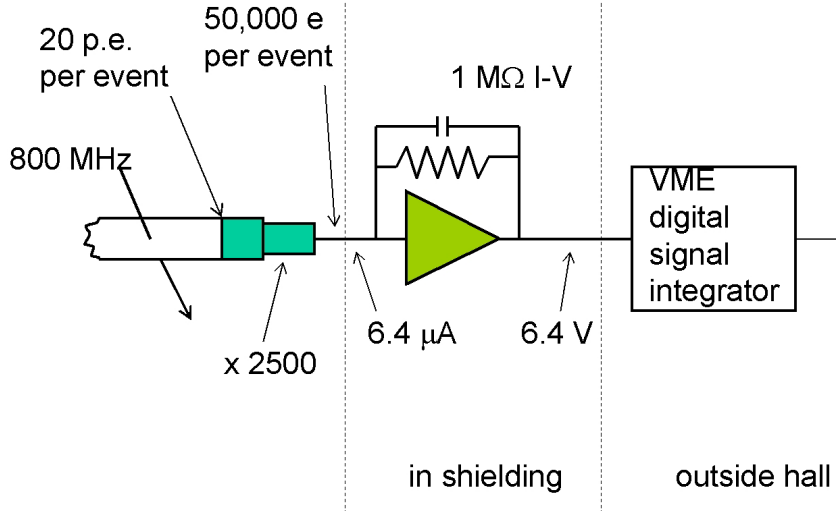}\\[2mm]
\caption{{\em  Nature of the \Qweak signals. During primary data-taking, the 6.4 $\mu$A current from the
photomultiplier tube anode is made up of rather large charge quanta of 50,000e. }} \label{fig:signals}
\end{center}
\end{figure}

\begin{table}[ht]
\begin{center}
\caption{Noise at integrator input for 6.4 $\mu$A from different sources, assuming a preamplifier with
$V_{out}/I_{in}$ = 1 M$\Omega$. The ppm column is as a fraction of 6.4 V.}
\begin{tabular*}{150mm}{@{\extracolsep{\fill}}lccr@{}}
\hline\hline

Condition   & charge quantum & noise                 & noise on 1 ms   \\
            &      (e)       & ($\mu$V/$\sqrt{Hz}$)  & integral (ppm)  \\
\hline

beam-ON shot noise                        & 50000  & 320 & 1120       \\
shot noise during LED tests               &  2500  & 72  & 250        \\
lowest possible shot noise on 6.4 $\mu$A  &   1    & 1.4 & 5          \\
preamplifier noise                        &        & 0.4 & 1.2        \\
digital integrator noise                  &        & 0.7 & 2.4        \\

\hline\hline
\end{tabular*}
\end{center}
\label{noisetable}
\end{table}

\paragraph*{Modulated Current Source:}

To assist in testing our full data acquisition system, TRIUMF is designing a Modulated Current Source. The
source will provide a reference current of nominally 5 $\mu$A, upon which a very small modulation is
superimposed to simulate the parity violating signal. The reference design specifies a switch-selectable choice
of 16 modulations from $10^{-6}$ to $10^{-9}$. These very small currents are formed by applying a voltage ramp
to a small capacitor. The module will respond to external spin state signals, or can run in stand-alone mode. By
placing such a source in Hall C, we will be able to show that we are able to detect a very small modulated
analog signal in the presence of all ambient sources of electronic noise.
\section{Tracking System Overview}

 The parity-violating asymmetry at \Qweak kinematics is directly
proportional to the momentum transfer $Q^2$; hence, it is essential that we make a precise determination of
$Q^2$. We need to determine the acceptance-weighted distribution of $Q^2$, weighted by the analog response of the
\v{C}erenkov detectors to within an accuracy of $\approx 1\%$. Recent simulations have shown that the anticipated
non-uniformity of light collection in the \v{C}erenkov detectors will shift the $Q^2$ by 2.5\%, demonstrating
the crucial need for a direct measurement.  This is the primary motivation for the tracking system; an
additional motivation is the measurement of any non-elastic backgrounds contributing to the asymmetry
measurement, such as inelastic events from the target, scattering from the target windows, and general
background in the experimental hall. Finally, since the hadronic structure contribution to the measured
asymmetry goes with higher powers of $Q^2$, the tracking system will be used to determine the important higher
moments of the effective kinematics, needed to correct for the hadron structure dependent terms.

For elastic scattering,
$$
\left. {Q^2} \right. = {{4 E^2\sin ^2{\theta
\mathord{\left/ {\vphantom {\theta 2}}
\right. \kern-\nulldelimiterspace} 2}} \over {1+2{E \over
M}\sin ^2{\theta \mathord{\left/ {\vphantom {\theta 2}}
\right. \kern-\nulldelimiterspace} 2}}}
$$
where $E$ is the incident electron energy, $\theta$ the scattering angle and $M$ the proton mass. In principle,
a measurement of any two of $E$, $\theta$, or $E'$ (the scattered energy) yields $Q^2$.  The absolute beam
energy will be known to $\leq 0.1$\% accuracy using the Hall C energy measurement system, corresponding to a
0.2\% error in $Q^2$. As the entrance collimator is designed to be the sole limiting aperture for elastically
scattered events, good knowledge of the collimator geometry and location with respect to the target and the beam
axis might seem to suffice for determining $Q^2$. The average radius of the as-built defining collimator will be
determined by CMM (Coordinate Measuring Machine) to better than 25 $\mu$m (0.01\% of the radius).  The distance
from the target center to the defining aperture will be determined using redundant survey techniques to better
than 3 mm (0.1\% of the distance). The purely geometrical contribution to the $Q^2$ determination is therefore
$dQ^2/Q^2 = 2 d\theta/\theta = 2\sqrt{ (dR/R)^2 + (dL/L)^2} = 0.2\%$.  The contributions from the beam energy
uncertainty and the angle uncertainty when combined in quadrature are therefore only 0.3\%.  However, we expect
that contributions such as the uncertainties in the detailed corrections for ionization and radiative energy
loss, collimator transparency, the angular dependence of the e+p elastic cross section, {\em etc.}, may
ultimately limit the $Q^2$ measurement to 0.5\%. Our significant investment in tracking detectors is expected to
help us quantify such subtleties, as well as to confirm the predicted inelastic contribution to the detected
electron flux. Last, but not least, we need to weight the experimental $Q^2$ distribution with the analog
response of the \v{C}erenkov detector in order to determine the effective central $Q^2$.

Rather than rely solely on a simulation to account for all of these effects, we choose to {\em measure} them
with a dedicated tracking system. These measurements will be made in special calibration runs in which the beam
current is reduced to less than 1 nA, allowing the use of the tracking system.  Recent tests have shown that at
currents as low as 100 picoamps the beam is still stable, and that adequate beam current and position
measurements can be made using special harp scans and a halo monitoring device. In this $Q^2$ measurement mode,
the \v{C}erenkov detectors will be read out in pulse mode and individual particles will be tracked through the
spectrometer system using a set of chambers (Region I, Region II, and Region III, described below). This
information will allow us to determine, on an event-by-event basis, the scattering angle, interaction vertex (to
correct $E$ for dE/dx and radiation in the target), $E'$ (to confirm elastic scattering) and location and
entrance angle of the electron on the \v{C}erenkov detectors.

The tracking system~\cite{Grimm:2005hf} consists of three regions of tracking chambers: the Region I vertex
chambers, based on GEM (Gas Electron Multiplier) technology, will have excellent position resolution and will be
located directly after the primary collimator.  The Region II horizontal drift chamber (HDC) pair will be just
before the entrance of the spectrometer magnet; together with Region I, they will determine the scattering angle
to high accuracy.  The Region III chambers, a pair of vertical drift chambers (VDC's), will be located just
upstream of the focal surface.  They will allow momentum analysis to ensure that the detected events are true
elastic electrons and will characterize the particle trajectories entering the \v{C}erenkov detector (and so
allow us to map out its analog response).  The tracking event trigger will be provided by plastic trigger
scintillators positioned between the VDC's and the \v{C}erenkov bars. Finally, a quartz ``scanner'' will be
mounted behind the \v{C}erenkov bars in one sector, to be used as a non-invasive monitor of the stability of the
$Q^2$ distribution during high-intensity production data-taking, and to verify that the distribution measured in
the low beam-current calibration runs is compatible with that measured at full beam intensity.

For each region, two sets of of chambers are being constructed, which will be mounted on rotator devices to
cover two opposing octants, and which will allow them to be sequentially rotated to map, in turn, all octants of
the apparatus.\\

\subsection{Region I - Gas Electron Multiplier  Chambers}

\begin{figure}[ht]
\subfigure[]{\includegraphics[width=0.55\textwidth]{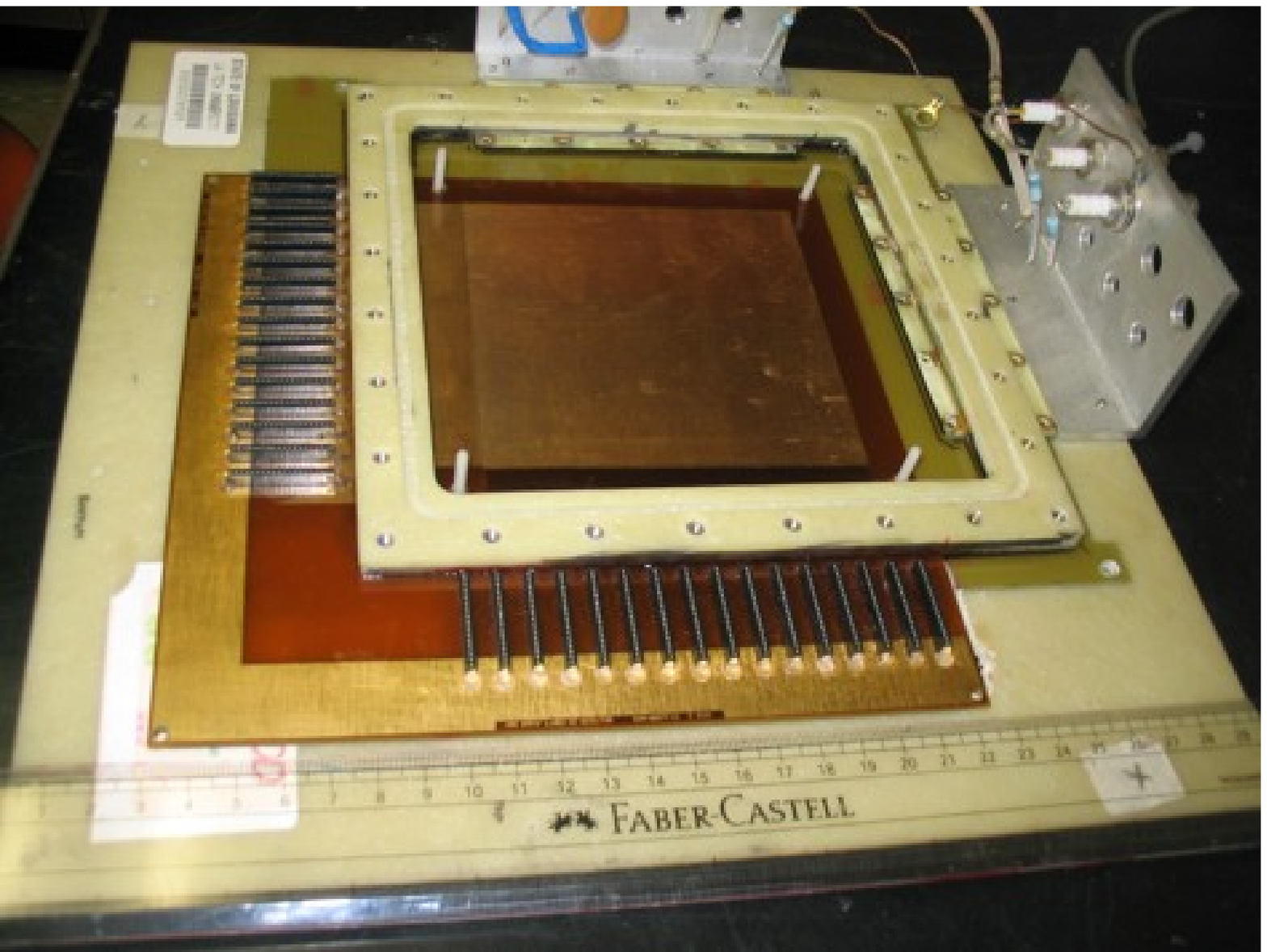}}
\subfigure[]{\includegraphics[width=0.38\textwidth]{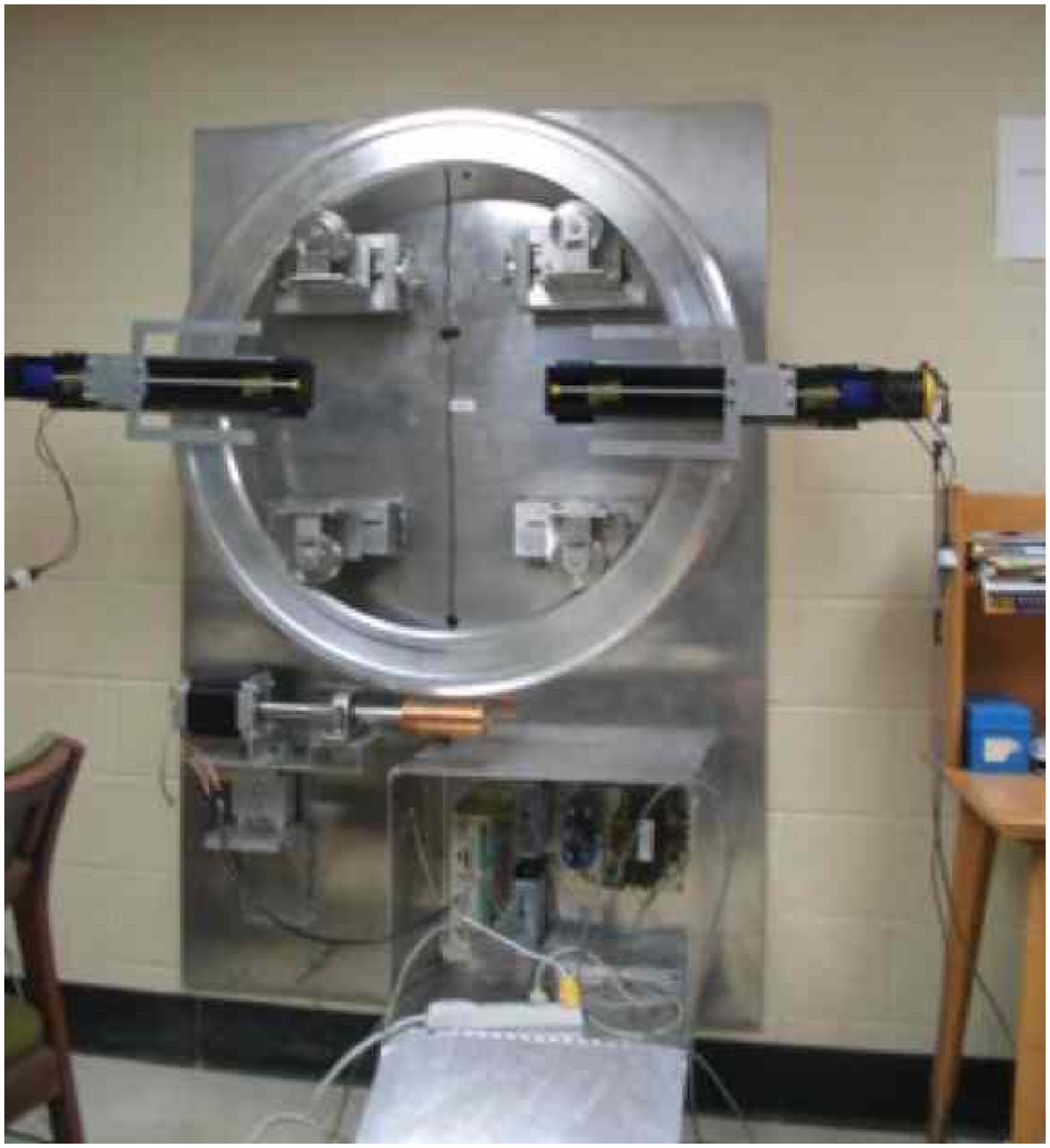}} \caption{{\em (a): Working
prototype GEM chamber. (b) Region I Rotator assembly. } \label{fig:GEMS}}
\end{figure}

The Region I tracking system is designed to track the scattered electrons less than 1 meter away from the target
and in opposite octants. The tracking system will be in a high radiation environment despite being behind the
first collimation element. This tracking system uses an ionization chamber equipped with Gas Electron Multiplier
(GEM) preamplifiers in order to handle these high rates as well as enable a measurement of an ionization event's
location within the chamber with a resolution of 100~$\mu$m. The system contains two ionization chambers located
180$^{\circ}$ apart on a rotatable mounting system which allows two opposing octants to be measured
simultaneously. The rotatable mounting system can move the chambers through a 180$^{\circ}$ angle such that
measurements may be made in all octants.

The ionization chamber final design has been completed, and the chambers are currently being constructed. The
GEM preamplifier foils have already been acquired, and the readout board is currently being manufactured at
CERN. The ionization chamber itself has been machined and will be assembled upon receipt of the readout boards,
which are currently scheduled for delivery in the first quarter of 2008. We anticipate that the detector will
have similar performance to the prototype detector (see Fig.~\ref{fig:GEMS}a) constructed previously.

The readout electronics for Region I use the VFAT board from CERN to digitize the analog signals on the readout
board and send them to a VME crate to be recorded. The control system (a ``gum-stick" microcomputer) for the
VFAT board has been acquired and tested. A signal junction box to transfer the control signals for 6 VFAT boards
to our control system has been designed. The 6 VFAT boards will digitize the analog output signals from a single
detector. The junction box will also collect digital detector signals and transfer them to a VME crate for
readout. A CAEN V1495 FPGA module has been purchased and is currently being programmed to transfer the digital
signals to CODA (the data acquisition).

The infrastructure to mount and rotate the chambers into position is 90\% complete (see Fig.~\ref{fig:GEMS}b).
The system uses four caster wheels to mount an aluminum ring that has teeth on its outer surface to mesh with a
worm gear in order to rotate the ring. One stepper motor is used on the worm gear to rotate the detector to
within 1 mrad. A stepper motor for each detector has been mounted on the ring itself in order to position each
detector radially using fixed stops. A controller for the all the stepper motors has been programmed to move the
detector between octants. A GUI is currently under development which will be used in the counting house to
position the detectors.

\subsection{Region II - Horizontal Drift Chambers}

This second set of chambers will be located just upstream of the QTOR magnet. Their purpose is to determine the
position and direction cosines of the scattered electrons as they enter the magnet, and, along with the Region I
vertex detectors, to provide an accurate measurement of the target vertex and scattering angle.  The Region II
drift chambers are horizontal drift chambers (HDC).  We are building two sets of two chambers, each set being
separated by 0.4 m to provide angular resolution of $\sim$0.6 mrad for position resolutions of $ \sim 200\
\mu$m. Each chamber has an active area of 38 cm x 28 cm, wire pitch of 5.84 mm, and six planes (in an
$xuvx'u'v'$ configuration, where the stereo $u$ and $v$ planes are tilted at an angle of 53$^{\circ}$).  There
are a total of 768 sense wires with a corresponding number of electronic channels.  The sense wires are being
read out with commercially available Nanometrics N-277 preamp/discriminator cards.

We have all systems in place to construct and test the chambers.  For construction, we have a wire stringing
area, wire scanning apparatus, and gas-tight high voltage test box.  For testing we have a cosmic ray test stand
with a DAQ system instrumented with the same JLab F1 TDCs that we will use in the experiment.  To date, we have
completed a prototype chamber and the first full production chamber.  Photographs of that chamber along with a
typical drift time distribution for one of the wires is shown in Figure~\ref{HDCfigure}.

\begin{figure}[h!]
\vspace*{0.2cm} \centerline{\includegraphics[width=6.5in]{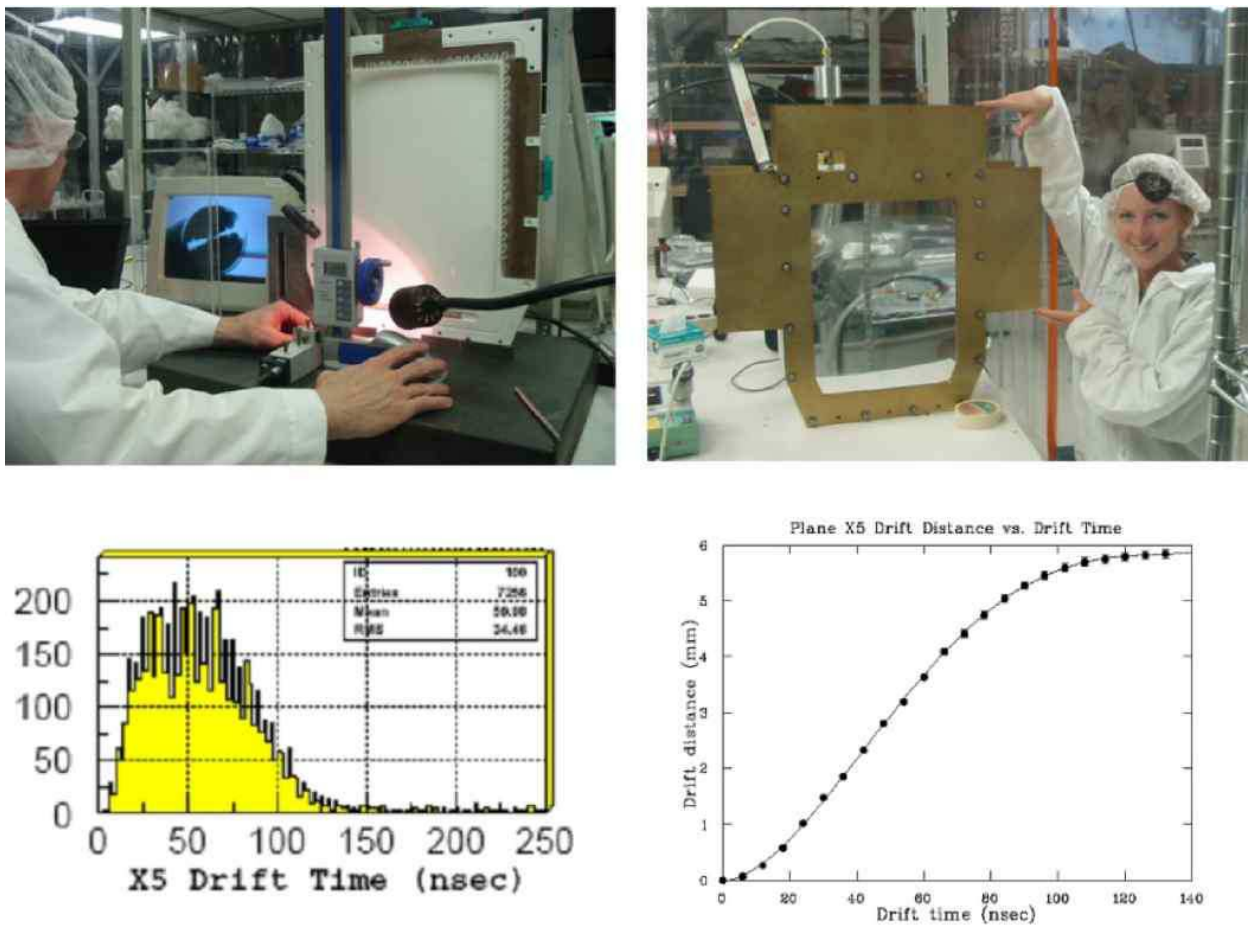}} \caption{\em Upper left: Norman
Morgan uses the wire scanning apparatus to measure wire positions.  Upper right: Undergraduate Elizabeth Bonnell
shows the first completed production chamber.  Lower left:  a typical drift time distribution from testing with
this chamber; it cuts off at about 120 nsec as expected for our drift cell size. Lower right: corresponding
drift distance to drift time correlation.} \label{HDCfigure}
\end{figure}

The chambers will be mounted on opposite sides of a rotator instrumented to be remotely rotated so that all
eight octants can be covered.  The rotator will be designed so that the chambers can be in an ``in-beam'' and
``parked'' position.  Design and procurement of that device will be done in collaboration with a Jefferson Lab
engineer.

\subsection{Region III - Vertical Drift Chambers}

The Region III vertical drift chamber design is based on and has evolved from the very successful VDC's used in
the Hall A High Resolution Spectrometers ~\cite{Fissum:2001st}. Over the last three years, we have made
considerable progress on the Region III project. A large clean room was constructed and is in use. The large
Region III rotating arm assembly, on which the detectors will be mounted, has been designed and constructed, and
is on site at JLab, awaiting final assembly (see Fig.~\ref{rotator}b). The rotator is a gym-wheel construction
with welded extrusion as a radial rail system holding the VDC's (see Fig.~\ref{rotator}a). The dual VDC's will
be moved along the rails for locking them into either an IN position (for tracking runs) or an OUT position (for
production runs).

\begin{figure}[h!]
\vspace*{0.2cm} \subfigure[]{\includegraphics[width=0.47\textwidth]{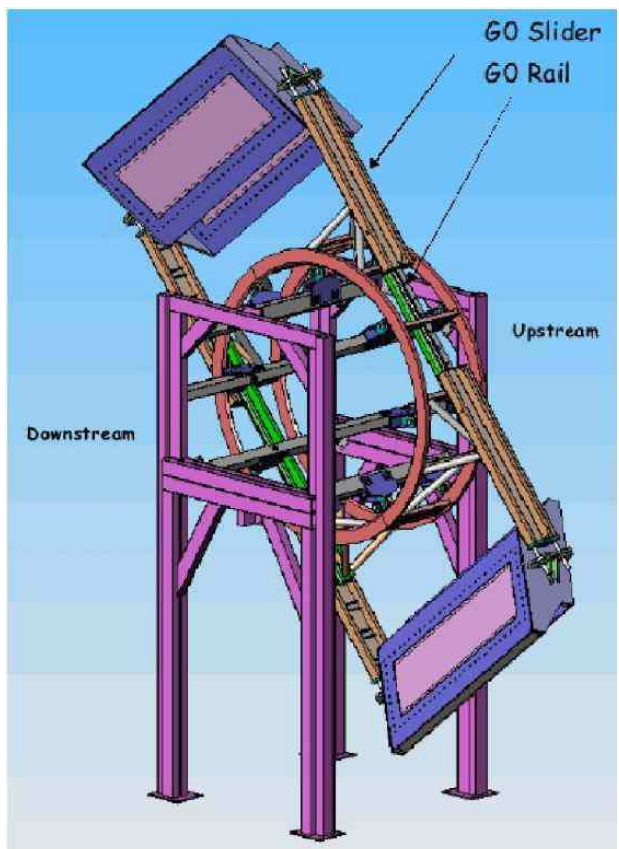}}
\subfigure[]{\includegraphics[width=0.52\textwidth]{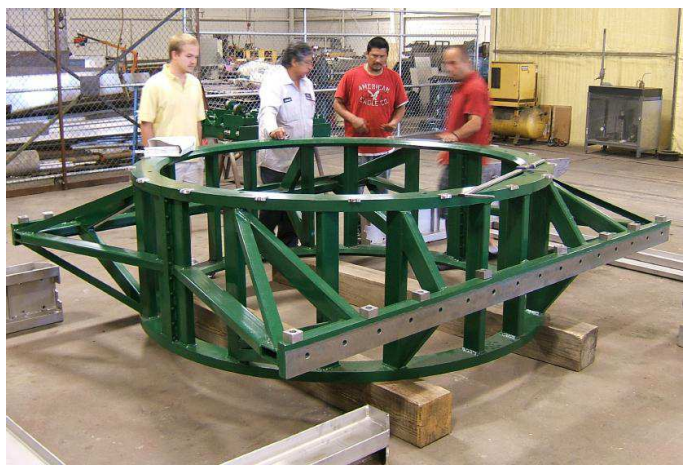}}

\caption{{\em (left) Design of the Region III support structure and rotator. (right) The completed ``gymwheel''
for the Region 3 rotator. }} \label{rotator}
\end{figure}

 The tracking response of the VDC's has been modelled using the GARFIELD
simulation package~\cite{GARFIELD}. We studied the correlation between the vertical distance (above/below the
wire) and the arrival time of the first drift electron~\cite{WECHSLER,ERIN}. A fast, accurate method was
developed for reconstructing the drift distance as a function of drift time and track angle. The projected
intrinsic position resolution per wire plane, for a track hitting at least 4 drift cells, is $\Delta$x$\sim$50
$\mu m$ and $\Delta$y$\sim$75 $\mu m$, more than adequate for this application.

The chamber design was finalized, and all parts of the G10 frame assembly have been machined and are in-house.
Each of the G10 frames was machined in four separate pieces (due to their large size, very few shops could
handle the machining in one piece), which need to be epoxied together. The epoxying technique has been
prototyped, and a complete chamber's worth of frames has been assembled.

Various chamber assembly jigs and tools have been designed and built, including a frame gluing jig, a wire
positioning jig, a ``wire scanner" system, and a tension measuring device.  These provide essential tools for
quality control and verification of the wire position and alignment, with a design precision of $\leq$50 $\mu
m$. The wire scanner (see Fig.~\ref{stringing}) moves a CCD camera over the drift chamber wire plane using a
precision translation stage. Wire scanner measurements of 70 wires test strung on the assembly jig have verified
that we can achieve this precision.  The tension measuring device uses the same assembly and the classical
technique of finding the resonant frequency for current-carrying wires oscillating in a magnetic field.

Stringing and final assembly of the chambers is awaiting the delivery of the electronics daughter boards, needed
to mate our wires to the electronics readout cards. Assembly will begin in the coming weeks and will take about
six months to complete.

\begin{figure}[h!]
\vspace*{0.3cm} \centerline{\includegraphics[width=4.5in]{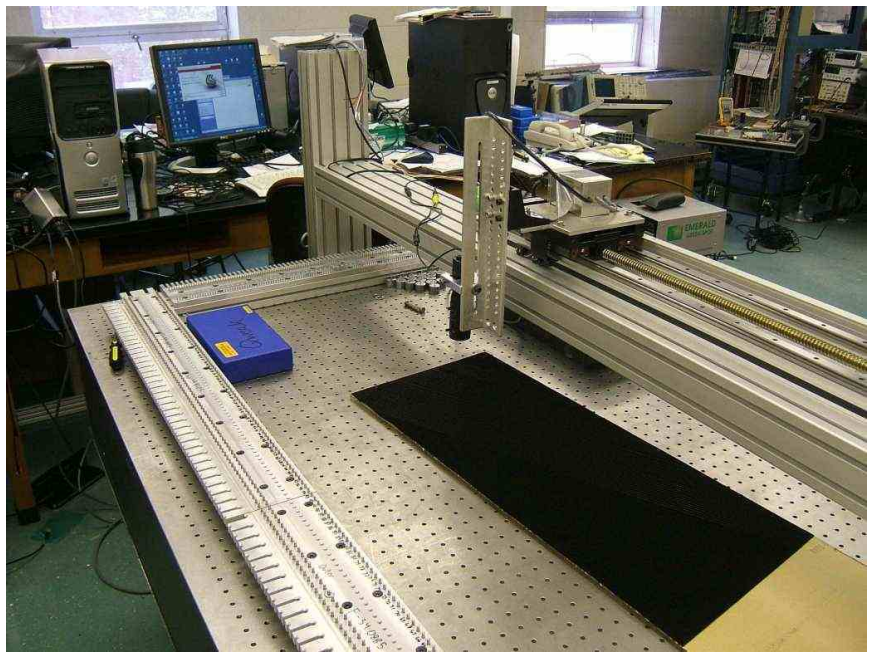}} \caption{{\em Region III wire
stringing jig and wire scanner system. }} \label{stringing}
\end{figure}

For the front-end electronics (preamp-discriminator) we have selected a system based on the MAD chip~\cite{MAD},
which was developed at CERN for modern wire chambers. We have tested these chips and they meet all our
specifications.  They will be mounted on circuit boards of a proven design already in use at Jefferson
Laboratory; the boards were sent out for manufacturing in mid-September, 2007.

To save the significant expense of instrumenting each of the 2248 sense wires in the 4 VDC chambers with an
individual TDC channel, we have adopted a delay-line multiplexing scheme (see Fig.~\ref{delayline}). The signals
from many (18) wires from separated locations in a given chamber are ganged together and put onto two signal
paths, leading to two individual TDC channels. The drift time is decoded from the sum of the two TDC times, and
the wire that was hit is identified via the difference between the two times, thus saving a factor of 9 in
number (and cost) of TDC's. We will use the new JLab standard 64-channel F1 multihit TDC for the final
digitization.

Our delay-line implementation adopts a novel technique - instead of the classical use of analog cable delay
(simple, but expensive and bulky) we will use a digital delay line, based on ECL gates as delay chips.  The LVDS
(low-voltage differential signal) signals from the MAD chips will be converted to ECL and duplicated in custom
conversion boards; the ECL signals will then be fed to a string of ECL gates which provide the quantized delays
(1.3 ns per chip).  The LVDS to ECL boards and the multiplexing boards have been completely designed by the JLab
electronics group in consultation with the W\&M group, and prototypes are under procurement, with testing
planned in the next few weeks.

\begin{figure}[h!]
\vspace*{1.0cm} \centerline{\includegraphics[width=6.2in]{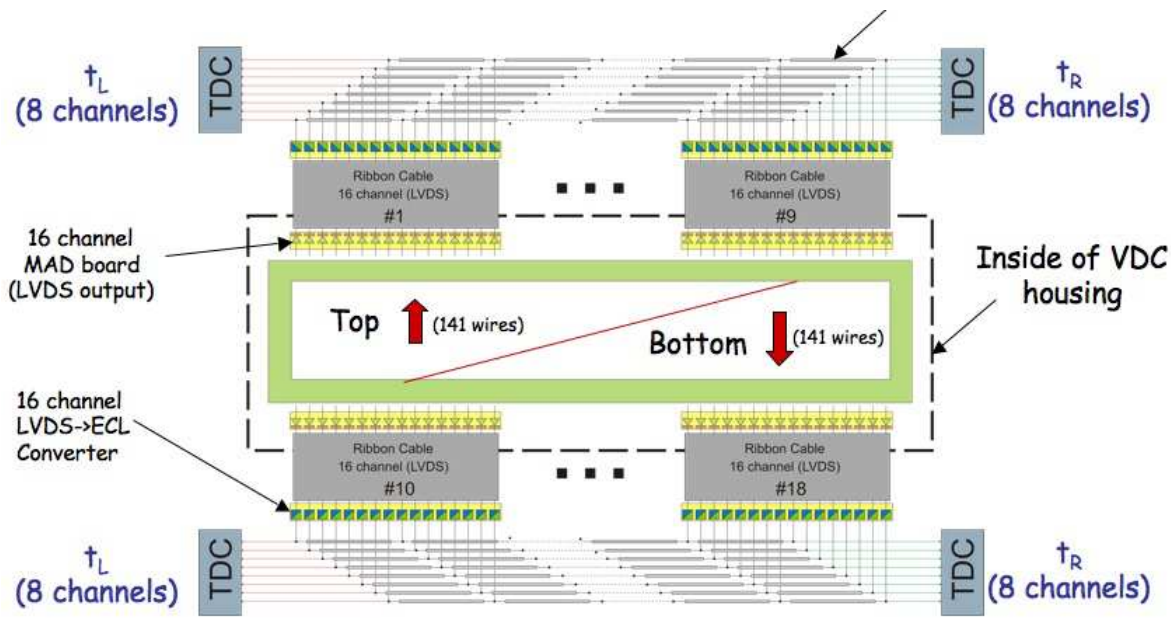}}

\caption{{\em Region III Delay-line readout schematic. }} \label{delayline}
\end{figure}

\subsection{Trigger Scintillator}
\label{trig_scint}

Scintillation counters will provide the trigger and time reference for the calibration system.  These are large
enough to shadow the quartz \v{C}erenkov bars, and tests with a prototype indicate that they have sufficient
energy resolution and timing capabilities to identify multiparticle events and veto neutrals. The scintillators
are long bars mounted between the \v{C}erenkov bars and the Region 3 chambers with a photomultiplier tube at
each end.

\begin{figure}[hbtp]
\centerline{\includegraphics[width=6.0in]{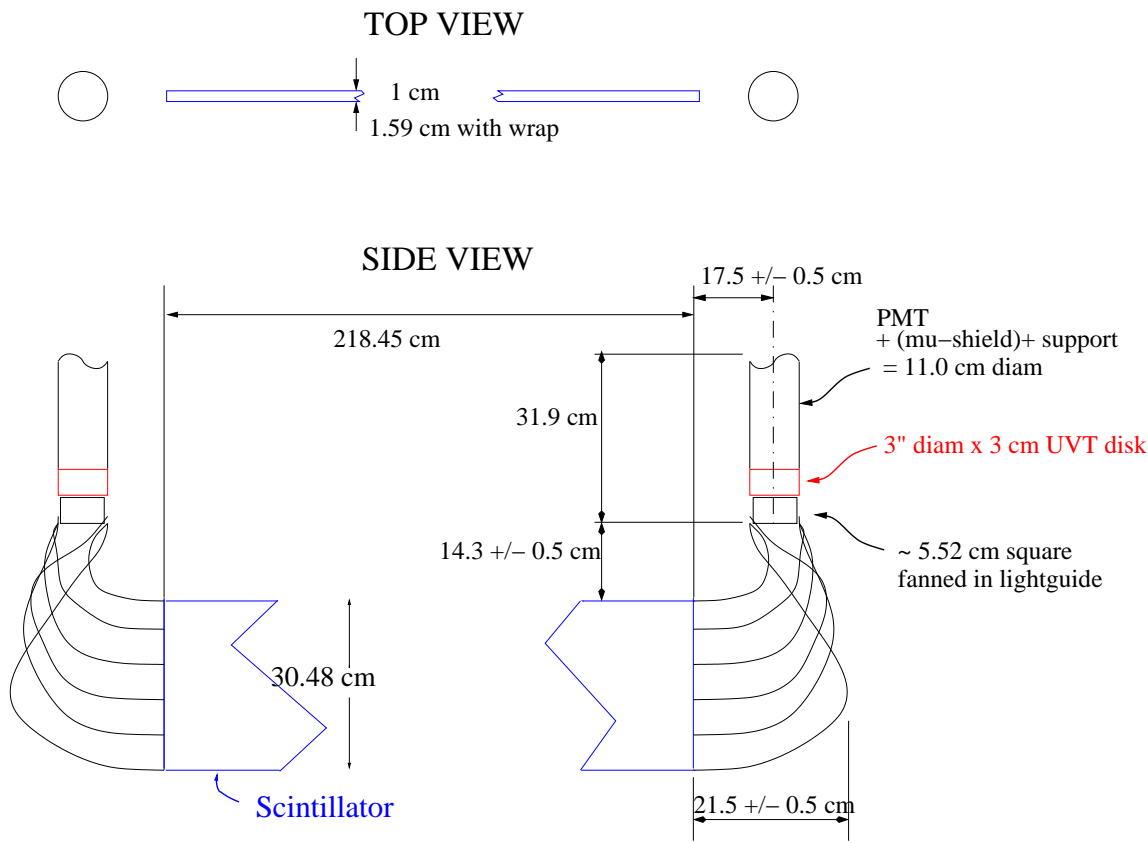}}
\caption{\em{Schematic of trigger scintillator and lightguides.
    }}
\label{fig_trig-scint}
\end{figure}

Each scintillator is made from BC408 (i.e. $\sim$ 3 ns time constant)
and is 218.45
cm long, 30.48 cm high, and 1 cm thick.
To minimize loss of light from the scintillator corners, each light guide is made of a row of ``fingers'' that
couple to the 30.5 cm $\times $ 1 cm ends of the scintillator and overlap each other to form a squared off
circle that is circumscribed within the PMT, as shown in Fig.~\ref{fig_trig-scint} and
Fig.~\ref{fig_trig-scint-pmt}, respectively.  We will use Photonis XP4312B 3 inch PMTs, which have a high gain
($\sim 3 \times 10^7$) and a uniform response over their photocathode areas \cite{XP4312B}. These three inch
PMTs have photocathodes that are 6.8 cm in diameter; this corresponds to a photocathode area of 36.3 $\rm cm^2$,
which will accommodate the 30.5 $\rm cm^2$ scintillator ends.  Tests with a scintillator and lightguide
prototype show that we can expect 70 to 210 electrons to be produced by the photocathode for every electron
going through the scintillator.  Combining this with a high gain PMT ($\sim 10^7$) yields 100 to 300 pC of
charge in the 3 nsec of the signal.

\begin{figure}[hbtp]
\centerline{\includegraphics[width=5.5in]{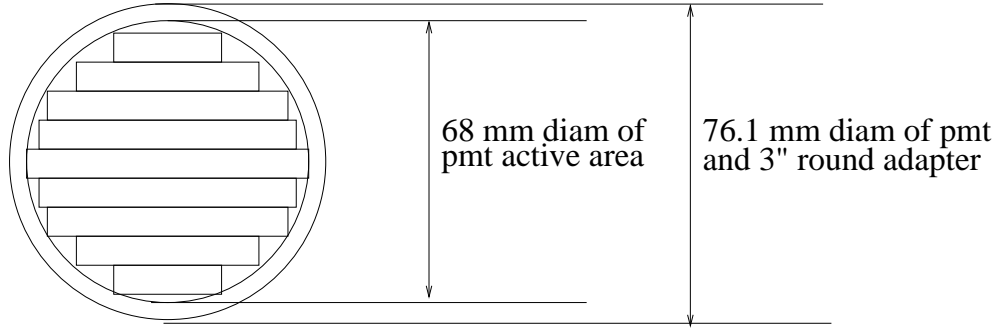}}
\caption{\em{Schematic of trigger scintillator lightguide coupling to PMT.
    }}
\label{fig_trig-scint-pmt}
\end{figure}

Saint-Gobain made, assembled, and wrapped the scintillators and lightguides and then delivered them to George
Washington University (GW) in October 2007.  GW did not have a working DAQ  system until recently, when we
purchased a VME crate, VME-USB interface, VME modules, and downloaded the recommended software for this VME-USB
interface system from the National Superconducting Cyclotron Facility at Michigan State University.  We can now
acquire and analyze ADC and TDC signals from our prototype detectors at GW.

Given the extreme ratio of thickness to length and weight of the PMT's, the scintillator counters and PMTs must
be supported by a frame which is then mountable to the rear VDC in the Region 3 package. The support frames were
designed and built by JLab staff and are ready for use. When the fall 2007 semester ends we will move the
support stands to GW, position the trigger scintillators on the stands, and verify that the response
characteristics of the trigger scintillors meet our design goals.  The scintillators will then be ready to be
moved to JLab for the experiment.

\subsection{Focal-Plane Scanner}

The tracking system in \Qweak\ will operated at low beam current in order to determine $\langle Q^2\rangle$ for
the light accepted by the main \v{C}erenkov detectors.  The parity-violating asymmetry measurement, on the other
hand, will be conducted at high beam current, where the tracking system will be inoperable.  Since our previous
Jeopardy Proposal, we have conceived of a new device, a focal-plane scanner.  The scanner is a tracking device
with the important property that it is operable in counting mode at both low and high beam currents. The scanner
has been fully funded by NSERC (Canada) and is nearing completion of the construction phase at University of
Winnipeg.

The focal-plane scanner consists of a quartz \v{C}erenkov detector with small active area which is scanned in
the focal plane of the spectrometer, just behind the main  detectors.  A scan consists of moving the detector
across the fiducial area of the main \v{C}erenkov detectors to make rate measurements.  The scanner would impact
knowledge of tracking results at high beam currents as follows.  First, comparison of tracking system results
with scanner results would be conducted at a low beam current acceptable to the Region III drift chambers. Then,
scanner results would be acquired at high beam current.  If the two scanner measurements would be found to
agree, then the tracking results would be believable at high current to high confidence.

A photograph of the scanner is shown in Fig.~\ref{fig:scanner}(a).  In the scanner, two pieces of fused silica
(synthetic quartz) are placed one in front of the other, with a 1~$\times$~1~cm$^2$ active area.
  Each piece of
quartz acts as a \v{C}erenkov radiator, and each is coupled to a photomultiplier tube (PMT) by an air-core light
pipe coated with specular, reflective Alzak (polished and chemically brightened anodized aluminum).
Fig~\ref{fig:scanner}(b) shows the pulse-height distribution in one of PMT's when the detector is calibrated
with cosmic-ray muons.  A 2D linear motion assembly is used to scan the detector.  The motion assembly consists
of two stainless-steel ball-screw driven tables, one mounted on the other, producing x-y motion. The linear
motion assembly is driven by servo-motors controlled remotely by a computer.

One complete scanner system is being constructed, to be mounted in one of the \Qweak\ octants.  It would be
possible to move the scanner to other octants by hand; however, it is not envisioned that this would be pursued
unless a systematic effect arose that was octant-specific.

Similar quartz-radiator scanner devices were used successfully in both
the E158 and HAPPEx experiments.  In E158, the device was found to
particularly important, as it was the only means with which to study
spectrometer optics and make studies of backgrounds.  It is envisioned
that the \Qweak\ focal-plane scanner would also be used to impact
questions of backgrounds, as well as spectrometer optics, particularly
the stability of such quantities with beam current during parity violation
measurements. \\

\begin{figure}[ht]
\subfigure[]{\includegraphics[width=0.45\textwidth]{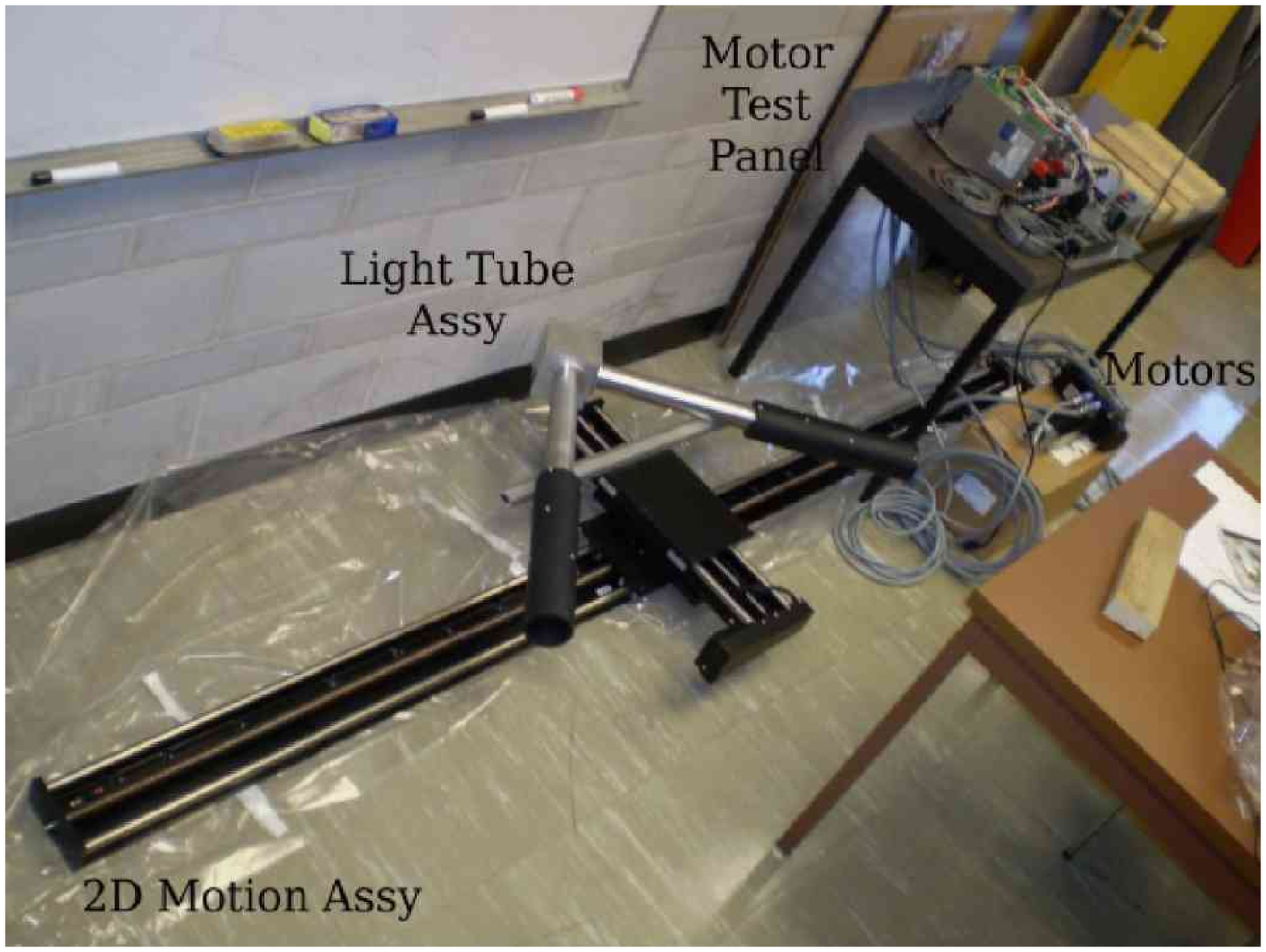}}
\subfigure[]{\includegraphics[width=0.54\textwidth]{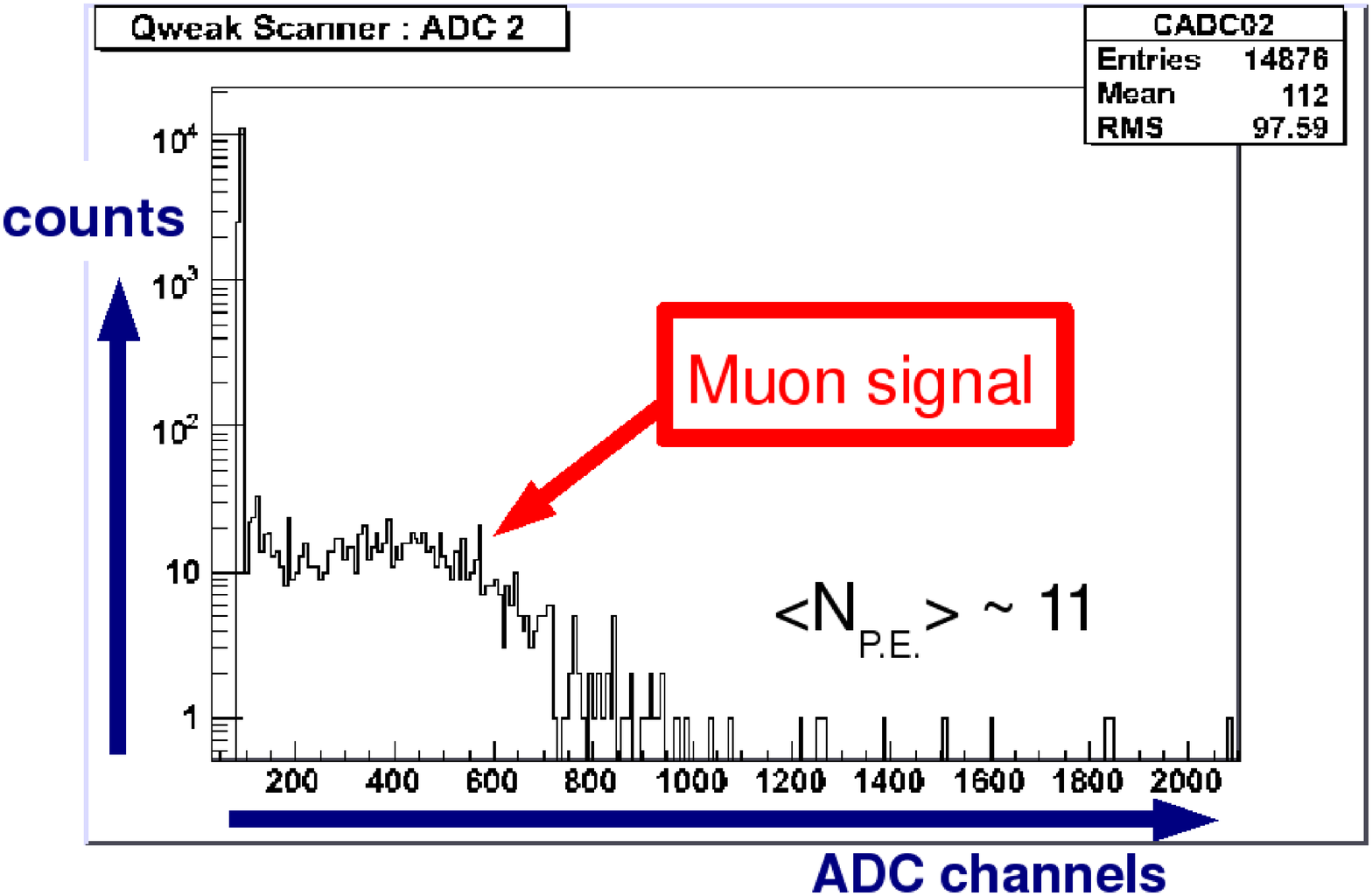}} \caption{{\em (a) Photograph of the
focal-plane scanner system under assembly at U. Winnipeg. (b) Pulse-height distribution in photomultiplier tube
when testing scanner with cosmic rays shows sufficient light yield.\label{fig:scanner}}}
\end{figure}

\subsection{Track Reconstruction Software}

The \Qweak\ Track Reconstruction software (QTR) must utilize the full capability of each detector region. We
need to to perform fast track reconstruction and to compile a statistics database, and the software must be
versatile so that it can be updated to perform other tasks, such as various detector calibrations.  The \Qweak
track reconstruction software is adapted from that used in the HERMES experiment, as discussed below.

The HERMES experimental setup had many similarities to that of \Qweak\ \cite{WANDER}.  In both experiments,
particles traverse two straight tracks separated by a curved track in a magnetic field.  In the initial phase of
HERMES, as in \Qweak\, no detectors were present within the magnetic field, which greatly simplifies the track
reconstruction.   Additionally, the HERMES software was designed to use many of the best tracking techniques
available \cite{BOCK,BLUM}. Thus, the HERMES track reconstruction software is an ideal model for \Qweak\. The
\Qweak tracking group has rewritten the HERMES reconstruction package into C++ and altered it to be more
object-oriented. We are developing the QTR package to be as versatile as possible.

One of the core components of QTR is the use of pattern recognition.  For Region III, patterns are generated for
a small subset of wires in one of the VDC planes.  The set of patterns represents all possible straight line
tracks of interest.  Using symmetry relationships, the pattern set can be used for the entire detector and can
easily be searched to be identified with a track. The ability to compare a set of hits in a detector to a known
good track allows for fast track identification, powerful noise and background rejection, and the ability to
easily resolve left/right wire ambiguities in drift chambers.  Additionally, initial track parameters can be
associated with the track to improve fit calculations.  Region I and II will use a single pattern database to
identify track segments upstream of the magnetic field.

The pattern recognition utilities have been completed, and a first-round pattern database algorithm has been
written for the Region III chambers.  Mock Region III data have been successfully fitted to track segments in
each plane and matched together. Currently, progress is being made on using the VDC straight-track segments to
compare to hits in the trigger scintillator and \v{C}erenkov bars which lie downstream of the VDC.  The software
package is also being refined for an initial release which will be used in association with Monte Carlo
simulations.

\section{Cryotarget}

\subsection{Specifications} The \Qweak $LH_2$ target must be 35 cm long and exhibit azimuthal symmetry. Density
fluctuations must not contribute significantly to the asymmetry width in the experiment, even though the target
will be used with up to 180 $\mu$A of 1.165 GeV beam. Sufficient cooling power must be provided to remove 2.5 kW
of power. The target must provide up to $\pm$2" of horizontal motion as well as enough vertical motion to
accommodate a dummy target plus several solid and optics target configurations. A CAD model of the target is
shown in Fig.~\ref{targetcad}.

\begin{figure}[hp]
\begin{center}
\rotatebox{0.}{\resizebox{3.0in}{!}{\includegraphics{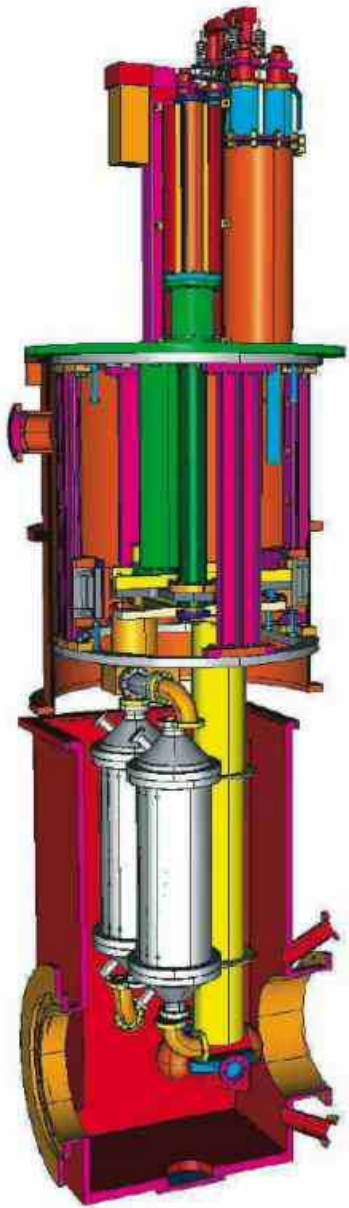}}}
\end{center}
\caption{{\em CAD model of the target as it stands in fall, 2007. The entire system is hung off the top plate,
which also supports the lifter motor, relief stack, and cooling connections. The target cell is located near the
bottom of the picture. Each leg of the target loop contains a heat exchanger, one for 4K and one for 15K Helium
coolant. The pump occupies one of the upper corners of the loop. }} \label{targetcad}
\end{figure}

\subsection{Cooling Power} In the fall of 2004, a scheme was worked out with laboratory management, the target
and cryo groups, which essentially guaranteed our experiment the cooling power it needs. Many options were
studied. Several of these were found to be viable. The default scheme agreed upon at these meetings was the one
which was easiest and cheapest to implement, and had the least impact on other halls and the FEL while still
meeting the minimum requirements of the experiment. This scheme requires us to design and build two independent
heat exchangers. One will remove heat from the target using 4K He coolant from the excess capacity of the CHL.
The other will make use of the more traditional 15K He coolant from the ESR. The 1.2 kW capacity of the ESR is
thus augmented by the CHL to meet the needs of the \Qweak experiment. M\o ller operation is taken into account,
as well as low power experiments in Hall A. An improvement to the scheme was undertaken in FY07, to build a
(portable) heat exchanger external to the target which will recover the unused enthalpy in the returning CHL
coolant and supply it to the ESR.

\subsection{Cell Design} Computational Fluid Dynamics (CFD) codes were employed (for the first time in the design
of a cryotarget, as far as we are aware) to study various cell designs. Variations on both longitudinal (G0 and
SAMPLE-like) cells as well as transverse cells were studied. Temperature and density profiles of the $LH_2$
flowing through the cells were obtained for realistic mass flow (1.1 kg/s), beam power deposition (2.5 kW in a
4x4 mm$^2$ raster area) and initial thermodynamic conditions (20 K $\&$ 50 psia). Window temperatures were
tabulated for each design where the beam enters and exits the cell to characterize the unavoidable film boiling
in those regions. Monte Carlo calculations were undertaken to assess the impact of basic cell geometries on the
backgrounds in the experiment. While some work in this area remains to be done, the CFD calculations have
steered us to a basic transverse flow cell design (see Fig.~\ref{targetcfd}) consisting of a conical shape which
puts all the scattered electrons of interest out normal to the exit window of the cell. The input manifold will
direct flow across both the entrance and exit windows as well as across the middle of the cell. The exit
manifold is a simple slot along the length of the cell. The cell volume is about 5 liters, and the head loss is
less than 0.4 psi for the design flow of 15 liters/sec (~$\simeq$1.1 kg/s mass flow).

\begin{figure}[hhhtttb]
\begin{center}
\rotatebox{0.}{\resizebox{4.0in}{!}{\includegraphics{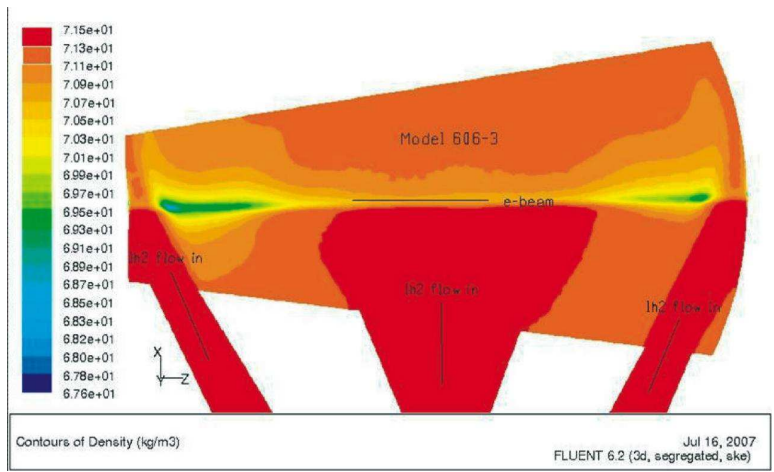}}}
\end{center}
\caption{{\em Computational fluid dynamics calculation showing the density profile of the $LH_2$ flowing in the
target (from the bottom to the top in the figure). Density units are kg/m$^3$.}} \label{targetcfd}
\end{figure}

\subsection{Heat Exchangers} The detailed heat exchanger design was approved by the cryo group as part of our
cooling power negotiations. All the parts for both the 4K and 15 K heat exchangers have been ordered and are on
site, including the pre-wound coils of 0.5" diameter finned Cu tubing (see Fig.~\ref{targethx}). The heat
exchangers were designed to provide a huge overhead in cooling power. We are now thinking about scaling back the
design to achieve a more modest cooling power overhead in order to reduce volume and with it $LH_2$ inventory.
Two schemes to do that are presently being considered, each of which makes use of the fin tube coils already on
hand. In parallel, a flow diagram is being prepared which describes the plumbing required for \Qweak in the
hall.

\begin{figure}[hhhtttb]
\begin{center}
\rotatebox{0.}{\resizebox{3.3in}{!}{\includegraphics{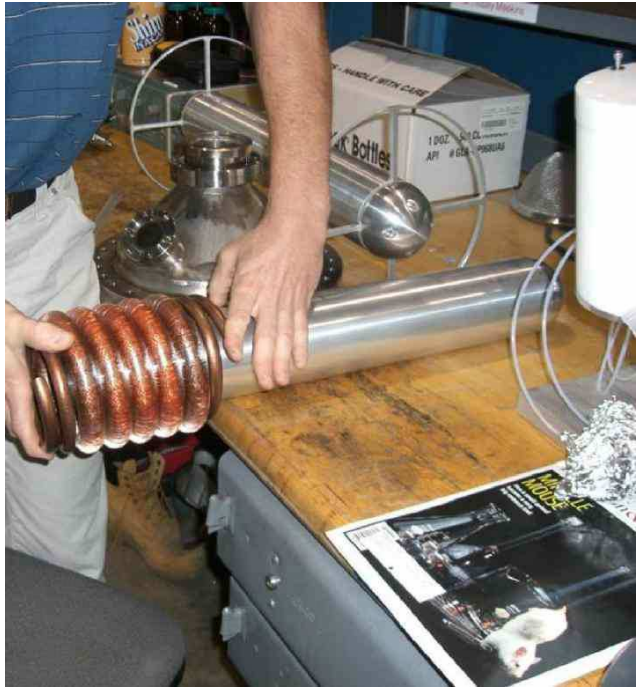}}}
\end{center}
\caption{{\em Photo showing the main componenents of the heat exchangers.}} \label{targethx}
\end{figure}

\subsection{Pump} Considerable attention was given to calculating realistic specifications for the pump: volume
flow and head loss around the entire target loop. Together these parameters give rise to viscous heating in the
loop which can quickly become a show-stopper in terms of the available cooling power if not kept in check. We
eventually settled on 15 l/s volume flow and 2 psi head for the pump specifications. Our actual calculated head
loss was only 1.3 psi for the loop, but given the difficulty of this type of calculation, and the assumptions
that have to be made about geometries that are still in a bit of flux, we settled on 2 psi as a conservative
head specification for the pump. This keeps viscous heating below a few hundred Watts. The calculation was first
baselined to the G0 target, and satisfactory agreement with measured head loss and flow in that target was
obtained.

The \Qweak flow and head parameters point squarely to a centrifugal pump design. The requirement that the target
be able to move horizontally as well as vertically favors a submersible design over one with an external motor.
Commercial pump vendors want more money than we have budgeted for a pump meeting our specs. As a result an
effort to build one in-house is just getting off the ground. Tests planned for the spring and summer 2008 will
tell us if our in-house effort has been successful or not. If not then we will still have time at that point to
go commercial, with a little help from the lab to push our budget envelope.

\subsection{Target Motion} The target motion systems have been designed and most of the parts have been procured.
A weight of 2000 lbs was assumed for the design of both systems. The horizontal motion system will provide
$\pm$2" of travel. The vertical lifter will provide 22" of travel, enough for the $LH_2$ target, dummy targets
for background subtraction, and several solid targets including optics foils, and a target out position. A
position repeatability of better than 13 $\mu$m can be achieved. As part of this effort, the scattering chamber
was also designed and most parts procured, as well as the $H_2$ relief system internal to the scattering
chamber.

The latter will consist of a 2 $\frac{7}{8}$" diameter cold, straight pipe inside a concentric heat shield,
which penetrates the top lid of the scattering chamber into a concentric bellows to accommodate the full range
of motion of the target lifter system. The straight pipe will tie in to the loop via a short length of 3" flex
hose at the top of the loop, which accommodates the small horizontal motion. The design further accommodates a
small diameter fill line connected to the opposite side of the pump. This small $\frac{1}{4}$ " line, needed to
measure the pump head, will be situated inside the larger return pipe and thus will share the return line's
thermal shield and bellows.

Both a thick and a thin dummy target are planned. Both are being designed such that scattered electrons which
reach the quartz do not pass through any of the other targets. The optics targets will be used primarily to tune
the vertex reconstruction from the region 1 and 2 chambers. Separate vertical and horizontal wire grids will be
placed at several z locations, with the raster system set up to illuminate all wires. The solid targets
envisioned include a hole target for beam position and halo studies, a BeO viewer, and a C target for basic
tuneup operations.

\subsection{Heater}
The first of two heaters has been successfully built and characterized in LN2 at Mississippi State. The heater
is wound in four parallel sections from 0.057" diameter nichrome wire. The
resistance at 80K was 1.226 Ohms. Based on these actual measurements, two 60V, 50A heater power supplies were
purchased.

\subsection{Relief/Safety Calculations:} This work has begun. The first step was to reproduce the calculations
that have already been performed for the existing Hall C standard pivot target. The code developed for that
check can now be considered a reliable template for the \Qweak application. Our goal is to continue to move
forward on this such that we are in a position to defend our design at a design and safety review by March,
2008.

\subsection{Target Boiling Considerations}

In order to mitigate the effects of target boiling on the measurement, the $LH_2$ must flow as fast as possible
across the beam axis. Alternatively, the beam must move more quickly across the target fluid. An effort to
double the existing raster frequency was completed successfully on the bench and will now become part of our
default experimental configuration for \Qweak.

Likewise, increasing the helicity reversal frequency from the standard 30 Hz to 250 Hz should help mitigate
boiling in two ways. First, the noise spectrum at 250 Hz is quieter than at 30 Hz. Second, the target boiling
contribution can be about three times larger than at 30 Hz without increasing the experiment's running time,
because the statistical width per quartet is three times greater at 250 Hz. Tests were completed in the spring
of 2007 which demonstrated that 250 Hz helicity reversal can be delivered by the accelerator.
\section{Simulations and Backgrounds}

In the 2004 PAC proposal~\cite{Carlini2} for \Qweak, we discussed initial simulations of backgrounds from a
variety of sources, including the background in the \qw measurement originating from the $LH_2$ target windows.
For 3.5 mil thick aluminum windows, we indicated an expected contribution of about 11\% of the free $ep$ elastic
asymmetry, which must be measured and corrected for.  Those background studies are explicitly included in our
\Qweak beam request.  Since the last PAC submission, we have done extensive studies with the GEANT-based
\Qweak{} simulation to identify other sources of backgrounds, as well as to quantify and reduce them by
optimizing the design of the \Qweak collimator system. This work is the focus of our proposal update discussion;
as our simulations have improved and the design of the collimator system has evolved, changes to reduce
backgrounds were not allowed to negatively affect the figure-of-merit (FOM) for the experiment.

The \Qweak simulation model was originally developed from the G0 simulation; both experiments use toroidal
magnets. The origin of our coordinate system is at the at the center of the QTOR, with z along the beam
direction, x vertical, and y toward beam-right, making a right-handed system. The simulation includes the \LH2\
target, target windows, the beamline, the acceptance-defining collimator that has a clean-up collimator upstream
and down stream of it, the QTOR magnetic field based on a recent calculated field map, QTOR coils, QTOR support
structure elements that are near the $ep$-elastic envelope, lintel-like photon shields, a shielding hut wall,
and quartz \v{C}erenkov bars. A GEANT-generated view of equipment that is typically used in investigations of
backgrounds is shown in Fig.~\ref{fig_geant-pers}. We track secondary electrons and photons down to 0.5 MeV
because \v{C}erenkov light production is barely possible with 0.35 MeV photons in fused silica, which has an
index of refraction of 1.48.

\begin{figure}[h!]
\centerline{\includegraphics[width=5.5in]{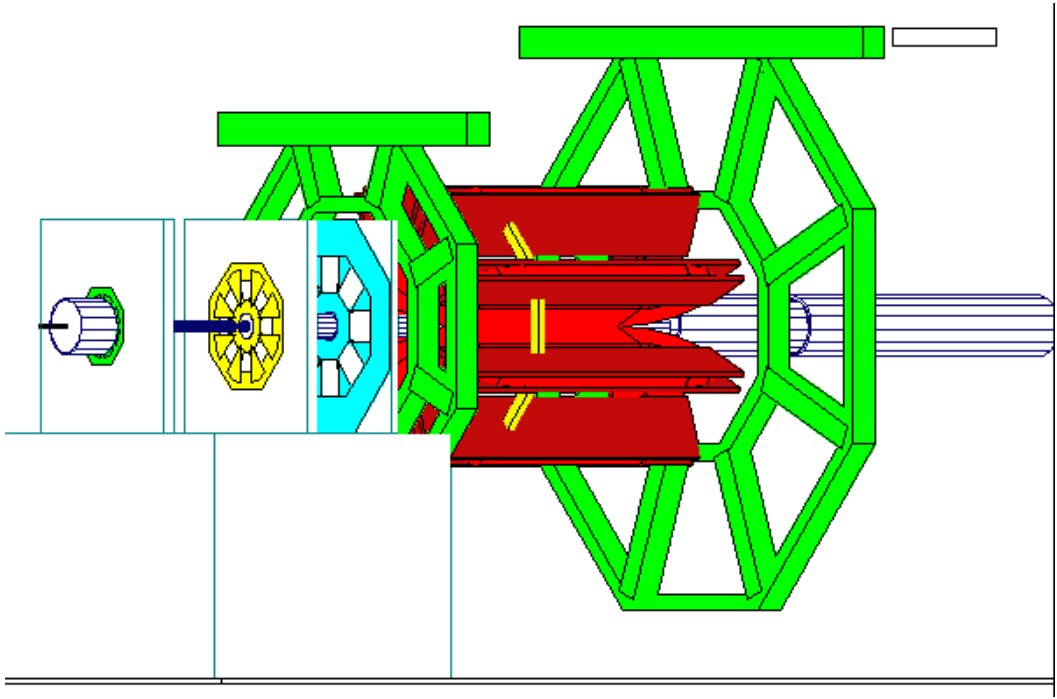}} \caption{{\em GEANT-generated view of
equipment elements used in a typical simulation to investigate backgrounds.  The upstream clean-up collimator is
shown in green, the acceptance-defining collimator is yellow, the downstream clean-up collimator is aqua, the
QTOR coils are red, the QTOR support structure is green, the lintel photon shields are yellow, and the
\v{C}erenkov bar is white.  The shielding hut wall has been removed from this figure.
    }}
\label{fig_geant-pers}
\end{figure}

\subsection{Collimator Design}
\label{sim-status}

The \Qweak collimator system will play a crucial role in defining the $Q^2$ acceptance and the figure of merit
for the experiment. Its geometrical symmetry and alignment with respect to the target and detector systems will
be major factors in determining the sensitivity to systematic errors associated with helicity correlated beam
motion. A very substantial effort has thus been spent on optimizing the collimator design using the \Qweak GEANT
simulation software.

When the defining collimator opening was finalized, the support structure for QTOR was already fixed.  This
defined the maximum size of the scattered electron envelope through the QTOR region. The initial step  in
optimizing the design was to choose the aperture and longitudinal location of the collimator to give the largest
possible acceptance while not interfering with the QTOR support structure.  Both upstream and downstream
locations were considered. The downstream option -- as close as possible to the entrance of QTOR -- gave the
larger acceptance; this is because the extended target becomes more ``point-like'' as the defining aperture is
moved downstream. Once this maximum aperture was determined, the collimator was ``trimmed'' further in order to
fit the scattered electron envelope onto quartz detector bar of reasonable size and shape at the focal plane.
Extensive studies were carried out to optimize the shape of both the collimator aperture and the detector to
minimize the overall error on \qw, while also keeping the contamination from inelastic events acceptably low.
The final collimator design consists of three sequential elements, the middle of which is the
acceptance-defining collimator, with the other two inserted for `clean-up' purposes.  A summary of the \Qweak
collimator geometries is given in Table \ref{tab:collimators}.

\begin{table}[h]
\begin{center}
\caption{ {\em \Qweak collimator system.   All elements are made of a machineable alloy consisting of 95\% Pb
and 5\% Sb}} \vspace{0.25in}
\begin{tabular}{|l|c|c|}
\hline Element & Upstream z (cm) & Thickness (cm)  \\ \hline Upstream  & - 583.4 & 15.2 \\
Acceptance Defining & -385.7 &  15 \\
Downstream clean-up & -271.9 & 11   \\

 \hline \hline
\end{tabular}
\label{tab:collimators}
\end{center}
\end{table}

While the first two elements of the collimator system will precisely machined, it is desirable that the
$ep$-elastic electrons that are detected by the \v{C}erenkov bars do not hit the upstream and downstream
collimators. If we had a point-like target and thin collimators, this would be a simple geometry problem; with a
35 cm long target and collimators tens of radiation lengths thick, the situation is less straightforward. We
used the JLab 3-D CAD code to design the upstream collimator aperture so that it clears the $ep$-elastic
envelope by at least 0.5 cm, and we verified this clearance with the GEANT simulation. We show the envelope of
the $ep$-elastic electrons that are detected by the \v{C}erenkov bar at the upstream and downstream sides of
collimators \#1 and \#2 in Figures \ref{fig:coll1} and \ref{fig:coll2}, respectively. Note that the envelope
fills the downstream side of the acceptance-defining collimator and easily clears collimator \#1. The image for
collimator \#3 is similar to that of collimator \#1.

\begin{figure}[hbtp]
\begin{center}
\subfigure[] {\includegraphics[width=3.0in]{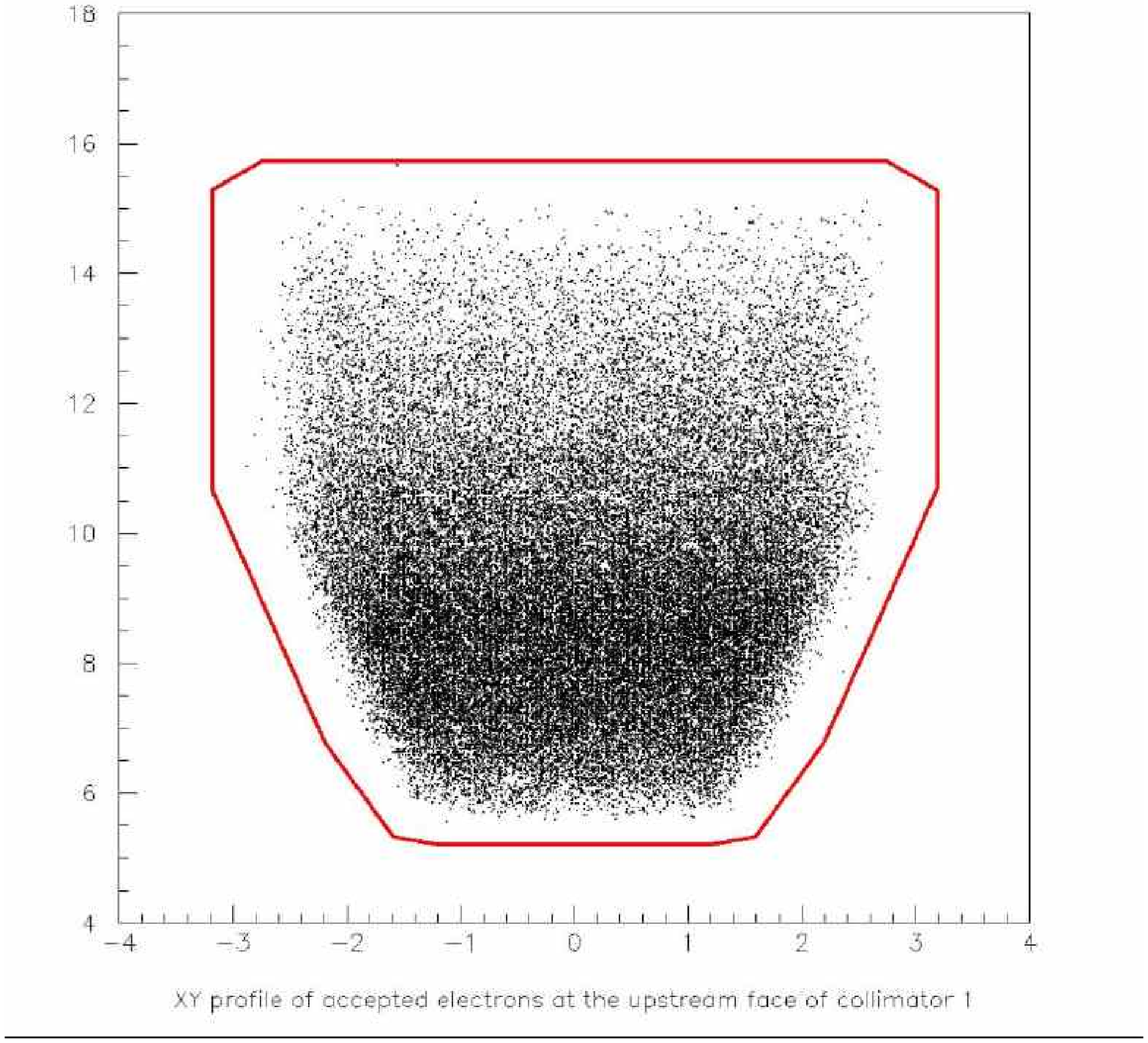}}
\subfigure[]{\includegraphics[width=3.0in]{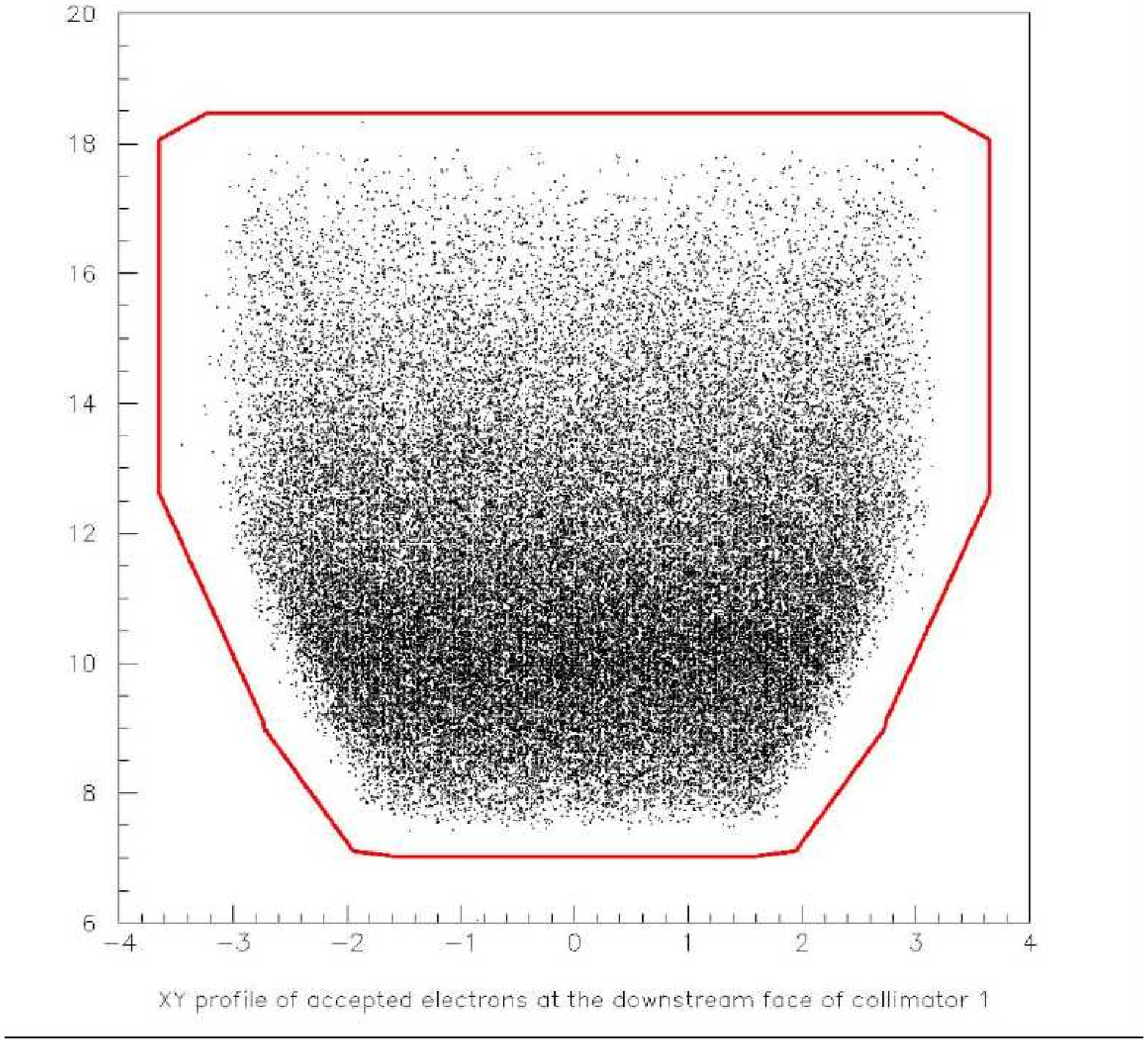}} \caption{{\em X-Y image of the ep-elastic electrons
that hit the  \v{C}erenkov bar at collimator \#1.  a)  upstream end;   b)  downstream end.  The aperture of
collimator \#1 is shown in red.
    }}
\label{fig:coll1}
\end{center}
\end{figure}

\begin{figure}[hbtp]
\begin{center}
\subfigure[] {\includegraphics[width=3.0in]{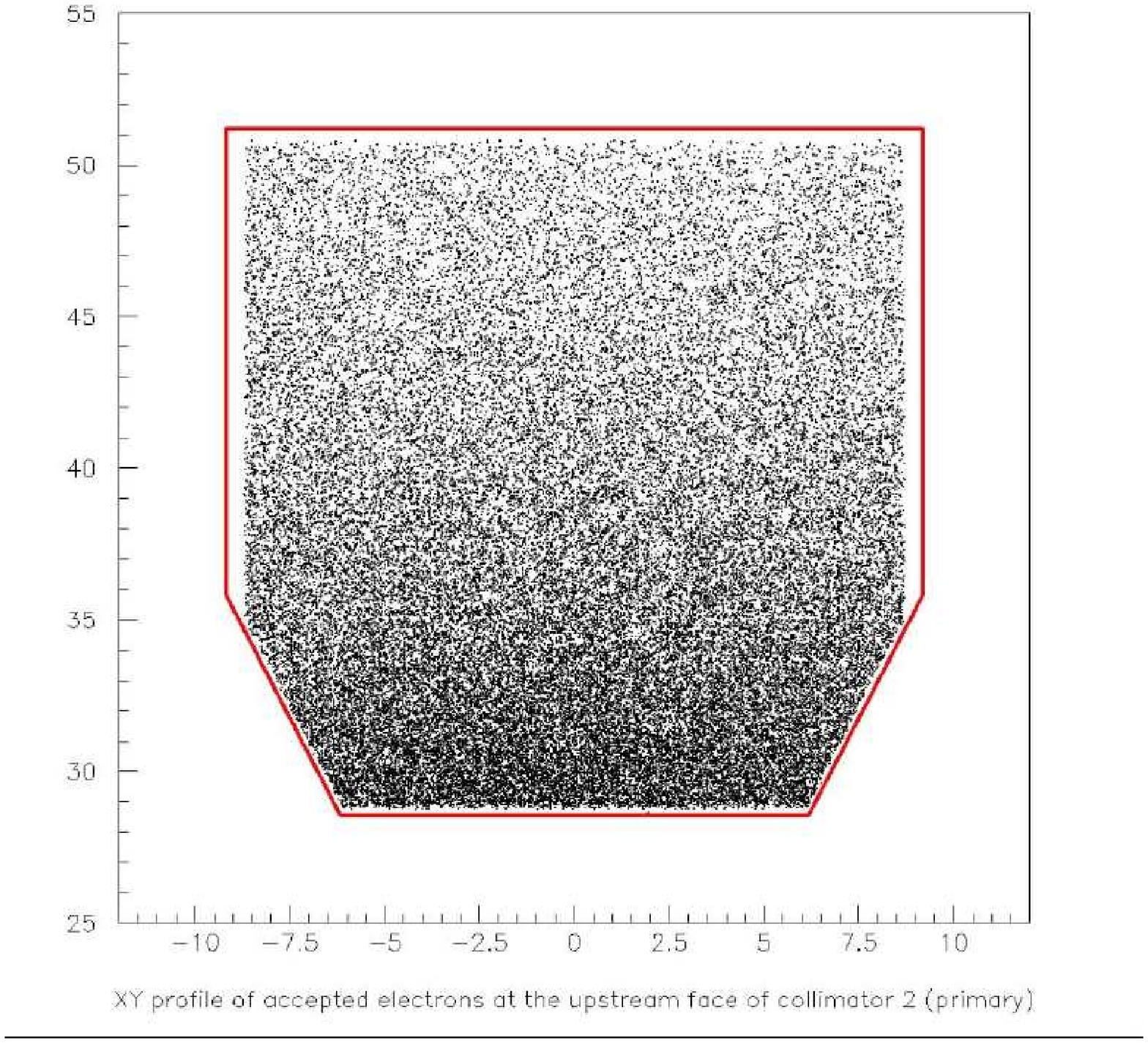}} \subfigure[]
{\includegraphics[width=3.0in]{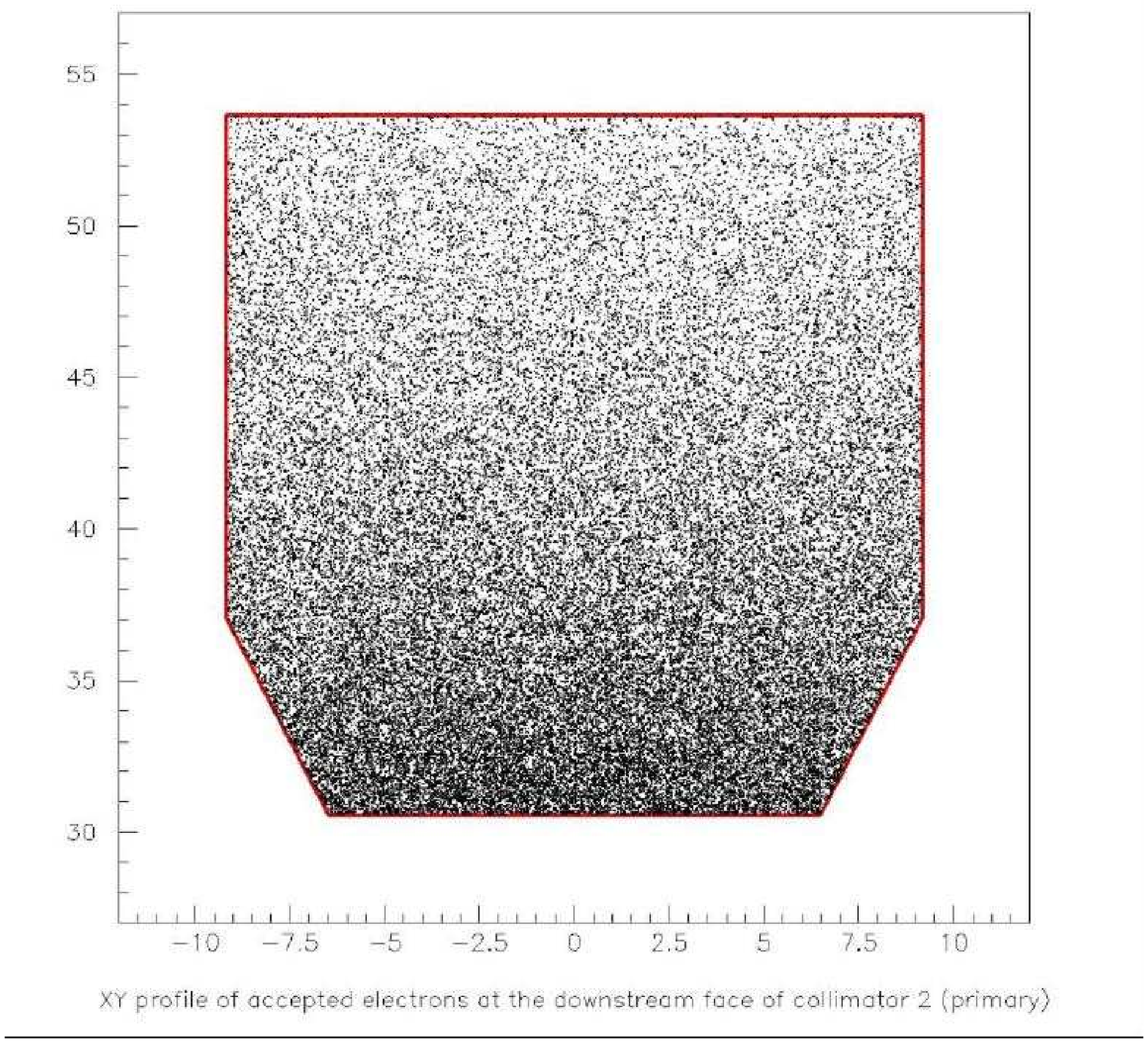}} \caption{{\em X-Y image of the ep-elastic electrons that hit the
\v{C}erenkov bar at  collimator \#2.  a)  upstream end;  b) downstream end.  The aperture of collimator \#2 is
shown in red.
    }}
\label{fig:coll2}
\end{center}
\end{figure}


\subsection{Backgrounds}

 After a decade of commissioning large experiments at JLab, it is clear to us that signals
 are often easy to calculate, but backgrounds are difficult to estimate and are therefore
 one of the most important factors in determining the success or failure of an experiment.
 Essential features of the  \Qweak apparatus  favoring a high signal to noise ratio are the
 use of a magnetic
 spectrometer to separate elastic from inelastic events, and the choice of main detectors that
 are located in a shielded detector hut and are sensitive only to relativistic particles.
 However, the integrating nature of our
 experiment and our 2\% asymmetry goal mean that we are potentially sensitive to percent-level
 soft backgrounds which may be difficult to measure and correct with high accuracy.

 The \Qweak collaboration has had a strong simulation team since the original proposal, and
 a significant part of this effort has been dedicated to background reduction.
 The experiment has adopted a 2-bounce design
 philosophy, which means that indirect backgrounds must have scattered at least twice after
 leaving the target before they reach the main detector.
 Where possible, the design strategy is
 2-bounce-plus-shielding.
 Our efforts over the last 3 years can be summarized as uncovering potential percent-level
 backgrounds, modifying the design of the experiment to reduce them by an order of magnitude,
 and then developing strategies to measure the remaining effects.

Since the last Jeopardy proposal, a potential neutral background of the 1-bounce type of $\cal{O}$ (1\%) was uncovered\cite{Liang}.
Insertion of a lead block in each octant will prevent $\gamma$ rays
created on the defining collimator from reaching the main detector, hence reducing this potential background by
almost an order of magnitude. We also discovered that a significant 1-bounce background can arise if the shield
house window aperture is too tight on the low energy loss side, causing showering into the main detector. The shield
house window will be designed to avoid this.

Our current background estimates are summarized in Table \ref{tab:back_rates}. Most of these rates will decrease
as we further optimize the design of the experiment and make the simulation more realistic. For example, the
most significant background in Table \ref{tab:back_rates} is partly from Compton scattering of M\o
ller-generated $\gamma$ rays in air. At 0.6\%, it seems unusually large for a 2-bounce background;  while this
background cannot be eliminated completely without replacing the air with vacuum, preliminary studies suggest
that it will be reduced by 2/3 with the incorporation of a shield house to reduce the solid angle acceptance for
this background source.  It will be reduced further if a thin dead-layer is placed in front of the main
detector bars to stop electrons below a few MeV.

We have a good understanding of expected 0-bounce backgrounds in \Qweak, $i.e.$ backgrounds coming directly from
the experimental target such as target window backgrounds and inelastic electrons from pion electroproduction.
There has been significant refinement in the inelastic studies, described in detail in the following section.

\begin{table}
\begin{center}

\caption{ {\em Background rates in the \Qweak main detectors (relative to the elastic rate and weighted by
relative asymmetry and light production) as predicted by the \Qweak simulation using the latest collimator
design plus an internal "lintel" collimator to block photons, but no detector shield house. The highest (M\o
ller) rates will be reduced by a shield house; the  next highest rate (inelastic electrons) is very sensitive to
the shielding house aperture, which remains to be optimized. It is important to note that these are background rate
estimates, and not the uncertainty to which they could be corrected - which will further suppress their contribution
to the final experimental uncertainty. }} \vspace{0.25in}
\begin{tabular}{|l|c|}
\hline Background & Rate \\ \hline M\o ller Electrons & 0.58\% $\pm$ 0.04\% \\ \hline M\o ller Photons &  0.21\%
$\pm$ 0.01\% \\ \hline Inelastic Electrons & 0.250\% $\pm$ 0.012\% \\ \hline
Elastic Photons & 0.15\% $\pm$ 0.005\% \\ \hline
Inelastic Photons & negligible     \\ \hline
\hline
\end{tabular}
\label{tab:back_rates}
\end{center}
\end{table}

\begin{figure}[hbtp]
\begin{center}
\rotatebox{0.}{\resizebox{6in}{6in}{\includegraphics{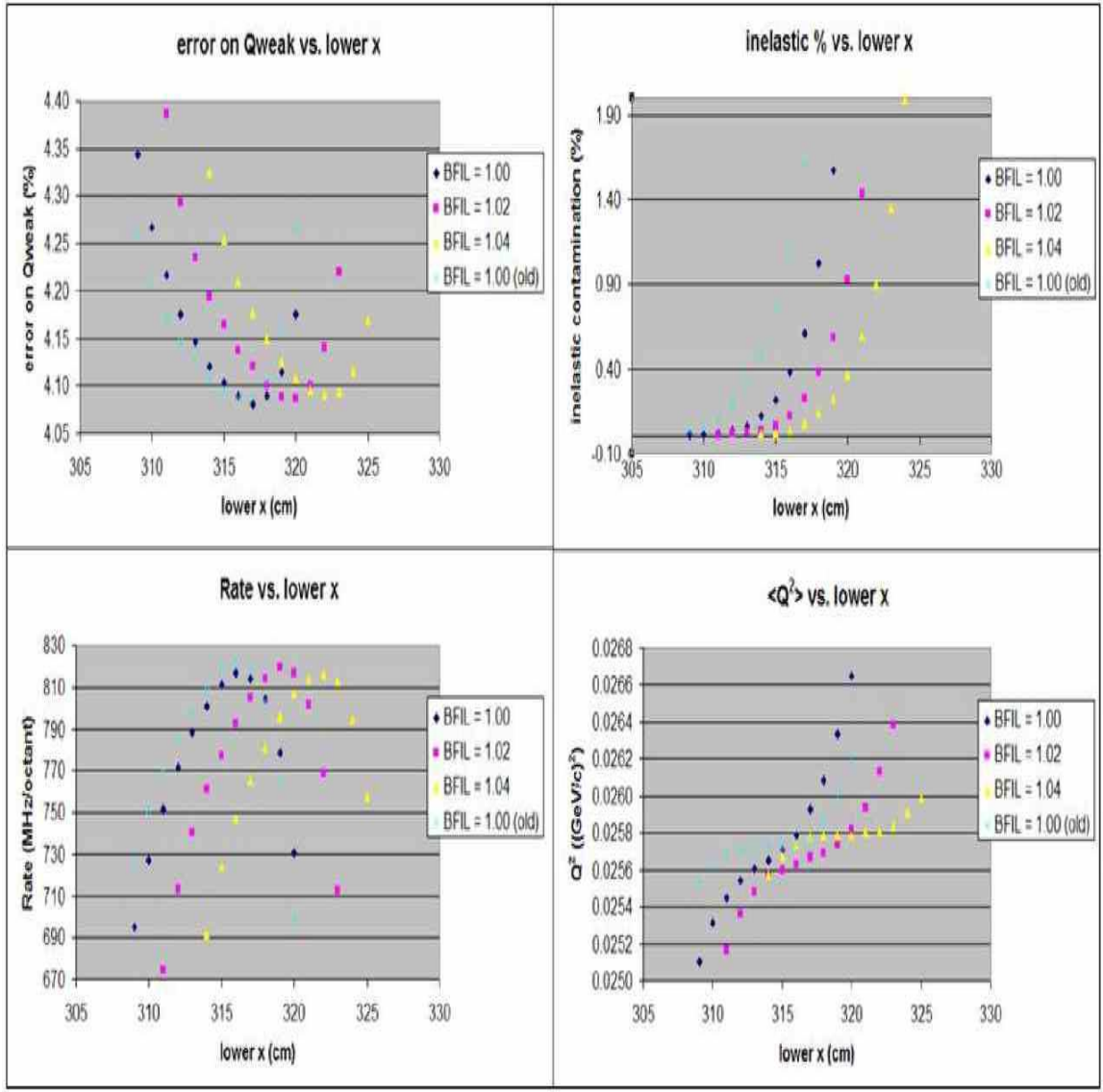}}}

 \end{center}
\caption{ {\em Systematic study for E = 1.165 GeV in which the QTOR magnetic field scale factor, BFIL, and the
position of a radiator bar are varied
 to optimize running conditions. The `x' coordinate refers to the radial distance from the beamline of the lower end of the top detector bar.
 Clockwise from the upper left panel:
statistical error on the proton weak charge, inelastic background correction
required (weighted
for both rate and asymmetry),
 the first moment of $Q^2$ neglecting detector bias, and the elastic rate
on a single bar. The nominal operating point we have selected is BFIL = 1.04 with the lower edge of the radiator
at 319 cm. } }\label{JulietteScan}
\end{figure}

\subsection{Inelastics from pion electroproduction}
\label{inelast-back}

Most of the inelastic electrons due to pion electroproduction are swept off the radiator bars by the QTOR
magnetic field. A seemingly negligible fraction, only 0.02\% by rate, strikes the outer radial edges of the
bars. However, since the inelastic astymmetry is expected to be an order of magnitude larger than the elastic
asymmetry\cite{hammer}, the correction for this inelastic background is estimated to be about 0.2\%.

Since the last PAC update, we have systematically examined the dependence of the \Qweak statistical error and
the inelastic background on the radial position of the radiator bars. As the lower edge of a bar is moved to
larger radius, the statistical error on \Qweak initially decreases, as shown in the upper left panel of Figure
\ref{JulietteScan}. However, if the radial coordinate increases too much, the statistical error increases again
because the elastic locus begins to slip off the lower edge of the bar. Furthermore, an increasing radius
corresponds to larger energy loss, resulting in a rapid increase in the inelastic contamination with increasing
radial position, as shown in the upper right panel of Figure \ref{JulietteScan}.

In this trade-off between between statistical and systematic errors, we have conservatively assumed a relatively
large uncertainty on the inelastic background.  In the lower right panel of Figure \ref{JulietteScan}, one notes
a few cm wide plateau in which the average $Q^2$ is fortuitously stationary, which would allow us to reduce
another potential systematic error. For these reasons, we have chosen a nominal radius of 319 cm until we can
confirm the simulations during commissioning.

It is expected that our tracking detectors will permit a clean separation of elastic scattering and pion
electroproduction near threshold, allowing us to determine the relative rate of inelastic tracks to high
accuracy. As for the inelastic asymmetry, since it is expected to be relatively large,  it should be possible to
measure it in current mode with small statistical errors in only a day of beam time. This will require lowering
the QTOR field, which will change its focusing properties and will dump elastic electrons onto the front of the
detector shielding wall. The uncertainty in the inelastic asymmetry measurement will be dominated by
systematics, such as in-showering from elastic electrons striking the inner radial edge of the shield house
windows. Data from the $G^0$ backward angle run will also provide a cross-check on predictions of the inelastic
asymmetry.

\subsection{M\o ller scattering ($e+e\rightarrow e+e)$}
\label{soft-back}

During the final optimization of the experimental layout, we moved our defining collimator downstream to improve
several important contributions to the figure of merit. However, it then became possible for the main detector
to directly view an illuminated portion of the defining collimator, producing an $\cal{O}$(1\%) soft background.
M\o ller electrons have fairly low energy in the neighborhood of 100 MeV, so they are not transmitted through
the QTOR field and into the main detector region.  In simulations, these low energy electrons produce a rather
distinctive ``fountain'' as they are repelled by the QTOR field.  However, since the M\o ller cross section is
roughly 1000 $\times$ larger than the $ep$ elastic cross section at our kinematics, dumping them on the 2nd,
acceptance-defining collimator then produced  a $\gamma$ flux through the main detectors which rivaled the flux
of elastic electrons.

Increasing the spectrometer bend angle would in principle allow the 3rd collimator to block all of the hot spot
on the 2nd, defining collimator. However, as this would move the elastic focus too far upstream,  it was
rejected as an undesirable option. Our solution is to simply insert a single lead baffle into each octant of
QTOR to block the $\gamma$ rays directed toward the main detector, as illustrated in figure \ref{fig:sideview}.
This additional element is 15 radiation lengths of lead (8.4 cm) thick, 16 cm high, and 62 cm wide; it is
referred to as the ``lintel'' collimator as it sits between QTOR coils like a lintel. Alignment requirements for
the lintel are relatively loose: there will be 1 cm separation between the nearest edges of the elastic electron
envelope and the $\gamma$ rays we wish to block.

\begin{figure}[ht]
\centerline{\includegraphics[width=6.0in]{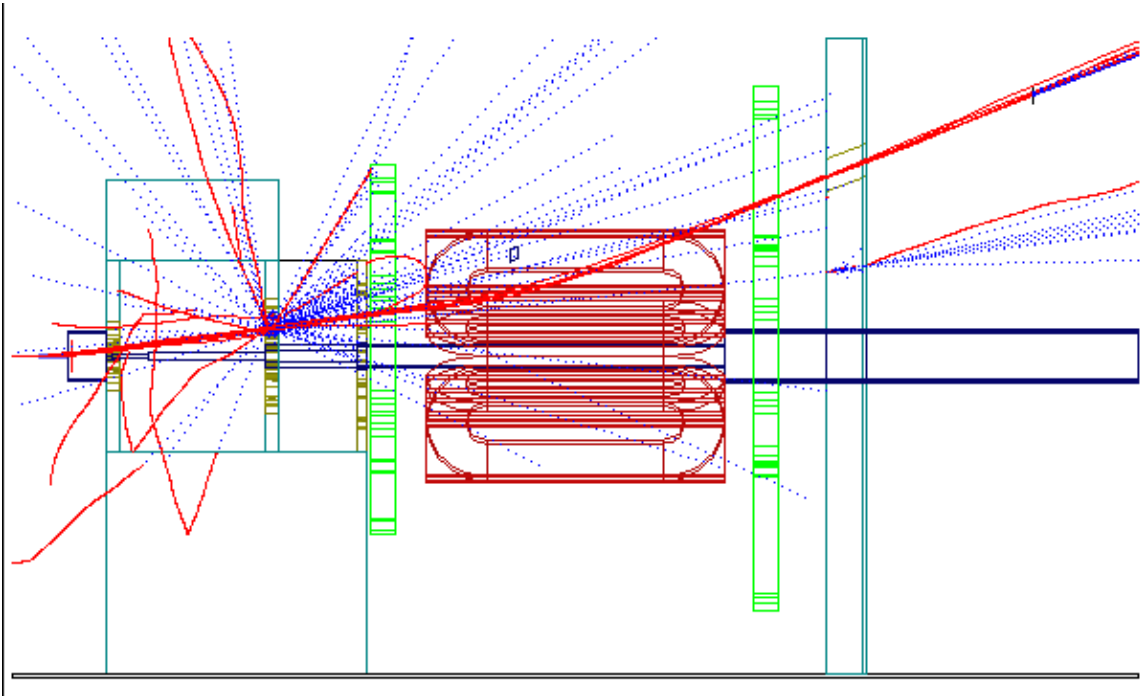}} \caption{{\em Side view of the \Qweak
apparatus with simulated events.  Electrons, shown in red, are selected by collimator \#2 and are deflected by
QTOR toward the main detector \v{C}erenkov bar. Some electrons hit the collimator \#2 aperture and produce a
photon shower that is in direct line of sight to the \v{C}erenkov bar.  These photons are blocked from hitting
the bar by a lintel collimator shown in white in the top octant.
    }}
\label{fig:sideview}
\end{figure}

We also checked and found the lintel itself to be a negligible source of additional background. It is located
deep enough inside QTOR that most of the M\o ller electrons do not penetrate to that distance, but not so deep
that much dispersion has built up for low energy loss electrons. Thus, only a narrow range of high energy loss
electrons are intercepted by the lintel, and the lintel is thick enough to stop most of the shower products.
Most of the small, remaining M\o ller-generated photon background (0.2\%) originates from azimuthally defocused
M\o ller electrons which strike the QTOR coils before they are ejected by the magnetic field.


\subsection{Beam-defining collimator}

As part of our overall strategy to minimize background in the main
detector, we plan to insert a beam-defining collimator downstream of the
target. Without this, scattered events from the target would
directly illuminate the entire beampipe from shortly downstream of the
target to the exit of the QTOR beampipe, presenting a difficult shielding
problem.
 Of course, events can still strike this
region of the beamline after one bounce from the beam-defining collimator. The beamline inside the detector hut
is therefore a source of 2-bounce background, but it is separated from the main detectors by 4'' of lead
shielding. Simulations still have to be done to verify whether this is sufficient. It would be expensive and
cumbersome to upgrade the beamline shielding, but on the other hand, there is plenty of room for additional
local shielding of the main detector modules.

To describe the plan in more detail, small angle scattered particles (in the 0.5 - 4.5$^{\circ}$
range) will be blocked by a $\sim$ 20 radiation length tungsten collimator
about 1 meter downstream of the target.  This collimator is designed so
that any scattered particles that do not interact with it experience their
first interaction in the beampipe far downstream of the main detectors.
The collimator will be fabricated from a ductile, sintered W-Cu mixture to
prevent shattering in the event of a beam strike. The site boundary dose
is being calculated, as is the power deposition in the collimator. The
results of these calculations will allow us to finalize the design of the
beam-defining collimator and its local shielding. In a small region just
downstream of the target, residual activities are expected to quickly
exceed the threshold for a High Radiation Area (1 R/hour on contact).


\subsection{Neutrons and other soft backgrounds}

A \v{C}erenkov radiator medium only produces light due to the passage of relativistic, charged particles.
Furthermore, our Spectrosil 2000 radiator material consists of almost pure $SiO_2$, so it contains no free
protons and has a very small scintillation coefficient. Our main detector is therefore insensitive to neutrons,
but light production by neutrons is still possible by multi-step processes, and the backgrounds may be
significant if the neutron field is sufficiently intense. We are steadily acquiring the tools to simulate this
difficult problem, and since any simulations would have to be
 benchmarked, a proposal to put some of our detector elements into an epithermal neutron beam
at LANSCE for benchmarking purposes has been submitted.

Below we summarize the scope of an initial neutron simulation and how neutron backgrounds can
be quantified.

The principal light production mechanism by neutrons in our detectors will be from the production of an excited
compound nucleus via neutron absorption, which decays to the ground state, emitting $\gamma$-rays. The
$\gamma$-rays then Compton scatter from the atomic electrons. Depending on the particular element, the isotope
may then decay further emitting either a $\beta$ or $\alpha$. However, both $^{29}Si$ and $^{17}O$ are stable
isotopes, so that the latter is not the case to first order.  Because low energy neutron capture cross sections
are proportional to $1/v$, with $v$ being the velocity of the neutron, upcoming simulations will focus on
thermal to epithermal neutrons which we refer to henceforth as slow neutrons.  Our first concern will be the
$SiO_2$ radiator material due to its large volume and direct optical coupling to the PMT's. Simulations will
focus on $Si$ because its slow neutron capture cross sections are many orders of magnitude larger than those on
$O$ and yield a fairly hard $\gamma$-ray spectrum.

 Backgrounds due to activation of detector materials with $>>$ 1 second decay time constant
will show up as an apparent slow shift in the detector pedestal. Provided that we monitor this pedestal shift,
there will be no signal dilution. Since the accelerator routinely trips off every 10 minutes or so, and an
accurate pedestal reading takes far less than a second, we will have plenty of data to quantify such long
time-constant backgrounds no matter where
 the neutron capture occurs (the radiator, the PMT's, detector support structures, shielding
concrete, etc.) Solid-state relaxation phenomena in the fused silica radiators such as long-lived luminescence
can be studied in the same fashion.

Any soft background with $<<$ 1 second decay time constant cannot be treated as a pedestal shift and so will
produce a dilution of the elastic electron signal. To help quantify these effects, we plan to move soft
background detectors to various locations inside the detector shield house. These signals will be acquired with
the preamplifers set to 50 times higher gain than nominal to provide sensitivity of better than 0.1\%. Three
movable types of background detectors (a complete detector assembly, a PMT in a small dark box, and a
preamplifier) will help us understand the source of any background. This information can be combined with
simulations and dosimetry information from standard TLD's to suggest shielding improvements as needed.

The soft background detectors described above are most useful for diffuse backgrounds.
However, the main detectors can also see soft backgrounds which are beamed through the
 window in the shield house wall. In principle, this can be quantified during pulsed-mode
running by triggering on the main detector with a very low (0.5 photoelectron) threshold and looking for an
appropriate minimum ionizing hit in the overlapping trigger scintillator. The bias would be small, but it might
be difficult in practice to set the threshold that low. A more robust and completely unbiased method will be to
take continuous blocks of main detector  and scintillator data using 250 MHz flash ADCs, and correlate the two
signals offline. Appropriate prototype modules from the JLab Electronics Group are at least a year overdue, but
may be available soon. If it looks like production versions of the JLab flash ADC's won't be available in time,
then we will purchase much more expensive but off-the-shelf versions from Struck.

Since our last Jeopardy proposal, we understand better how the very high single photoelectron rates from our S20
photocathodes will affect the soft background measurements described above. Such dilution will be completely
negligible during production running. However, during pulsed mode running, reducing the dark rate dilution to
O(0.1\%) will require Region III rates of 0.5 MHz, hand-picking our lowest noise PMTs, and a well
air-conditioned operating environment.
\section{Systematic Errors and Polarized Source Requirements}
\label{systematics}

Changes of beam properties with helicity can lead to false parity asymmetries. Parity violating scattering
experiments generally have dealt with this by keeping helicity correlations as low as possible, by measuring residual
correlations and by making corrections for them based on measured sensitivities. The measured parity asymmetry,
$A_{meas}$, is written in terms of the physics asymmetry, $A_{phys}$, in the following way for sufficiently small
helicity correlations:
\begin{equation}
A_{meas} = A_{phys} + \sum_{i=1}^{n} \Big(\frac{\partial A}{\partial P_i}\Big)\delta P_i,
\label{eq:corr_procedure}
\end{equation}
where beam parameter $P_i$\ changes on helicity reversal to $P_i^{\pm} = P_i \pm \delta P_i$. The detector
sensitivities $\partial A/\partial P_i$\ can be determined preferably by deliberate modulation of the relevant
beam parameter or from natural variation of beam parameters. The helicity-correlated beam parameter differences,
$\delta P_i$, are measured continuously during data-taking. From estimates of the sensitivity of our apparatus,
we can set requirements on how accurately beam parameters have to be measured and upper limits on acceptable
helicity-correlated beam properties.

The 2004 update proposal described GEANT simulations that led to predictions of the sensitivity of our apparatus
to helicity-correlated beam intensity, position, angle, and size modulations, and set limits on both DC and
helicity correlated beam property values aimed at constraining individual beam-related false asymmetries to be
no larger than 6 $\times 10^{-9}$, i.e. the same size as the statistical error in the parity asymmetry
measurement.  Ancilliary measurements and diagnostic apparatus must be sufficiently sensitive to permit
systematic error corrections to be made to $\pm$ 10\% of this upper limit in each case.

These sensitivity estimates have been updated as the designs of the target, collimator and detector systems have
evolved to their current, final specifications. In addition, considerable experience has been gained from recent
PVES experiments at JLab, and it is now both advisable and feasible to aim for more stringent control of most
beam-related systematics for \Qweak. Accordingly, our goal is now to keep individual beam-related false
asymmetries to be no larger than 6 $\times 10^{-10}$.  The only exception is for helicity-correlated size
modulation, where we are developing techniques to measure small modulations for the first time at JLab -- in
this case, we retain our previous criterion for the maximum false asymmetry to be no larger than 6 $\times
10^{-9}$ to set an initial goal for source and instrumentation development to address this challenging
systematic issue.

\subsection{Summary of Beam Requirements}

Table  \ref{Table:BeamSpec}  gives limits on allowable beam properties and detector asymmetry which should keep
any false asymmetry generated by helicity correlations in the beam to less than $6\times 10^{-10}$. Column 2
gives the limit on DC values while column 3 shows the limit on the helicity-correlated properties averaged over
the whole run.  Column 4 shows the allowable random noise in a measured beam parameter which is consistent with
meeting our systematic error goals.

\begin{table}[h]
\caption{{\em Summary of systematic error requirements for \Qweak.}\label{Table:BeamSpec}} \vspace*{0.2cm}
\begin{tabular}{l|c|c|c} \hline
Parameter       & Max. DC value               & Max. run-averaged           & Max. noise during \\
                &                             & helicity-correlated value   & quartet spin cycle \\
                &                             & (2544 hours)                & (8 ms) \\ \hline
Beam intensity      &      & $A_Q < 10^{-7}$ & $< 3 \times 10^{-4}$ \\ \hline
Beam energy     & $\Delta E/E \leq 10^{-3}$ & $\Delta E/E \leq 10^{-9}$  & $\Delta E/E \leq 3 \times 10^{-6}$ \\
                & ($Q^2$\ measurement)        & 3.5 nm @ 35 mm/\%            & 12 $\mu$m @ 35 mm/\%  \\ \hline
Beam position   & 2.5 mm                      & $\langle \delta x \rangle < 2$\ nm         & $7 \ \mu$m \\
\hline Beam angle      & $\theta_0 = 60\ \mu$rad     & $\langle\delta\theta \rangle < 30$\ nrad  & $100\ \mu$rad
\\ \hline
Beam diameter   & 4 mm rastered               & $\langle\delta\sigma \rangle <$0.7 $\mu$m                  & $< 2$ mm \\
                & ($\simeq 100\ \mu$m unrastered) & (unrastered)            & \\ \hline \hline

\end{tabular}
\end{table}
\vspace*{0.2cm}

\subsection{Parity Quality Beam at JLab}

\paragraph*{Beam Intensity:}

The \Qweak experiment must control the integrated beam intensity asymmetry between the beam helicity states to
be smaller than 0.1 ppm.  The HAPPEx and G0 collaborations have demonstrated control of this asymmetry at the
level of 0.2 ppm, via careful setup  and implementation of a feedback system. The intensity asymmetry was
measured continuously using beam charge monitors in the experimental halls. The resulting values were  used to
determine the necessary corrections, which were applied at the polarized source.

The dominant cause of intensity asymmetry in the polarized source is a difference between the small residual
component of linear polarization in the nearly (99.9\%) circularly polarized laser beam for each helicity state.
This difference interacts with the intrinsic analyzing power in quantum efficiency of the strained GaAs
photocathode to modulate the electron beam intensity, correlated with helicity. While it is possible to
compensate for this in a feedback loop by selectively attenuating the laser intensity, this approach does not
eliminate the difference in linear polarization which gives rise to the effect in the first place. Linear
polarization components also contribute to changes in the electron beam trajectory and spot size, and in fact
the laser attenuation system itself can also contribute to changes in the electron beam trajectory.

A preferred approach to reduce intensity asymmetries is to correct the difference in linear polarization of the
laser beam between helicity states by applying offset voltages to the Pockels cell used to create the circular
polarization. Electron beam intensity asymmetries can be adjusted using these voltage offsets, which are
commonly referred to as ``PITA'' voltages. In beam studies, helicity-correlated changes in electron beam {\em
position} have consistently been seen to be reduced when the charge asymmetry was corrected using the PITA
voltage. This effect is not typically observed with intensity-attenuation methods. While both correction
mechanisms have been successfully employed in feedback control of electron beam intensity asymmetries, the PITA
mechanism is preferred for the reasons explained here.

For \Qweak, we anticipate using a similar feedback system to that used by G0 and HAPPEx-II, but updating PITA
voltage corrections on the time scale of 10-100 seconds, with only small upgrades required to accommodate the
faster helicity flip rate.

\paragraph*{Beam Position and Angle:}

The \Qweak experiment must control the helicity-correlated asymmetry in beam position to 2 nm and in angle to 10
nrad.  The HAPPEx-II collaboration, working with the electron gun group, was very successful at controlling
these position differences at the source through a combination of carefully selected laser optics components and
novel alignment techniques~\cite{KDP2007}. In addition, significant work was performed by CASA physicists to
maintain the electron beam optics throughout the machine close to design specification, thereby avoiding
phase-space correlations which might exaggerate intrinsically small helicity-correlated effects. As a result,
helicity-correlated position differences, averaged over the HAPPEx-II run, were held to $< 2$~nm and angle
differences to $<1$~nrad, without active feedback on the beam trajectory.

In contrast, to control helicity-correlations in beam position, the G0 experiment used the ``PZT system'',
which consists of a mirror in the laser beam path mounted on a piezo-electric transducer.  The laser beam
position could be adjusted in a helicity-correlated way to compensate for any helicity-correlated beam position
measured in the experimental halls.  While this system did achieve the desired specifications, it was difficult
to maintain for two reasons. The response of the system would change with the tune of the accelerator, so the
system had to be recalibrated every 2-3 days. Secondly, there was a significant coupling between
helicity-correlated position differences and intensity asymmetries due to scraping at apertures in the injector.

For \Qweak{}, we are developing the capability to control the position and angle of the beam using corrector
coils either in the Hall C beamline or in the 5 MeV region of the injector, which is downstream of where most of
the significant interception of the beam on apertures occurs.  This will eliminate the problem of the coupling
between helicity-correlated position differences and intensity asymmetries.  If the corrector coils are
implemented in the Hall C line, then the calibration of the system should be much more constant and independent
of the accelerator tune.  Finally, the current PZT system only really allows adjustment of helicity-correlated
position differences at the experimental target.  A system based on correction coils can be used to
independently null both helicity-correlated position and angle differences at the \Qweak{} target.

It is worth noting that the estimated sensitivity of the individual HAPPEx detectors to helicity-correlated beam
position differences is approximately the same (within a factor of two) as that for the individual \Qweak
detector elements. The symmetric cancellation between the left and right High Resolution Spectrometers used by
HAPPEX-II was imperfect, and led to only a factor of $\sim 5$ reduction in sensitivity to beam motion.  In
addition, with  only two independent detectors, it was difficult to demonstrate the precision of the final
correction to better than, approximately, the size of the correction itself. \Qweak, with an 8-fold symmetric
detector system, will be capable of more complete cross-checks of the applied corrections;  a factor of 30
reduction in sensitivity to beam motion is expected by averaging over all 8 detector elements in \Qweak.

A more subtle advantage of \Qweak over the HAPPEx-II effort lies in its comparatively longer running time.  The
small averaged helicity-correlated position changes observed during HAPPEX-II were all consistent with the
statistical noise expected from the observed magnitude of beam jitter. That is, the approximately 1~nm observed
position difference represents an upper limit on the true systematic change in beam position under helicity
reversal. Assuming that the random jitter in beam trajectory is not larger for \Qweak than for HAPPEx-II, the
longer \Qweak running time will allow a measurement of systematic position differences, at the level specified
for systematic error control, early in the running period.

\paragraph*{Beam Size:}

The \Qweak experiment requires that the beam spot size must not change by more than 0.7 $\mu$m ($\delta \sigma$)
upon helicity flip.  While this effect was estimated to be negligible in previous measurements at Jefferson Lab,
it is potentially important for the high precision of the \Qweak experiment.   Studies of elements of the
polarized source optics made in preparation for the SLAC E-158 experiment have suggested that spot size
asymmetries larger than $10^{-3}$ (or 0.1 $\mu$m) are unlikely. Further studies are planned using a test bed
being developed at the University of Virginia, with the goal of placing a firm upper-limit on the possible
helicity-correlated spot size asymmetry which could be generated in the polarized electron source.

\paragraph*{Transverse Beam Polarization:}

The two photon exchange terms in the elastic scattering process produce a transverse beam spin asymmetry of the
order of $10^{-5}$ to $10^{-6}$ \cite{Carlson:2007sp,Afanasev:2004hp_v2}, comparable to the PV asymmetry. A residual
transverse component of the beam polarization will result in contamination of the PV asymmetry measurement with
the beam normal single spin asymmetry, which will largely cancel when averaged over the 8 independent main
detector elements. A reasonable limit for beam spin alignment for a precision PV measurement would be to limit
the transverse polarization component to 5\%, which corresponds to a beam spin alignment of about 3\deg to
longitudinal.

To verify the beam spin alignment in Hall C, it will be necessary to conduct a mini-spin dance during the
commissioning period, after major accelerator reconfigurations, such as energy changes, and after extended
accelerator down periods. In a mini-spin dance, the longitudinal polarization of the beam in Hall C is measured
as a function of the angle setting of the Wien filter in the injector to determine the Wien filter setting to
maximize the longitudinal polarization.  Typically data are taken at four or five Wien filter settings,
requiring about one shift of beam time.

It should be noted that azimuthal asymmetries measured with the \Qweak luminosity monitors can serve as a
monitor of the transverse component of the beam spin during standard running periods.  This technique was used
successfully in the G0 backward angle measurement.  Because the luminosity monitors can accept a mixture of M\o
ller and $e-p$ electrons, the transverse asymmetry can be difficult to calculate; however, luminosity monitor
measurements taken during either a mini-spin dance, or during dedicated transverse asymmetry measurements, will
allow us to calibrate the luminosity monitor azimuthal asymmetry as a monitor of the transverse component of the
beam polarization. A part of the spin dance program brief runs will need to occur utilizing fully vertical and horizontal transverse 
polarization to cross-check our ability to measure azimuthal asymmetries.

\subsection{Recent Progress in the Polarized Source}

In addition to the reduction of helicity-correlated position differences, efforts have been invested in making
the Jefferson Lab polarized source increasingly robust in high-current operation. The strained-superlattice GaAs
cathodes are now considered standard photocathode material at CEBAF, with demonstrated rugged and reliable
performance at polarization ~85\%. The modelocked Ti-Sapphire lasers, which could be balky and difficult to
maintain, have been replaced with reliable fiber-based drive lasers. The purchase of a powerful fiber amplifier
is planned for \Qweak to provide more laser headroom for longer periods of uninterrupted operation. A new
``load-locked'' photogun has been installed at the CEBAF photoinjector during 2007 summer shutdown.  This new
gun was commissioned at the Injector Test Cave and demonstrated improved high current performance compared to
the ``vent/bake'' guns that have been used since 1998.  Besides improved vacuum, the gun design accommodates
rapid photocathode reactivation and replacement, to minimize accelerator downtime.

It has been standard practice at Jefferson Lab to flip the helicity state of the beam at a rate of 30~Hz.
Much more rapid helicity reversal, in the range of 125-500~Hz, is planned for \Qweak. A new Pockels cell high
voltage switch has been developed to provide 500~Hz (1 ms) helicity flipping.  The Pockels cell voltage is flipped
using LED-driven opto-couplers placed directly on the cell, thereby eliminating the capacitance of a long cable.
Rise/fall times around 50~$\mu$sec at 500~Hz flip rate were measured in bench tests. This new switch was
installed at the CEBAF photoinjector, with beam tests to happen soon.

In addition we anticipate that the newer available reversal rates of 125~Hz (up to 500~Hz) will improve many of
the helicity correlated properties of the beam. The higher reversal will also provide oversampling capability to
observe 60~Hz and multiples of the line frequency. Ideally, oversampling and real time analysis should allow us
to provide guidance to the source group as to when line noise is unacceptably large, as well as a capability to
reduce 60~Hz noise at the front end of the accelerator prior to our production run. These efforts are underway,
and significant progress has already been achieved in minimizing line correlated beam modulations.\\

\subsection{Low Current Operation During Calibration Running}

During the $Q^2$ calibration running with the tracking system, the beam current will be reduced to achieve
acceptable rates to run in pulse counting mode. The drift chambers that set the upper limit on the beam current
in order for the full tracking system to run are the region 2 horizontal drift chambers.  Due to the high flux
of low energy M{\o}ller electrons, the beam current needs to be $\sim$0.15~nA in order for the count rate in
these chambers to be tolerable (approximately 400~kHz).  In March 2007, the collaboration performed beam tests in
collaboration with the accelerator division to show that  this low beam current could succesfully be delivered
to and monitored in Hall C.

\newpage

A procedure has been developed for establishing low beam currents using the polarized source laser attenuator
and the injection region chopper slits.  The procedure was robust enough that the accelerator operator on shift
was easily able to adjust the beam current for us over a large dynamic range (0.15 nA - 5 $\mu$A) on demand with
only minimal ($<$ 10 minute) wait times.  The stability of the low current 0.15 nA beam in Hall C was measured
using a lucite {\v C}erenkov detector detecting scattered electrons at small angles from an aluminum target.
Figure~\ref{beamstability} shows a typical stability plot measured over 1000 seconds.  The observed $\pm$10$\%$
stability is adequate for our purposes.

\begin{figure}[htbp]
\centerline{\includegraphics[width=5.5in]{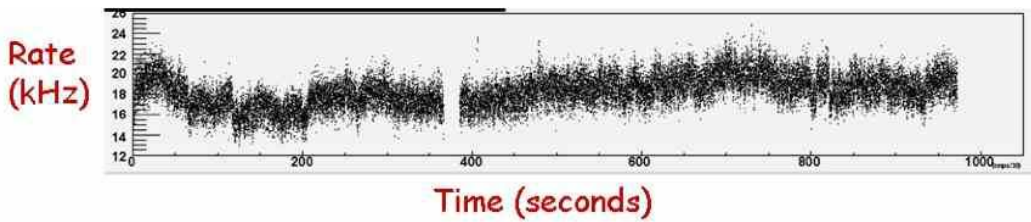}} \caption{\em Rate in lucite detector versus time
for 0.15 nA beam current running.} \label{beamstability}
\end{figure}

Most important for the $Q^2$ determination is the stability of the beam position, angle, and size.  This was
monitored by monitoring the count rate in the lucite detector as a superharp monitor with tungsten wires was
slowly scanned through the beam.  Typical results of such a scan are shown in Figure~\ref{beamposition}.  The
conclusion of several scans like this over several hours was that the beam positions varied at most by 0.3 mm
over that time period.  This is more than adequate for our purposes.  Simulations have shown that a 0.3 mm
position shift corresponds to a worst case of $<$0.18$\%$ shift in the measured value of $Q^2$.

\begin{figure}[h!]
\centerline{\includegraphics[width=5.5in]{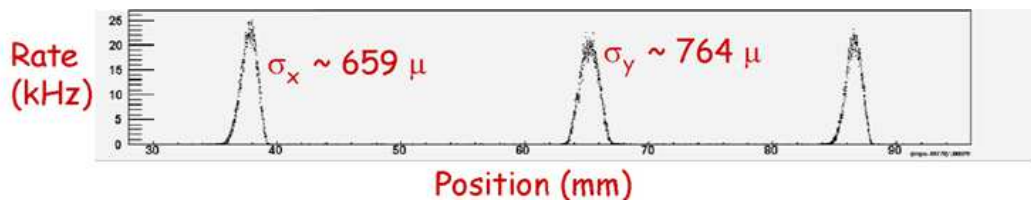}} \caption{\em Rate in lucite detector versus wire
position during a slow superharp scan.} \label{beamposition}
\end{figure}
\section{Beam Diagnostics}

\subsection{Beam Intensity and Position/Angle Monitoring}

As with all parity-violation experiments, highly linear, low noise beam property measurements are needed in
order to correct helicity-correlated false asymmetries.  The \Qweak{} experiment will use the beam monitoring
already installed in Hall C that was successfully used for the G0 experiment.  This includes four microwave
cavity beam charge monitors, a total of twelve 4-wire SEE beam position monitors, and two microwave cavity beam
position monitors.  This equipment is described in more detail in the 2004 \Qweak{} jeopardy proposal.  An
important upgrade that will be made to the readout of the equipment is the replacement of the TRIUMF 2 MHz
voltage-to-frequency converters with the TRIUMF 18 bit, 500 KHz, sampling ADC's which have effectively 27 bit
resolution. The beam position and angle can be deliberately varied using an air-core coil beam modulation system
that already exists in the hall from G0.  Another change since 2004 is the addition of new collaborators
from University of Virginia and Syracuse University who have significant experience in these areas from running
parity-violation experiments in Hall A with similar beam monitoring equipment and controls.

\subsection{Higher Order Beam Moments}

        Our last proposal was written shortly after simulations showed a small sensitivity
to beam spot size modulation. Since then, we have made some progress in understanding how spot size changes
would influence our azimuthal dependence, how such changes might be produced, and how to measure them. So far,
beam spot size changes are the only higher order moment of a beam parameter which is appears potentially large
enough to pose a challenge for \Qweak. However, we note that our solution for monitoring beam spot size changes
can be applied to any beam parameter which can be rotated into coordinate space somewhere along the beamline.

\paragraph*{Spot size systematics:}

In our azimuthally symmetric detector, after corrections for changes in the first moments of beam properties,
the experimental asymmetry becomes

$$
A(\phi) = A + B \cos(\phi) + C \sin(\phi) + D\cos(2\phi) + E\sin(2\phi)
$$

$\bullet$ The A term is dominated by parity violation (PV) but may contain relatively small contributions from
several classes of beam spot size changes. It can also contain a small contribution from leakage from the
$\phi$-dependent terms when the latter are large.

$\bullet$ The B and C (``dipole'') terms are due to the product of residual transverse electron polarization and
two-photon exchange. Unless one carefully nulls the transverse beam polarizations, the natural magnitude of
these parity conserving (PC) dipole terms will be similar to that of the PV asymmetry of interest.\cite{Pitt} In
principle, the contribution of these terms will vanish when averaged over the 8 detector bars, but broken
symmetries in the apparatus may cause a small leakage of the PC asymmetry into the PV
 asymmetry, which will be challenging
to accurately measure and correct. Conservatively assuming that our apparatus is only symmetric to about 1\%, we
therefore plan to suppress the magnitude of the dipole terms by feeding back to the injector to null the
transverse beam polarization. If the magnitude of the B and C terms is significantly smaller than the PV
contribution to A, this potential PC leakage issue can be ignored.

$\bullet$ Beam spot size changes along a single preferred axis could make small contributions to {\em both} the
offset (A) and the quadrupole terms (D and E). The contribution to A could be corrected in principle, but raises
the possibility that even if there are no beam spot size changes, a dipole with weak statistical significance
would lead us to shift the central value of our PV asymmetry by half the error bar.
When the modulation has no preferred axis (a radial breathing mode), it contributes a false asymmetry to A, but
does not provide any helpful diagnostic contribution to the $\phi$-dependent terms. An interesting but harmless
special case occurs if the major axis of the beam spot is first aligned along the x-axis for one spin state,
then the y-axis for the opposite spin state; such a toggling modulation would generate a pure quadrupole.

Laser table studies are planned to set upper limits on helicity correlated spot size changes. However, even if
the spot size never changes on the laser table, there are still downstream elements (the vacuum window and the
gun cathode) which could produce spot size changes. These effects, too, can be studied though with increasing
difficulty: window effects can potentially be studied with a suitably stressed mock-up, and cathode changes can
be studied by moving the laser spot on the cathode.

The insertion of a half-wave plate effectively reverses the physics asymmetry. Any helicity-independent false
asymmetries, such as electronic pickup from the injector, can therefore be cancelled exactly by subtraction
provided the offsets have not  drifted between slow reversals. On the other hand, false asymmetries which
reverse along with the physics asymmetry are problematical. Unfortunately, we believe that most classes of beam
spot size changes will be caused by spatial nonuniformities in bi-refringence, and therefore will flip sign with
the polarization change.

To summarize, not all classes of potential beam spot size changes which cause a false asymmetry will provide a
diagnostic $\cos(2\phi)$-like dependence in the detectors. Most potential mechanisms for spot size changes
cannot be cancelled using the half-wave plate. Laser table studies are beginning which will help us understand
the potential phenomenon better. Clearly, we need to measure, or at least bound, such potential effects on the
beamline inside Hall C. Plans for an appropriate detector are discussed in the next section.

\paragraph*{Beam spot size monitor:}

A non-intercepting beam monitor whose signal is linear in the position coordinate can only return the first
moment, $<x>$. Hence, a principal requirement for accessing the second moment, $\sigma_x$, is a monitor with
nonlinear response. Another important requirement is that the monitor be insensitive to beam position changes,
so that the small expected signal for spot size modulation is not swamped by normal position jitter.

In our 2004 proposal, we suggested using an offset pair of 4-wire BPMs to measure helicity-correlated size
changes, but subsequent simulations showed the sensitivity to spot size modulation was extremely small. The good
news is that the position information derived from a 4-wire BPM is essentially free of contamination from higher
order moments.

\begin{figure}[h]
\begin{center}
\rotatebox{+90.}{\resizebox{1.5in}{5.in}{\includegraphics{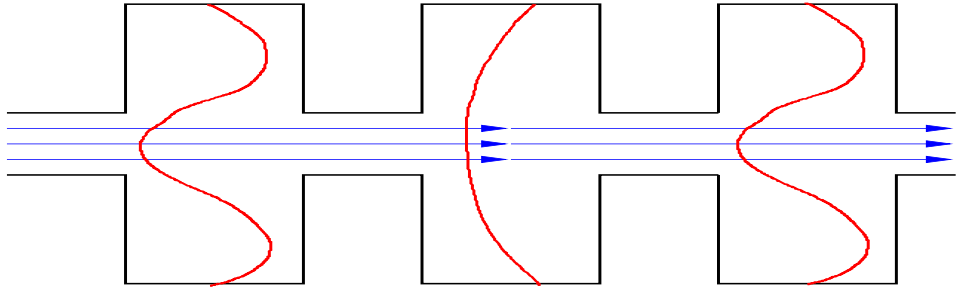}}}
 \end{center}
\caption{{\em Sketch of the 3-cavity concept for measuring helicity correlated beam spot size changes, $\Delta
\sigma_x$ and $\Delta \sigma_y$. The curves are an attempt to represent $E_z(x,y,t)$.}} \label{Concept}
\end{figure}

We finally arrived at the scheme requiring 3 cavities sketched in Figure \ref{Concept}. The only way power can
be coupled from the beam into the cavity is via the interaction of the bunch charge with $E_z(x,y,t)$. Hence the
larger the longitudinal electric field, the larger the signal. The basic idea is that the beam spot size is
sensed by the higher curvature in $E_z(x,y,t)$ in two of the cavities, while the lower curvature cavity
normalizes the beam current. Note that none of these cavities has any first order position sensitivity when the
beam is perfectly on axis, so one expects that the position sensitivity would still be weak for a realistic
alignment scenario. This is confirmed analytically below.

We will use a standard BCM pillbox cavity resonating in the $TM_{010}$ mode at $3\times f_{rep}$=1497 MHz to
normalize the beam current, while a pair of rectangular cavities resonating in the $TM_{310}$ mode at $6 \times
f_{rep}$ = 2994 MHz will measure changes in $\sigma_x$ and $\sigma_y$.
The ratio of the rectangular to cylindrical cavity signals would be proportional to $1+\epsilon*\sigma_i$ in
first order. The contribution of the finite spot size to this ratio is only about 0.1\%, but small spot size
{\em changes} will reveal themselves as a helicity correlated asymmetry calculated using the cavity signals.
Calibration will be done by modulating quadrupoles or the dipoles of the fast raster system.

In Reference \cite{MackCavity}, we performed detailed analytic calculations, and confirmed some important
features with MAFIA. Here, we repeat a small part of that work. Using the $TM_{010}$ mode from a JLab
cylindrical BCM as a current measurement, we define the normalized $TM_{310}$ signal as $ V_{310} \equiv
\frac{V_{310}}{V_{010}}$ where the $V$'s are downconverted signal voltages converted to DC, and the helicity
correlated asymmetry for the $TM_{310}$ mode is defined as
\begin{eqnarray}
A_{310} \equiv \frac{\bf{ V_{310}^+ -  V_{310}^-}}{\bf{ V_{310}^+ +  V_{310}^-}}
\end{eqnarray}
If the helicity correlated width and offsets are $ w^{\pm} = w_0 \pm \Delta w/2$ and $ x^{\pm} = x_0 \pm \Delta
x/2$ respectively, then it can be shown that
\begin{eqnarray}
A_{310} = -\frac{3}{8} (\frac{\pi}{a})^2 w_0 \Delta w -\frac{9}{8} (\frac{\pi}{a })^2 x_0 \Delta x ,
\end{eqnarray}
where the first term contains the spot size changes of interest and the second term is the background due to
beam position changes.
These results are given in Figure \ref{Sensitivities} for an interesting range of parameters. For $a$ = 20 cm,
$w$ = 0.4 cm, and the smallest size modulation to which the experiment is sensitive ($\Delta w$ = 0.1 microns),
then the spot size asymmetry is $A_{310}$ = 37 ppb. For reasonable alignment tolerances, the corresponding
correction for position jitter would be smaller by an order of magnitude.

The measurement time will be determined by the poorly
 known electronic noise floor. If the TRIUMF sampling ADC's are the limiting factor, then it
will take only minutes to establish whether the spot size is changing by an amount large enough to affect the
experiment.

\begin{figure}[h]
\begin{center}
\rotatebox{0.}{\resizebox{6.0in}{3in}{\includegraphics{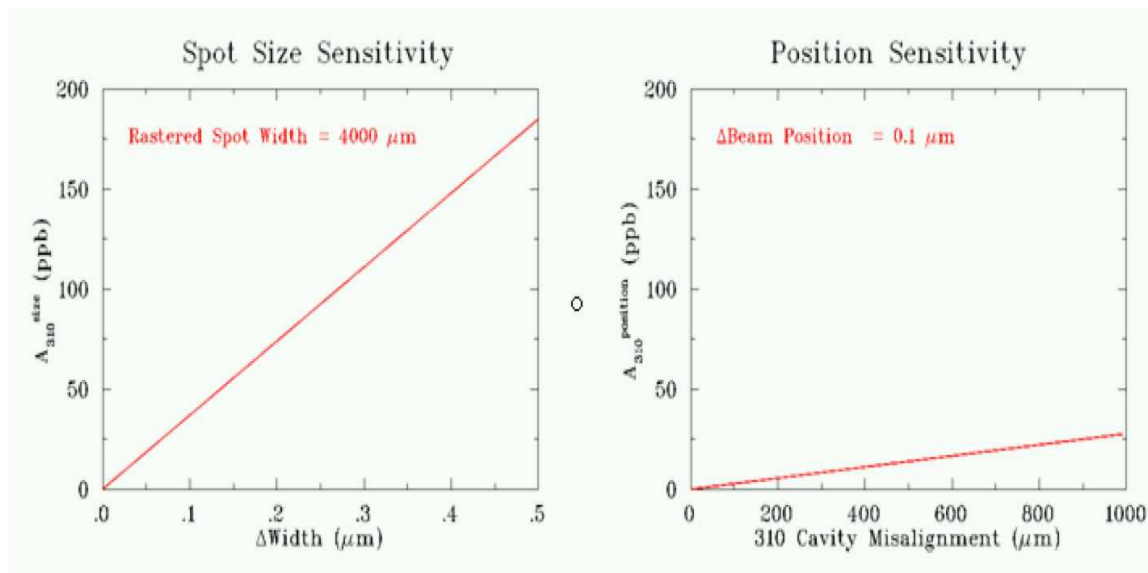}}}
 \end{center}
\caption{ {\em Left: Expected asymmetry from the normalized spot size monitor cavity as a function of the
helicity correlated change in width. The threshold of concern for \Qweak is about 0.1 microns. Right: Expected
asymmetry from the normalized spot size monitor cavity as a function of the beam-cavity misalignment. The
position regressions are generally relatively small.}} \label{Sensitivities}
\end{figure}

\subsection{Luminosity Monitors}

The luminosity monitors are deployed in locations where their count
rates will much higher than the main detectors, so the resulting
statistical errors are small.  They will be used for two purposes.
Since they have a much smaller statistical error per
measurement period than the main detector, they are much more
sensitive to the onset of target density fluctuations.  Second, the
luminosity monitor will be used as a valuable ``null asymmetry
monitor'', since it is expected to have much smaller asymmetry
than the main detector; thus if its asymmetry is non-zero it could
indicate the presence of a false helicity-correlated effect in the
experiment.

Since 2004, we have expanded our plans for the luminosity monitors, and
they will now be deployed in two locations - an upstream location
on the front face of the primary defining collimator and a downstream
location about 17 meters downstream of the target.  The upstream set
will primarily detect M{\o}ller scattered electrons at about 6 degrees;
this cross section is insensitive to beam energy and angle changes,
so this set will be ideal for monitoring target density fluctuations.
The downstream set will be located at a scattering angle of about
0.5$^{\circ}$ and will be equally sensitive to M{\o}ller and e-p
elastic electrons.  This set will be equally or more sensitive to
helicity-correlated beam properties than the main detector.

The detectors will be {\v C}erenkov detectors with quartz (Spectrosil
2000; the same grade as the main detectors) as the active medium
- 3 cm x 5 cm x 2 cm for the downstream version and 7 cm x 25 cm x 2 cm
for the upstream version.  Light from the detectors will be transported
via air-core light guides coated with polished and chemically
brightened anodized aluminum.  Figure~\ref{lumifigs} shows the collected
light per cosmic ray event from a prototype of the downstream luminosity
monitor.  The collected light yield is adequate for our purposes, and
the observed value of the fractional photoelectron fluctuations
$\sigma_{pe}/\langle pe \rangle \sim 4/7$ implies only a 15\% increase
beyond counting statistics in the luminosity monitor per measurement
period.  This is acceptable because the luminosity monitor is simulated
to have a count rate that implies a factor of at least 6 smaller statistical
error then the main detector from pure counting statistics.

\begin{figure}[htbp]
\centerline{\includegraphics[width=6.in]{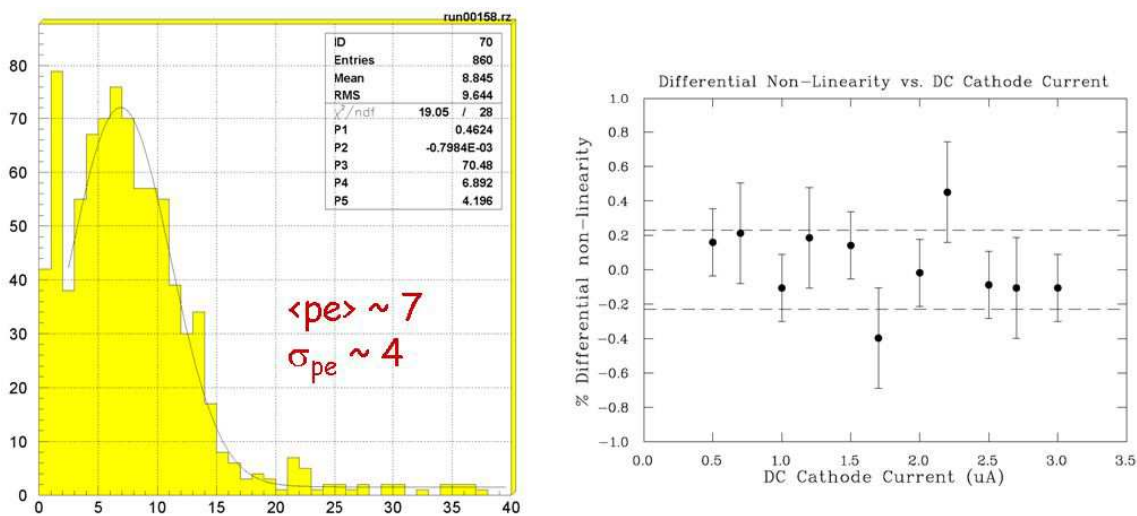}} \caption{\em a) Observed photoelectron yield per cosmic
ray event in a prototype detector with quartz and 30 cm long air lightguide. b) Observed differential
non-linearity of phototube cathode current output versus DC current level.} \label{lumifigs}
\end{figure}

The photodetector will be a Hammamatsu R375, 10 stage, multi-alkali
photomultiplier read out in ``photodiode mode'' (all dynodes tied
together).  We have made studies of the linearity of this scheme
using a setup with a DC and AC light emitting diode, a picoammeter
to read the photodector cathode current, and a lock-in amplifier
to monitor the photodector AC response as the DC light level is
varied.  Some typical results are shown in Figure~\ref{lumifigs}.  The
observed differential non-linearity ($< 0.2\%$) is below our
required 1$\%$ over our operating range.
\section{Precision Polarimetry}

The dominant experimental systematic uncertainty in \Qweak will result from corrections due to beam polarization
($\frac{\delta P_e}{P_e} = 1 \%$).
 Since the previous \Qweak proposal update, work has continued with the goal of improving the performance of
the existing Hall C electron polarimeter, the Basel M\o ller Polarimeter, in particular in the area of extending
the operability of the M\o ller to high currents. In addition, design of a new Compton Polarimeter for Hall C is
also proceeding, with the aim of determining the incident beam polarization to the 1\% level statistical
uncertainty on the timescale of one hour, and monitored on a continuous basis.  While the Compton polarimeter is
being designed with the goal of achieving 1\% systematic precision in mind, we will rely on cross--calibration
with the Hall C M\o ller during the initial phases of its use and use the Compton primarily as a relative
monitor of the polarization.

\subsection{M\o ller Polarimeter Operation at High Currents }

Since the submission of the last \Qweak update, more studies have been performed attempting to extend the
operation of the Hall C M\o ller Polarimeter to high currents. The nominal operating current of the Hall C M\o
ller is $\approx$~2~$\mu$A, this limit set by the need to keep foil heating effects (and hence target
depolarization) low.  Studies have been underway to extend the operating current of the Hall C M\o ller to
$\approx$~100~$\mu$A using a fast magnetic beam kicker in conjunction with a thin iron strip or wire target. The
short duration of the kick (on the order of $\mu$s) ensures minimal target heating effects.

As of the submission of the previous update, preliminary tests of a first generation kicker had been performed
on a 20$\mu$m diameter iron wire target. These tests were partially successful, but pointed out the need for a
different kind of target to keep instantaneous rates low and random coincidences under control. In December
2004, a second round of tests were performed with a 1~$\mu$m thick iron strip target. The results of these tests
are shown in Fig.~\ref{kicker_dec04}~\cite{kicker_proceedings}. In this case, the kicker scanned the beam across
the iron foil for 10~$\mu$s at a frequency of 5 to 10 kHz for beam currents up to 40~$\mu$A. Higher currents
were not accessible due to beam loss issues, likely due to an unoptimized beam tune. As can be seen from the
figure, the technique worked in a global sense, but the polarization measurements were not stable enough to
prove stability at the 1\% level. In particular, control measurements made at 2~$\mu$A with no kicker at the
beginning and end of the test run varied as much as 3\%. The source of these fluctuations is unclear since these
tests were performed during a running period when the beam polarization was not being regularly measured in Hall
C. Nonetheless, these results were taken as proof of concept. Further tests were planned for the G0 Backward
Angle run in 2006, but were not possible due to the extremely low beam energy. We hope to make further
measurements during the currently running $G_E^P$ experiment in Hall C.

For reference, we show the table of kicker performance properties needed to make polarization measurements at
various currents (Tab.~\ref{kicker_op}). In particular, we wish to note that a kicker magnet capable of the
shortest kick interval (2~$\mu$s) has been constructed and is ready for installation.

\begin{table}[htb]
\caption{\em Operating parameters for a planned beam--kicker system that will allow operation of the Hall C M\o
ller Polarimeter at high currents. $\Delta t_{kick}$ refers to the total interval of time for which the beam
will be deflected from its nominal path onto a half--foil or strip target. In order to keep beam heating effects
to a minimum, the kick interval must be shorter at higher currents.} \label{kicker_op}
\begin{center}
\begin{tabular}{|c|c|c|}
\hline $I_{beam}$ ($\mu$A)  &  $\Delta t_{kick}$ ($\mu$s) & f$_{kick}$ (Hz) \\ \hline
200          &       2             &   2500  \\
100          &       4                     &   2500  \\
50           &       8                     &   2500  \\
20           &       20                    &   2500  \\\hline
\end{tabular}
\end{center}
\end{table}

\begin{figure}[h!]
\begin{center}
\vspace*{0.3cm}
\includegraphics[width=4.5in]{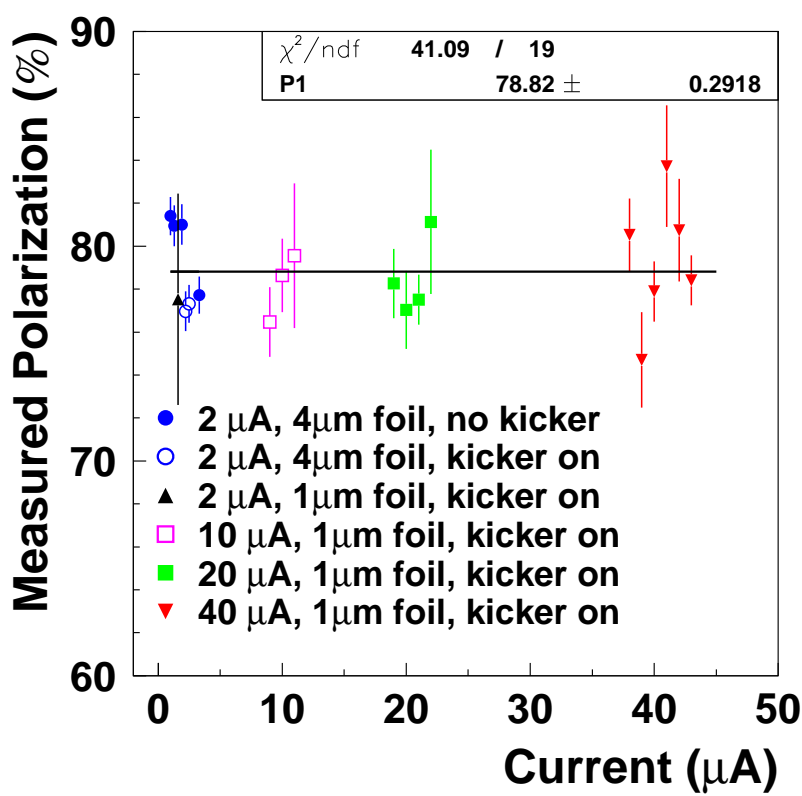}
\end{center}
\caption{\label{kicker_dec04} \em Polarization measurements using the Hall C M\o ller Polarimeter and a second
generation kicker magnet and iron strip target. Globally, the technique yields asymmetries independent of beam
current at the several percent level. However, instabilities of unknown origin (likely beam related) make it
impossible to show that the kicker system yields measurements stable to 1\%.}
\end{figure}


\subsection{Hall C Compton Polarimeter}

In Compton polarimetry, circularly polarized photons from a laser are scattered from polarized electrons in the
electron beam.  Scattering rates measured in electron and photon detectors determine the cross-section asymmetry
and hence polarization.

A schematic diagram of the Compton polarimeter is shown in Fig.~\ref{fig:chicane}.
\begin{figure}[b]
\begin{center}
\vspace*{0.3cm}
\includegraphics[width=\textwidth]{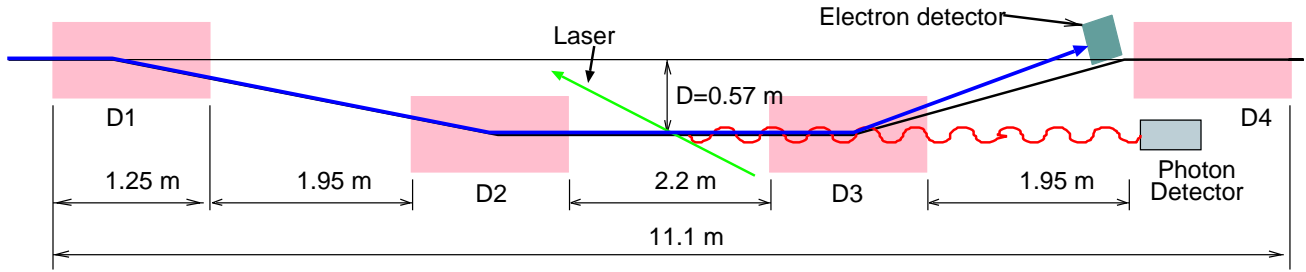}
\end{center}
\caption{{\em Schematic diagram of the Compton polarimeter chicane.}} \label{fig:chicane}
\end{figure}
A four-element vertical dipole chicane is used to displace the Compton interaction point from the beam axis,
allowing the scattered photons to be detected in a PbWO$_4$ calorimeter.  The scattered electrons will be
momentum analyzed in the 3rd dipole of the chicane, and will be detected by a diamond strip tracker comprised of
four planes.  The diamond strip tracking detector is a new development, and will be constructed in collaboration
between groups from the Universities of Winnipeg, Manitoba, TRIUMF,  and Mississippi State University. The
design of this detector, with a focus on systematic uncertainties in polarization extraction, was studied in a
recent honours thesis\cite{Storey}.

The development of diamond detectors for minimum ionizing radiation has been led by the CERN RD42 collaboration.
H.\ Kagan (Ohio State U., OSU) and W.\ Trischuk (U.\ Toronto), who are collaborators in CERN RD42, have assisted
us in the initial fabrication and testing of diamond detectors.  As a first step, two test detectors were
fabricated with a 6~mm diameter electrode of Cr-Au was sputtered onto each face of each diamond sample.  A
picture of a resultant two-electrode detector, and pulse-height spectra using the detector with a
minimum-ionizing beta-source ($^{90}$Sr) are shown in Fig.~\ref{fig:phs}.  The results show a charge-collection
depth of 230~$\mu$m when the detector is biased to 1000~V, consistent with typical good results achieved by CERN
RD42. We are in the process of building test setups similar to the OSU test setup at both U.\ Winnipeg and at
Mississippi State U.

We are also in the process of designing and fabricating prototype strip detectors at both OSU and at the
Nanosystem Fabrication Laboratory at U. Manitoba.  At the time of writing, a half size (10 $\times$ 10 mm$^2$)
prototype detector with 15 strips has just been completed at OSU, as illustrated in Fig.~\ref{fig:OSU2}, and is
currently being tested.  Each step of the fabrication process is well-understood at this time and progress is
continuing smoothly. In addition to the funds already obtained from NSERC and from DOE, the Canadian group has
requested from NSERC funds to complete the vacuum chamber, the motion mechanism, and funds for a spare set of
detector planes.

\newpage
\vspace*{-0.5cm}
\begin{figure}[t]
\subfigure[]{\includegraphics[width=0.54\textwidth]{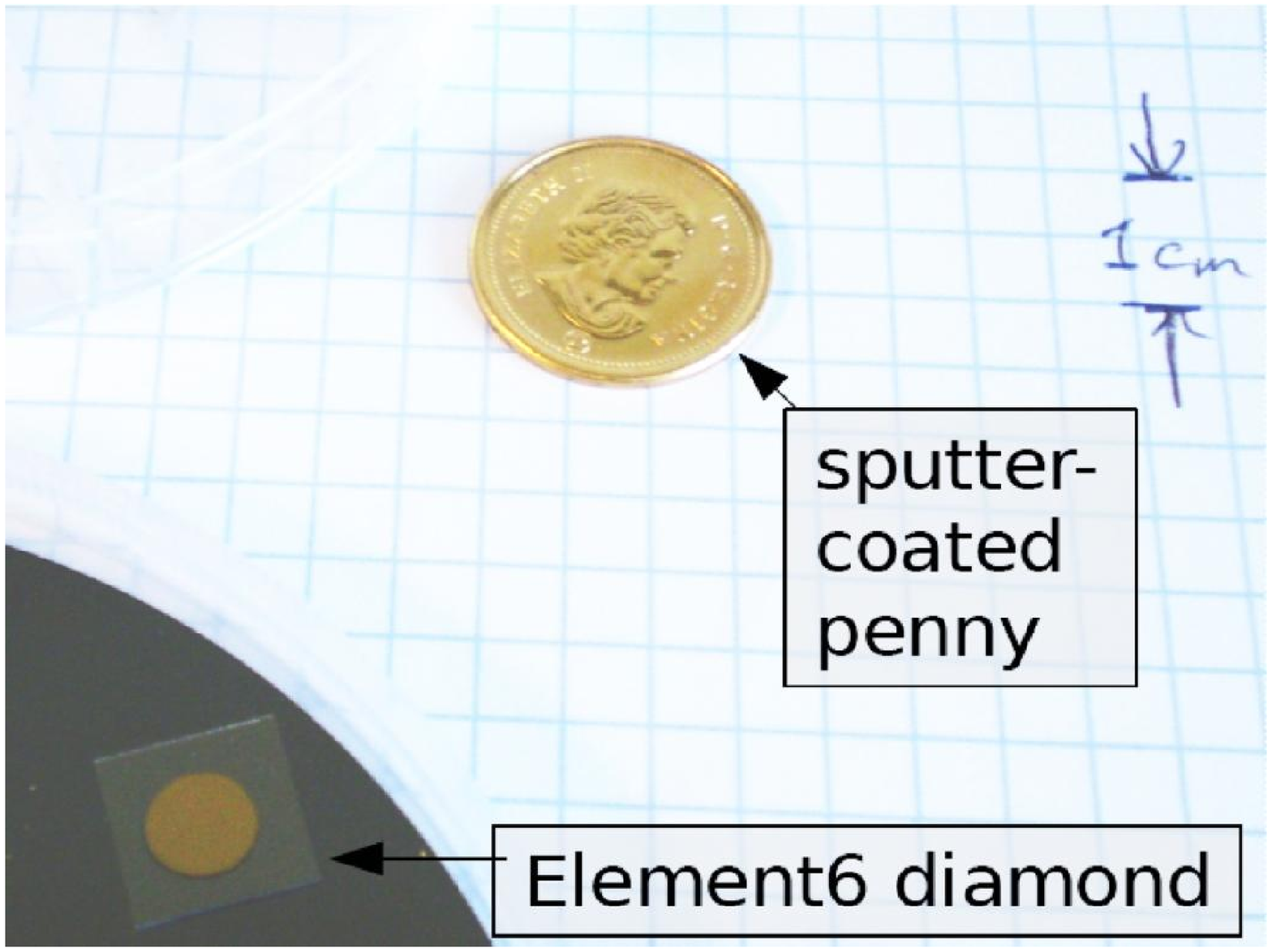}}
\subfigure[]{\includegraphics[width=0.45\textwidth]{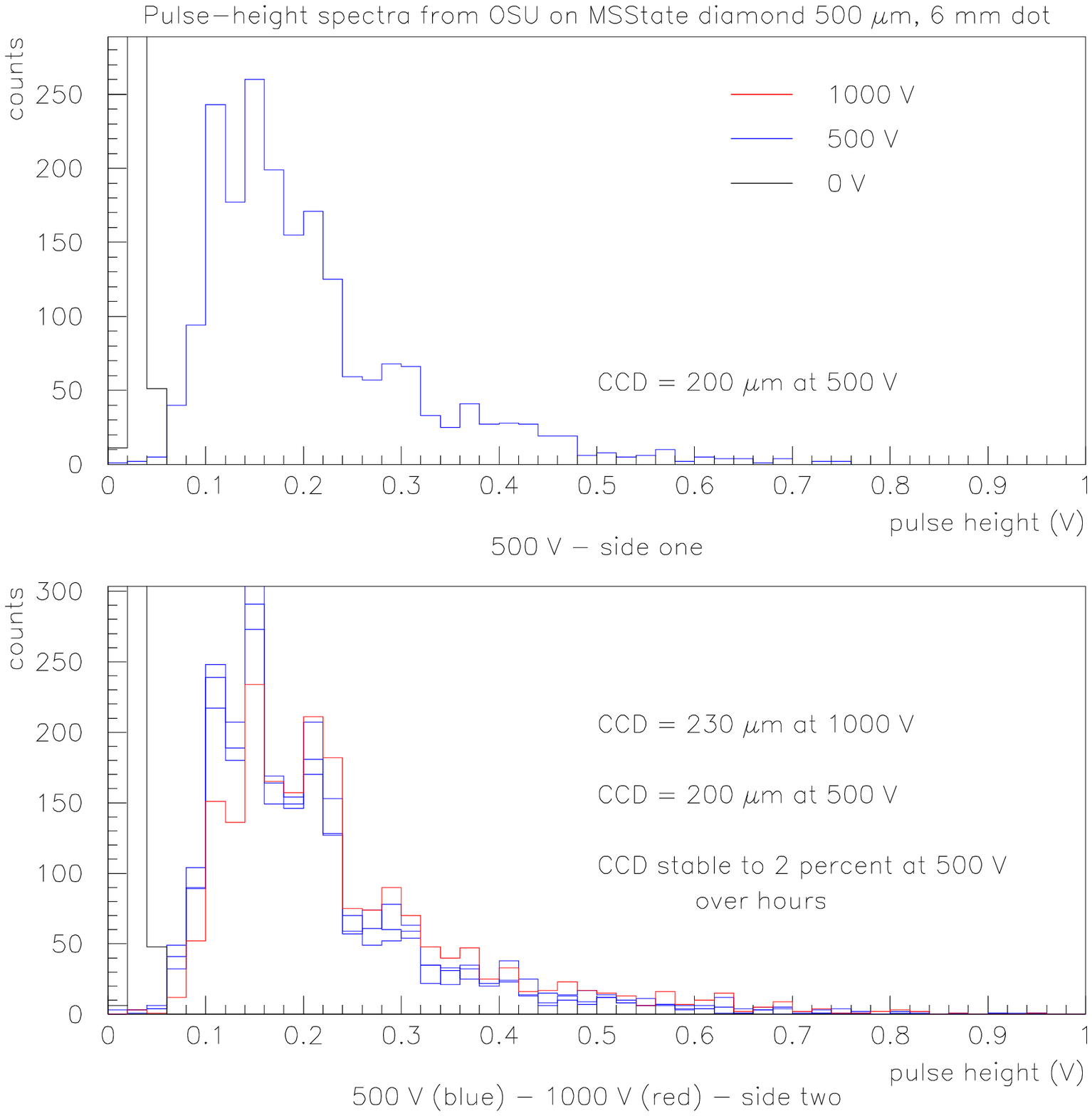}} \caption{{\em (a) Photograph of two-electrode
prototype diamond detector, after successful metallization.  (b) Pulse height spectra for two-electrode
prototype sensing minimum ionizing betas, showing energy deposition well-separated from pedestal. Upper and
lower plots show consistent results achieved when biasing each side of the detector. Various curves at 500~V
show stability of the response over hours. \label{fig:phs}}}
\end{figure}
\vspace*{0.1cm}

\begin{figure}[h!]
\begin{center}
\subfigure[]{\includegraphics[width=0.45\textwidth]{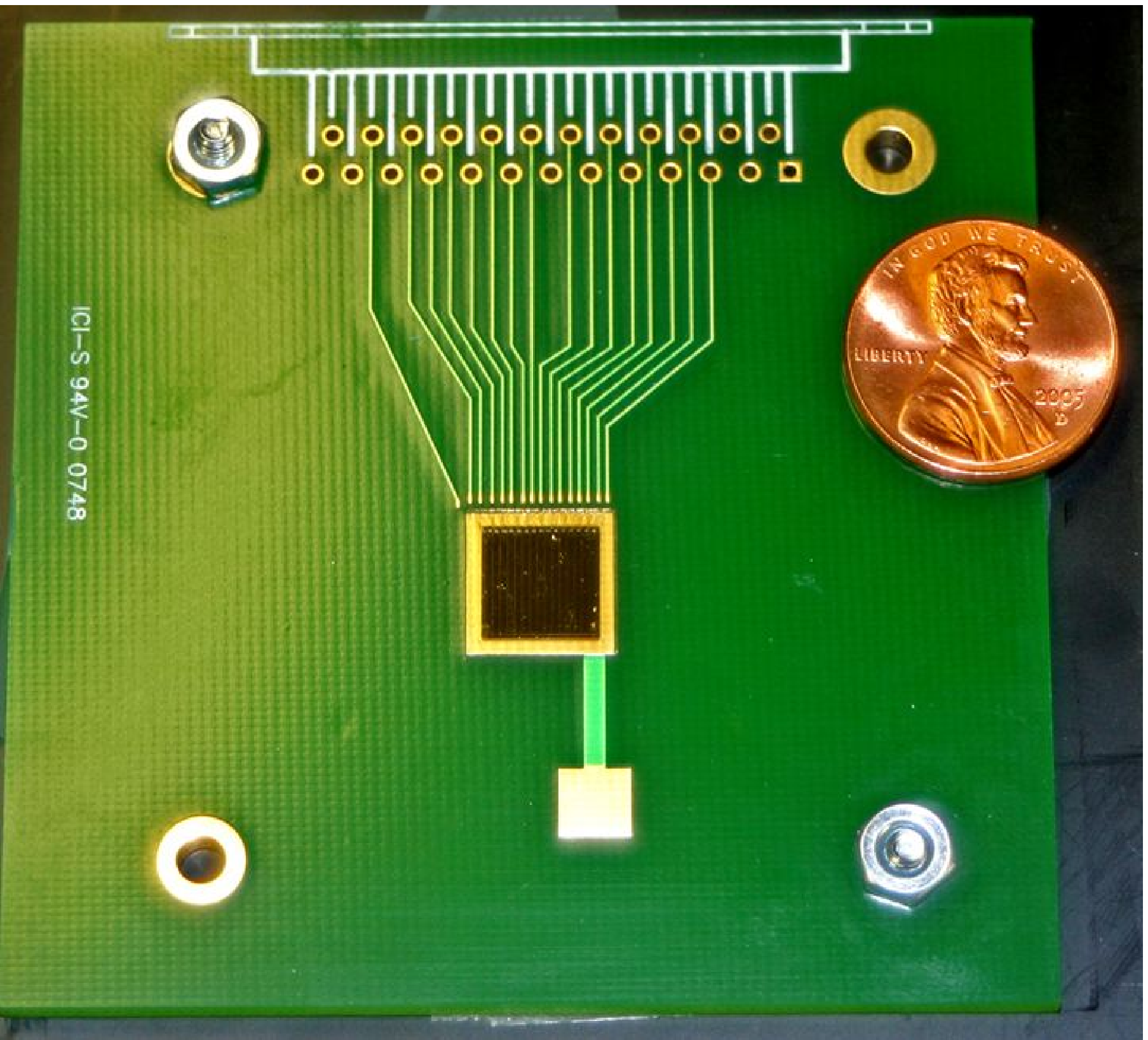}}
\subfigure[]{\includegraphics[width=0.38\textwidth]{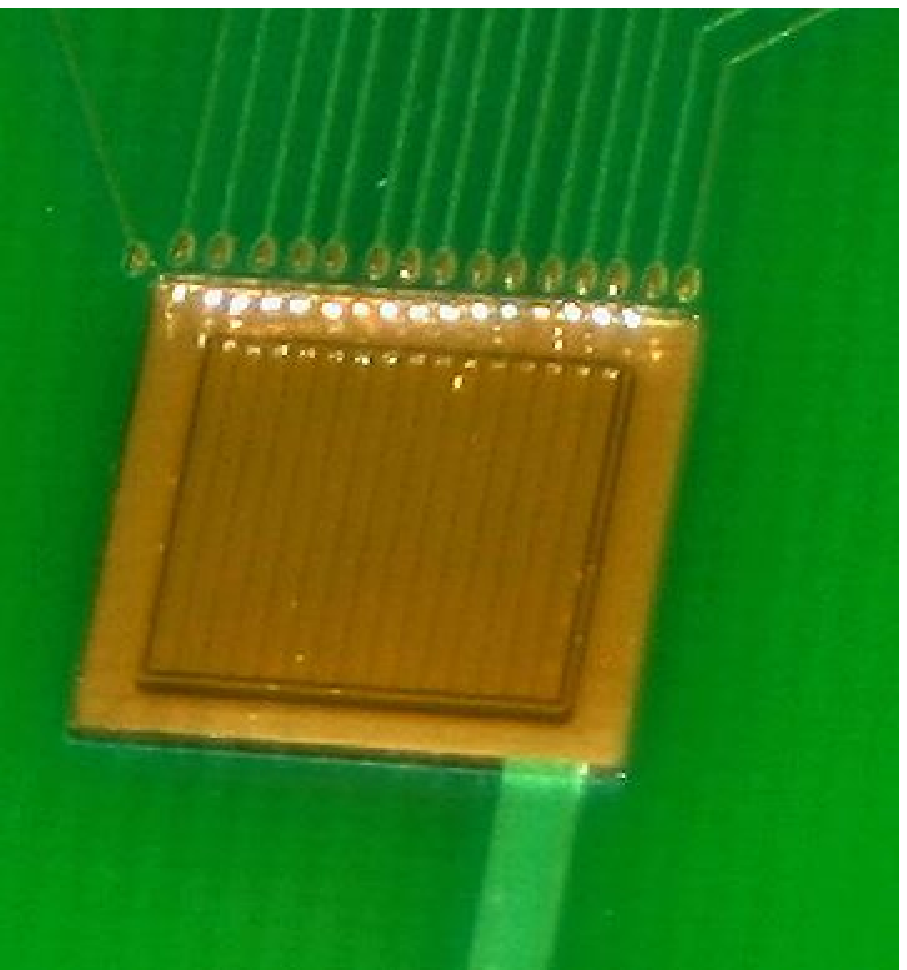}} \caption{{\em Two views of the first
15-strip diamond detector prototype, recently fabricated at OSU. \label{fig:OSU2}}}
\end{center}
\end{figure}

As noted in the previous \Qweak{} update, the construction of a Compton Polarimeter in Hall C will require a
substantial re--work of the Hall C beamline. Our plan in 2004 had been to insert the 4--dipole chicane for the
polarimeter downstream of the existing M\o ller Polarimeter. This plan has been re--evaluated by CASA and a new
design has been developed that inserts the Hall C Compton upstream of the M\o ller Polarimeter, shifting the M\o
ller and all other downstream beam elements closer to the Hall C pivot~\cite{benesch_report}. This is shown
schematically in Fig.~\ref{beamline_2007}. This beamline concept has been vetted by CASA, the \Qweak{}
collaboration, and Hall C physics staff and is currently in the design and engineering stage. It also worth
noting that the modified beamline design calls for slightly longer dipoles than originally proposed (1.25~m as
opposed to 1~m). This modification will facilitate operation after the 12~GeV upgrade. The design and
procurement of these dipoles, as well as the associated stands and vacuum chambers is being done by MIT--Bates
under the auspices of an M.O.U. between Jefferson Lab and MIT--Bates.

\begin{figure}[ht]
\begin{center}
\includegraphics[width=0.5\linewidth,angle=-90]{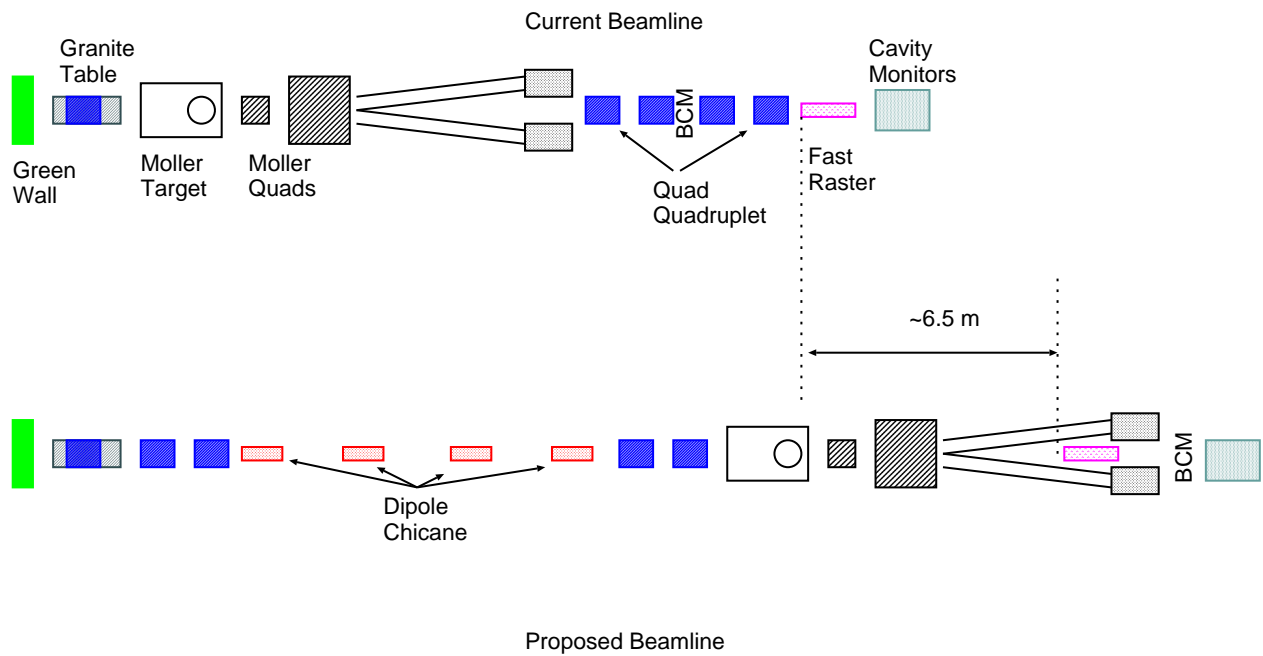}
\end{center}
\caption{\em Schematic of the modifications required to insert the Compton dipole chicane in the Hall C beam
line. Additional quadrupoles will be required to achieve a tightly focused beam at the Compton interaction
point. The M\o ller Polarimeter will be shifted $\approx$ 11 m closer to the Hall C target position, while other
beamline elements, the fast raster for example, will move less by making use of the space between the M\o ller
``vacuum'' legs.} \label{beamline_2007}
\end{figure}

Finally, at the suggestion of the JLab Polarized Source Group, we are pursuing the implementation of a high
power fiber laser for the Compton Polarimeter. The Polarized Source Group has recently begun using ``fiber''
lasers commonly used by the telecommunications industry. These systems amplify relatively low power diode seed
lasers to tens of Watts. While the potential average power is still relatively low ($\approx$20 W of green light
after frequency doubling the 1064 nm output of the fiber laser system), these lasers have the advantage that
they can be pulsed at the same repetition rate as the Jefferson Lab electron beam. 
For a laser
pulsed at the same repetition rate as the electron beam (499 MHz) with a narrow pulse structure ($\approx$ 35
ps), the increase in luminosity as compared to a CW laser system is approximately,

\begin{equation}
\frac{\mathcal{L}_{pulsed}}{\mathcal{L}_{CW}}  \approx \frac{c}{f\sqrt{2\pi}} {1 \over \sqrt{ \sigma^2_{e,z} +
\sigma^2_{\gamma,z} + \frac{1}{\sin^2{\alpha/2}}(\sigma^2_e+\sigma^2_\gamma)}},
\end{equation}
where $f$ is the laser/electron repetition rate and $\sigma_{e,z}$ ($\sigma_{\gamma,z}$) represent the
longitudinal size of the electron (laser) pulse. For typical values of the electron/laser beam sizes and pulse
widths, this ratio is approximately 20. This means that an RF pulsed laser with 20 W average power represents an
``effective'' laser power of 400 W.

Currently, we are actively pursuing the fiber laser option as our laser system of choice. Such a system poses
some small risk since a fiber laser system of this precise configuration has not been built at Jefferson Lab
before and is not completely a ``turn-key'' system. However, the likelihood of success is high due to the
extensive experience of the Polarized Source Group with fiber lasers, and additional on-site experience with
high power frequency doubling. The high power fiber amplifier has been ordered, and we should know within months
whether such a system is tenable or not. If this system proves unworkable, the commercial pulsed green laser
option discussed in the previous \Qweak update~\cite{coherent_evolution} is a safe fall--back option.
\section{Data Acquisition}

\label{DAQ}

The \Qweak experiment requires two distinct modes of data acquisition: the current mode measurement of the
quartz bar signals, and the low current tracking mode measurements in which individual particles will trigger
the DAQ. These two DAQ schemes will be implemented as two essentially independent systems with separate crates
and DAQ/analysis software with some sharing of beam line instrumentation electronics.

\subsection{Current mode DAQ}
The experimental asymmetry measurements will be made with the current mode data acquisition. The core of this
system is the readout of the TRIUMF ADC modules. These ADC modules integrate the current from each quartz bar
photomultiplier tube. In normal operation, the ADC's allow a four-fold oversampling of the planned $250~{\rm
Hz}$ helicity readout (4~ms helicity windows), and should allow the same oversampling of a higher helicity
readout rate of up to $1000~{\rm Hz}$ (1~ms helicity windows). With additional timing signals from the polarized
source, the DAQ would be able to take events with greater than four-fold oversampling of the helicity windows.
In addition to digitizing the current from the quartz bar PMT's and several shielded background detectors, the
same type of ADC will be used to digitize information from both the injector and Hall C beam line monitors. This
beam line information will include charge cavities, BPM's and luminosity monitors. The ADC's for these signals
will be located in a separate crate so that any small helicity correlations in beam parameter signals will not
be present in the same crate as the detector ADC's.

Since the previous proposal submission, a quartz scanner has been added to the experiment, to allow measurements
of the scattered electron profile at the full beam current. The two PMT's of the quartz scanner will be
instrumented by both a scaler counting discriminated pulses and by a scaler counting pulses from a
voltage-to-frequency converter to allow charge integration of the PMT signals over the measurement interval. A
small subset of the pulses in the quartz scanner will be measured with a conventional ADC to monitor the gain of
the PMT's. The PMT's will have additional instrumentation as part of the low-current mode DAQ, which will be
described in the following section.

The rate and volume of data for current mode acquisition is comparable to what has been demonstrated to work
with the typical DAQ and analysis capabilities. In the four-fold oversampling mode, each ADC channel produces
six 32-bit data words per event. With a total of 168 channels (for the detector, injector beam line, and Hall C
beam line), and allowing a 50\% overhead for headers, an event size of 6048 bytes is estimated.
With a readout rate of $250~{\rm Hz}$, the \Qweak data rate will be about 1500 kBytes/second, comparable to the
data rate of the G0 forward angle DAQ, in which the DAQ was able to easily operate with 0\% deadtime. At this
rate, a 2200 hour run would produce a data set of about 12 TB. Scaling from G0 analysis rates, an analysis on
this data set using a single fast CPU would take about 3 months.
A readout rate of $ 1000~{\rm Hz}$ would lead to a correspondingly higher data rate and may require using a
subset of the readout channels. Data acquisition test stands have been in use at Ohio University and at JLab
since early 2007. The TRIUMF ADC module performance has been tested at the planned $250~{\rm Hz}$ helicity
reversal rate, and limited tests have been performed at rates of up to $1000~{\rm Hz}$. Several minor issues in
the module readout have been identified and corrected during these tests. Figure~\ref{F:VQWK_asymmetry} shows
the distribution of quartet asymmetries during an 80-minute battery test at Ohio University.  The battery was
used to generate a 6~$\mu$A input current for a TRIUMF preamplifier, which produced a 4.82~V input signal for
the ADC (6~V above the baseline).  The ADC accumulated 2000 voltage samples over four 4~ms integration windows
each separated by a holdoff of 0.2~ms, for a quartet measurement interval of 16.8~ms. The sigma of the asymmetry
of 2.3~ppm corresponds to a sigma on the voltage 11.~$\mu$V over the 16.8~ms of the quartet integration,
yielding an 1.4~$\mu\mathrm{V}/\sqrt{\mathrm{Hz}}$, or 240~ppb/$\sqrt{\mathrm{Hz}}$ as compared to the 6~V
signal.  As a test of the module's performance in the hall environment, it will be connected to beam line
instrumentation channels during beam tests at JLab in January 2008.

\begin{figure}[tbh]
\vspace*{0.3cm} \centering
\includegraphics[width=0.75\textwidth]{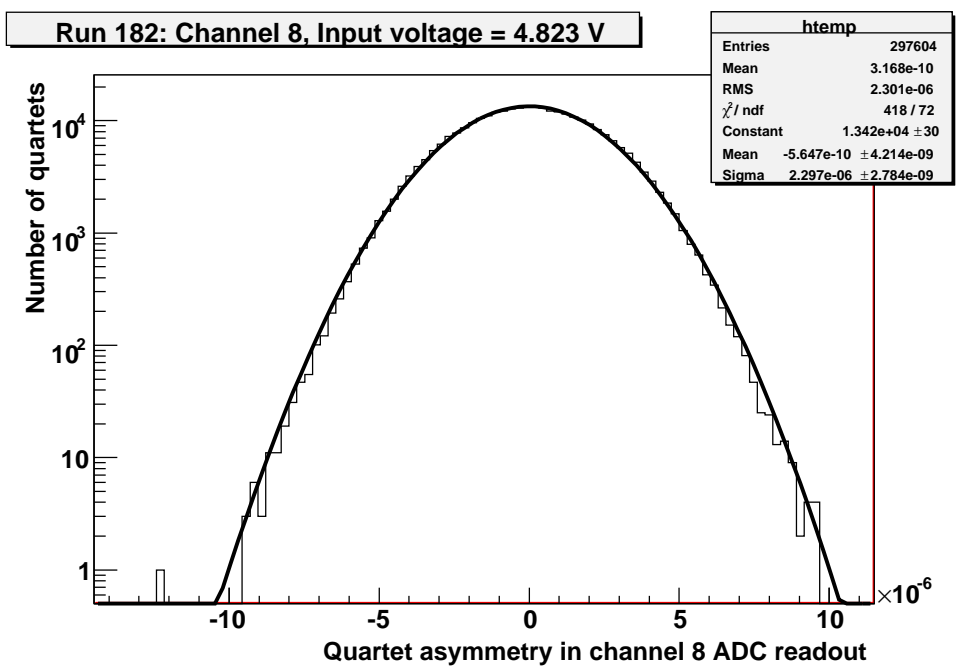}
\caption{{\em Battery test asymmetry distribution for ($+--+$) quartets with a 4~ms integration window and
4.2~ms gate period \label{F:VQWK_asymmetry}}}
\end{figure}

\subsection{Low current tracking DAQ}
The \Qweak{} apparatus will be partially instrumented with tracking detectors in order to study optics and
acceptance. Measurements with the tracking system will be done at low beam current, so that individual particles
can be tracked through the magnet. For this mode of measurement, the quartz bar photomultiplier tubes will be
instrumented with parallel electronics so that the timing and amplitude of individual particles can be recorded.
The tracking data acquisition will operate like a conventional DAQ, triggering on individual particles. The
front end electronics will be all VME, using the JLAB F1TDC for wire chambers and timing signals, and commercial
VME ADC modules for the GEM detectors and PMT amplitudes. As the hardware needed for tracking measurements is
different from the current mode hardware, the tracking DAQ can be operated as a distinct system, allowing
development of the two DAQ modes to proceed in parallel. The tracking DAQ will have the option of reading beam
line information from the same VME crate used for this purpose in the current mode DAQ. The quartz scanner PMT's
will be instrumented with ADC and TDC channels during the low current running, in addition to the integrating
electronics described in the previous section. This will allow tracking of events which hit the scanner, and
will allow comparison of the integrating mode readout at low and high currents.

\subsection{Beam Feedback}
A real-time analysis, similar to that used for G0, is planned. In addition to providing prompt diagnostic
information, this analysis will calculate helicity correlated beam properties such as current, position and
energy. The results of these calculations can be used for feedback on the beam.
\section{Infrastructure}

\subsection{QTOR Power Supply}

The new 2MVA power supply for the QTOR magnetic spectrometer has completed its final tests at the vendor and
will be packed and shipped to the MIT/Bates facility in December, 2007.  Jefferson Laboratory operations funds
in the amount of \$54,000 have been awarded to MIT/Bates under a contract to perform full power tests and
mapping of the magnet including paying for the required AC power cables and electricity. This pre-ops work at
MIT-Bates should allow the assembly and testing of the magnet at Jefferson Laboratory to proceed efficiently
during actual installation in Hall C. New 2MVA lines have been installed in Hall C such as to allow the final
installation of the power supply to be straightforward.

\subsection{Support structures}

The magnet is now assembled in its support structure at MIT-Bates. The large rotation system components for the
region three drift chambers have been delivered to Jefferson Laboratory and are awaiting trial assembly in the
JLab Test Lab building. The CAD assembly drawing of the experiment is shown in Figure~\ref{fig:CAD_of_Qweak} and
continues to be detailed as sub-systems are engineered and delivered to the laboratory.

\begin{figure}[h]
\vspace*{0.2cm}
\begin{center}
\includegraphics[width=16.5cm,angle=0]{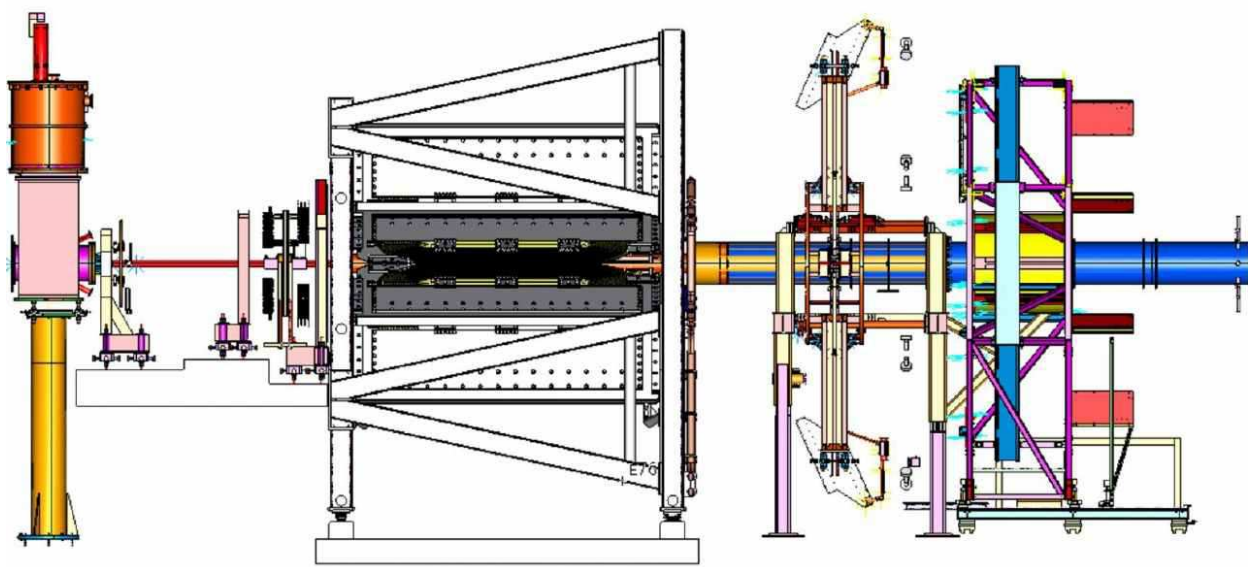}
\caption{{\em CAD layout drawing of the \Qweak experiment without shielding.}}
 \label{fig:CAD_of_Qweak}
\end{center}
\end{figure}

\subsection{Collimators and Shielding}

\begin{figure}[h!]
\vspace*{0.2cm}
\begin{center}
\includegraphics[width=12cm,angle=0]{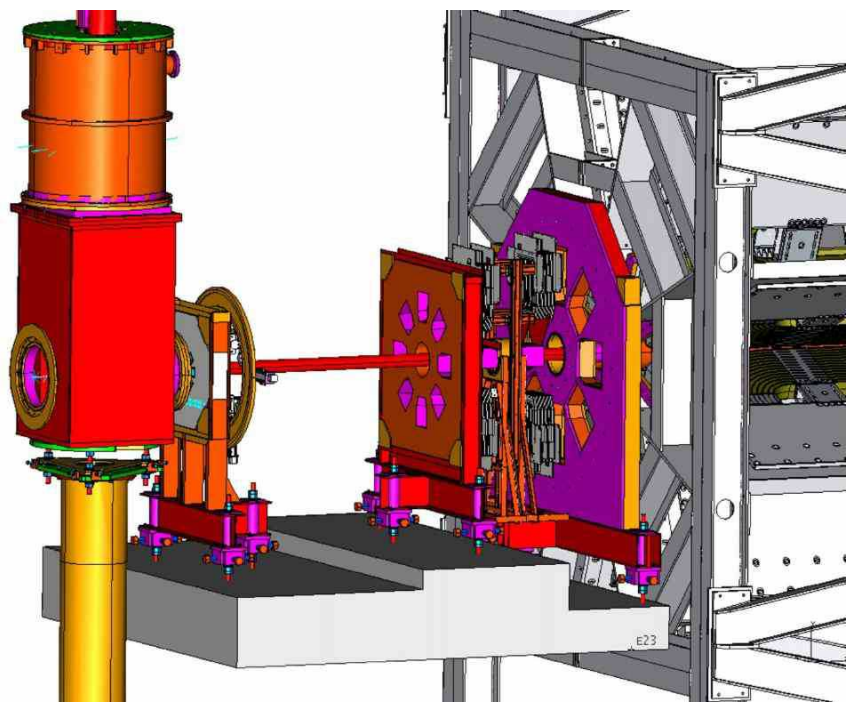}
\caption{{\em Close up of the \Qweak triple collimator system.}}
 \label{fig:CAD_of_Collimators}
\end{center}
\end{figure}

Figure~\ref{fig:CAD_of_Collimators} shows a close up view of the collimator assembly, but without the transverse
concrete shielding that will form a shielding vault between the first and second collimator assemblies. The
design of collimator support / adjustment structures are also shown. The defining collimator will be located
directly downstream from the concrete shielding vault. There will be small gaps between the vault and primary
collimator to allow access for air core light pipes from an upstream luminosity monitor which will be mounted on
the lower upstream face of the primary collimator.

Line-of-sight background particles from the $LH_2$ target region to the quartz detectors are blocked primarily
by the first and last collimator bodies. Line-of-slight backgrounds generated from the tungsten collar which is
located around the beam pipe at the first clean-up collimator must penetrate two Pb secondary collimator bodies
to reach the quartz detectors. This tungsten primary beam collimator is necessary to shield the QTOR magnet
region from small angle target scattering. The maximum beam pipe diameter through the QTOR magnet is limited by
the magnet design and kinematics constraints.  Of the potential additional large angle backgrounds generated by
the tungsten, neutrons are the most penetrating.

Background simulations have been performed for the basic experiment geometry shown above plus the beamline to
assess secondary electromagnetic backgrounds. A simplified model to study only the neutron source terms has also
been run. This work is ongoing by the \Qweak simulation group and the JLab RADCON group, respectively;  it is
presently anticipated that backgrounds will be within acceptable limits.  The detailed configuration of
transverse shielding around the target area and first cleanup through the primary collimator region required to
achieve optimal shielding from a site boundary perspective is still under investigation by the JLab RADCON
group.

After obtaining budgetary quotes from a number of vendors and further discussions with our simulation group, the
decision was made to construct all collimators from a Pb alloy containing 5\% Sb (a hardening agent).
Because of their smaller dimensions, the central defining and upstream clean-up collimators can be affordably
cut using precision electrical discharge machining (EDM).  This is the most cost-effective high precision
manufacturing technique we have found and should allow tolerances approaching 130 microns on all critical
surfaces. These mechanical tolerances are conservative, based on our models of how false asymmetries can be
generated by geometrical misalignments and helicity correlated beam properties. The final, large clean-up
collimator will be cast, as it does not require precision tolerances, and its support structure will be
relatively simple.  Preliminary manufacturing prints for all collimator assemblies have been sent out for
budgetary quotes. The latest round of these quotes indicates that we should be able to stay within our budget
goals for the collimators and achieve or do better than all required tolerances.
\section{Beam Time Request}

\subsection{Basic Operational Requirements and Constraints}
\label{Basic Operational Requirements and Constraints}

The first part of this section briefly summarizes some of the requirements important for the \qw measurement. A
number of factors constrain the range of acceptable incident beam energies and currents.  The nominal beam
energy was very carefully selected based upon extensive optimization of kinematics, hadronic backgrounds,
control of beam properties (requirements scale inversely with the asymmetry), practical solid angle acceptance
issues, detector dimensions, deliverable cooling for the target and magnet power supply issues. More details can
be found in the 2004 Proposal, Technical Design Report and numerous other progress reports that are available on
the \Qweak document server which is accessible from the web site ``http://www.jlab.org/qweak/".

The second part of this section is our formal summary of the requested beam time broken down by activity. Our
beam time request covers all aspects of conducting the measurement, including hardware commissioning,
systematic studies, background measurements, calibration measurements and production running. Since the
experiment will probably be run in no more than three blocks of almost contiguous time, our ability to deliver a
fully analyzed 8\% initial measurement prior to beginning the final 4\% production running is severely
handicapped. However, we plan to generate the 8\% measurement from the first several weeks of running of
sufficient quality. This will mostly likely be the running just after commissioning has been completed. Due to
schedule uncertainties, this period may or may not end up being contiguous with the commissioning period. In
either case, we will analyze this first set of data as rapidly as possible in order to
obtain the maximum information concerning the quality of all parameters prior to committing to our final
production running configuration.

\subsubsection{Parity Quality Longitudinally Polarized Beam}

A quick summary of the \Qweak experiment requirements includes:  85\% polarized parity quality beam delivered at
of 125 Hz up to possibly 500 Hz (1 ms) pseudo random helicity reversal with the polarization settling periods
contributing no more than 5\% gating dead time. The faster rapid helicity reversal rates of either 125 Hz or 500
Hz are to be compared to the present much slower reversal rate refered to as ``30 Hz reversal". 
Due to significant recent advances by the
JLab polarized source group, these requirements now appear straightforward. The final decision on how fast to
flip the spin during production running will depend on results from ongoing source performance tests, the
practical experience of the upcoming Pb parity experiment which has adopted our higher flip rate scheme, and
noise studies on our LH$_2$ cryo-target during the \Qweak commissioning phase. Changing the flip rate and
reversal pattern is now relatively simple for the polarized source group to implement. Although the Pb parity
experiment will likely have been run in Hall A just before \Qweak starts, it is important that polarized beam
experiments running concurrent with \Qweak make the necessary preparations to handle the faster reversal rate.

\newpage

We also require that there be no significant (less than $10^{-9}$) pickup of the coherent reversal signal
through electrical ground loops as observed by our integrators due to leakage of the prompt reversal signal at
the injector. We have worked with the polarized source group on this and other issues, and they have implemented
hardware changes which should accomplish this requirement. The measurement also requires stable source and
accelerator operation, with modest trip rates, energy and position feedback locks, acceptable halo, beam motion,
size and current modulation within specified limits.  Additional detailed requirements are given elsewhere in
this document and in the 2004 proposal.

Although the measurement's sensitivity to residual transverse polarization is strongly suppressed by both
kinematics and our 8 fold detector symmetry, the acceptable upper limit is about 5\% on residual transverse
polarization. Therefore we  require that the polarized source be configured to deliver ``full" longitudinal
polarized beam to Hall C, and that the experiment be allowed to adjust or implement feedback as necessary to
keep the residual transverse polarization level acceptable. We will also need the capability to implement
precision feedback to control the helicity correlated beam energy, intensity, and position. Basically, these are
the standard complement of precision beam controls afforded to parity measurements at JLab.

There are so called "magic" energies which allow Hall C to have full longitudinal polarization while preserving
excellent polarization in Halls A and B.  These are illustrated with a few examples in Table
\ref{Magic_Energies}.

\begin{table}[htb]
\centering\caption{ {\em Examples of Polarized Beam "Magic" Energies}}
\label{Magic_Energies}
\begin{tabular}{lllll}
                        &               &               &                \\
                $E_{linac}$               & $E_{A or B}$         &  ~~$E_C$       & Available $P_e$                    \\
                 (GeV)                & (GeV)         &  (GeV)       & ~~A/B/C                    \\     \hline
                 \\
               \\
               1.069        & 5.405            & 1.129         & 98\% / 100\% / 100\%   & \\
               1.088        & 5.503            & 1.150         & 96\% / 100\% / 100\%    &\\
               1.108        & 5.602            & 1.170         & 94\% / 100\% / 100\%    &\\

                                 &       &               &      \\ \hline \hline
\end{tabular}
\end{table}

\subsubsection{Energy, Precision and Background Tradeoffs}

The nominal design beam energy for the \Qweak experiment is 1.165 GeV. We have modelled the experiment to
determine how the uncertainties change as a function of incident beam energy. Table
\ref{Kinematics_and_Backgrounds} considers three extreme cases, for incident energies of 1.095 GeV, 1.165 GeV
and 1.240 GeV. In the simulations, the magnetic field was scaled by the incident beam energy.

\begin{table}[htb]
\centering\caption{ {\em Examples: Kinematics and Background Tradeoffs }}
\label{Kinematics_and_Backgrounds}
\begin{tabular}{lllll}
                   &          &        &                    \\     \hline
\\
    Beam Energy (GeV)                & 1.095            & 1.165             & 1.240   & \\
         Rate (MHz)                      & 911          & 810           & 702    & \\
         Average $Q^2$ (GeV/$c)^2$              & 0.0229            & 0.0258                 & 0.0292 & \\
         Statistical uncertainty (\%)   & 3.37          & 3.20          & 3.05  & \\
         Hadronic uncertainty (\%)           & 1.37         & 1.51          & 1.68  &  \\
         All other errors (\%)              &  2.00         & 2.09          & 2.18  &  \\
         Relative Error on $Q^p_W$ (\%)                & 4.15          & 4.11          & 4.11  &  \\

        &       &               &      \\ \hline \hline

Note: The quoted errors are the errors on $Q^p_W$.

\end{tabular}
\end{table}

Although, the overall figure-of-merit (and therefore the total uncertainty on \qw ) is relatively flat with
incident beam energy the average $Q^2$ raises from about 0.023 (1.095 GeV) to 0.029 (1.240 GeV). The 0.029 case
is undesirable as the error contribution due to the residual hadronic background increases. Table
\ref{Kinematics_and_Backgrounds} shows only the average  $Q^2$, when in reality we have a tail of higher  $Q^2$
events within our acceptance, so we desire to keep the average $Q^2$ small.  Lower beam energies suppress the
hadronic uncertainty. However, if the incident beam energy is lowered significantly below ~1.1 GeV then the
overall figure-of-merit deteriorates rapidly because the asymmetry-weighted statistical error increases.
Therefore, from these arguments we  prefer an incident beam energy of greater than 1.1 GeV but less than 1.165
GeV.

\subsubsection{Maximum Magnet and Power Supply Capabilities}

The installed AC, DC power supply, cables and magnet have been designed to operate at 8615 Amps which corresponds to an
incident beam energy of 1.165 GeV. The maximum current capability of the power supply and the magnet (cooling
and forces) is 9500 Amps. The reserve is largely consumed by headroom needed for reliable and safe operation of the power supply-magnet system and to allow
for a modest current increase required because we had to slightly enlarge the radial separation of the coils as a result of tolerance considerations
encountered during the trial assembly at MIT-Bates. The net result leaves us with a modest 4\% reserve over and above operation at 1.165 GeV.
However, we will not know for certain how much reserve there is in the power supply or how high
in field we can actually energize the magnet until the field mapping has occurred at MIT-Bates in the late
Spring of 2008. It is possible that operation much above 1.165 GeV will not prove practical for other reasons.

\subsubsection{Cryogenics Available to Cool the Target and Accelerator}

The experiment requires sufficient cooling to operate the 35 cm LH$_2$ hydrogen target with high current beam
and keep the noise contribution due to ``boiling' and other density fluctuations (noise) significantly smaller
that the total statistical error per helicity pulse pair. This requirement and the suppression of other
helicity correlated beam ``residuals" are the primary reasons we will be flipping the beam helicity at the new
much higher rate. As the beam energy is increased above the nominal 1.165 GeV the accelerator rapidly requires
more cryogens,  and therefore less are available for our 2.4 kW target. This problem was addressed in 2004 with
extensive discussions between the \Qweak collaboration, the Accelerator Division, the FEL group, the Cryogenics
group, and the Physics Division leadership. These discussions occurred over about 6 months and culminated in an
agreement (``cryo-agreement.pdf" - available on the document server at ``http://www.jlab.org/qweak/") between
all parties concerning how the cryogenic requirements of the \Qweak experiment would be achieved and the
constraints that would be imposed on running in Hall A, Hall B and the FEL. The agreement was predicated on 5.5
GeV, 5-pass operation, and under the assumption a lower power target program in Hall A could be run concurrently
with \Qweak and the FEL. However, the agreement also showed that at 5.8 GeV, an additional 8 g/sec of CHL
reserve capacity would be lost. In such a case, it would be necessary to shut down either Hall A or the FEL load
in order to run \Qweak. At beam energies below 5.5 GeV, it is easier for the CHL to deliver the required
cooling.

\subsection{$Q^2$ Calibrations}

In this section, we discuss the aspects of the beam time request
related to the $Q^2$ calibration.

{\bf Commissioning}  For all of the commissioning activities, it is
assumed that the listed tasks will be appropriately interleaved
with other commissioning activities to insure that adequate time
can be given to analyzing the data and planning appropriately.  Below,
we list the tasks we anticipate carrying out during the $Q^2$ related
commissioning activities listed in Table~\ref{Beam_Request_Itemized}.

{\bf Commissioning: 0.1 - 5 nA beam and diagnostics - 4 days}  The goal of this
activity is to establish the procedures for routine ``on-demand''
delivery of beam in the 0.1 - 5 nA current range and to commission
all the needed hardware to monitor the intensity, position, and
size of the low current beam. Tasks to be carried out
during this activity are:
\begin{itemize}

\item Establish the laser and chopper slit settings need to deliver
the range of low current beam desired

\item Establish the gains and thresholds of the ``halo'' monitor detectors

\item Calibrate the beam current monitoring system (aluminum target plus
halo monitors) by cross calibration with the injector Faraday cup and
other techniques

\item Run several selected superharp monitors and optimize the beam
position/size measuring system (by determining which halo monitor
detectors give the best signal to noise)

\item After establishing the beam position/size measuring system,
determine if any further beam tuning is needed to optimize the beam
size and eliminate any halos

\item Run for several hours with low current beam (0.1 nA) to monitor
the stability of the beam properties using the optimized
detector/superharps chosen after analysis of the data from the previous
tasks

\end{itemize}

{\bf Commissioning: Region I, II, III tracking - 7 days}  The goal of this activity is to do beam-related
commissioning of the tracking system components (the three sets of tracking chambers, trigger scintillator, and
quartz scanner).  Tasks to be carried out during this activity are:

\begin{itemize}

\item Establish that insertion of the tracking system hardware can be done
in an acceptable amount of time ($<$ 4 hours)

\item At 0.1 nA, run with the full tracking system to establish the trigger
timing and measure the wire start times

\item Run the tracking system at 0.1 nA beam current with a variety of
targets: vertical/horizontal wire grids, carbon target, liquid hydrogen
target

\item At 100 nA with a liquid hydrogen target, run
with only
the Region III drift
chambers, trigger scintillator, main quartz detector, and quartz
scanner to establish the
relationship between the Region III and quartz scanner light-weighted
$Q^2$ maps

\item With a hydrogen gas target, study the rate dependent tracking
efficiency of the region II chambers (the most sensitive) by varying
the beam current over the range of 0.1 - 10 nA (corresponding to
rates ranging from 7 kHz - 700 kHz in the region II chambers)

\end{itemize}

{\bf Commissioning: Initial $Q^2$ Measurement - 4 days} The goal of this
activity is to perform an initial $Q^2$ measurement in each of the eight
octants (a total of four measurements since two octants can be done at
at time) and to measure the sensitivity of the $Q^2$ measurements to
several variables.

\begin{itemize}

\item For each octant pair, take enough data to satisfy our usual
requirements (about 2 hours per octant pair) with liquid hydrogen target

\item For a single octant pair, take data under a variety of conditions
with liquid hydrogen target -
vary magnetic field to move the elastic distribution across the
detector, vary beam position and angle

\item With a hydrogen gas target, take data for two different target pressures for one octant pair; the
difference will yield a $Q^2$ distribution for a hydrogen target with minimal external bremsstrahlung; this will
be useful in benchmarking our $Q^2$ simulations

\end{itemize}

{\bf Production: $Q^2$ measurements - 12 days}  The $Q^2$ related
measurements during production running will consist of two activities:

\begin{itemize}

\item Scans of a main detector {\v C}erenkov bar using the quartz scanner at the nominal running current of 180
(or 150) $\mu$A.  These can be done parasitically during regular production running.  It is anticipated that a
typical scan for a single octant will take  less than half an hour.

\item Dedicated calibration runs at low beam current ($\sim$ 0.15 nA).  The
total time needed for one of these measurements is expected to be about
1 day.  This includes 8 hours for setup/backout and a conservative 16 hours for
data-taking.  The minimum needed data taking-time per octant pair is $\sim$ 2 hours.
This amount of data will yield 1$\%$ relative statistical error
 per pixel assuming the quartz
bar is divided into 360 1 cm x 10 cm pixels for the $Q^2$ analysis.

\end{itemize}

The overall estimate of 12 days needed for this activity comes from assuming that
we will perform a $Q^2$ measurement roughly once per month during the $\sim$ 12
calendar months of production running.

\subsection{Requested Time}
\label{beamrequest}

Recognizing that there are unknowns with regard to the total cryogenic cooling capacity available, the need for
flexibility with respect to other experimental programs at JLab, and potential limits on the quantity and
quality of very high power polarized beam from a single gun, we present two scenarios for the \Qweak beam time
request. These are for the case when the high power production running and associated backgrounds measurements
are limited to 150~$\mu$A, and the ideal condition when the source and accelerator can deliver on demand the
full 180~$\mu$A of beam current. The difference in time requested is not as dramatic as one might initially
expect, as many of the commissioning and systematic measurements will be performed at reduced beam currents.

Therefore, as detailed in Table~\ref{Beam_Request_Itemized} we request approval for 198 (PAC) days if the
production running current is 180~$\mu$A or alternately 223 (PAC) days if the production running is limited to a
maximum of 150~$\mu$A. This requests covers the commissioning of the new beamline and Compton Polarimeter, all
experiment sub-systems, and the experiment-specific setup for the polarized source. It includes time for an
initial 8\% measurement and for the full production run associated with a 4\% measurement of the weak charge of
the proton. ``Production" refers solely to full current running on the LH$_2$ target. Allowable overhead includes
time for background measurements, $Q^2$ calibrations, beam polarization measurements, systematic checks, and the
configuration changes needed to accomplish these. We assume that time needed to optimize $P^2 I$ in the injector
will come out of the factor of two in scheduled days versus PAC days ({\em i.e.}, it is unallowed overhead).

In summary, since the 2004 proposal, our beam request for the experiment has been significantly refined and
takes into account recent experiences concerning the commissioning of major new hardware. Based on the recent
experiences of parity measurements at JLab, better estimates of the effort required to conduct systematics
studies and backgrounds measurements are now incorporated. The request allows for serious commissioning of the
most critical sub-systems for \Qweak -- specifically, our high power 2.4 kW cryo-target system, extensive new
tracking hardware, a new Hall C beamline and polarimeter.   The request accounts for expected inefficiencies due
to the compression of the entire measurement into a very tight calendar period just prior to the end of the 6
GeV program at JLab.

\begin{table}[p]
\centering\caption{ {\em Itemized beam request for the experiment.}}
\label{Beam_Request_Itemized}
\begin{tabular}{lll}
                                        &               &                \\
                 Category               & Time                    &       Time                         \\
                                                &  ($I_{max}$ = 180~$\mu$A)         &        ($I_{max}$ = 150~$\mu$A)                         \\  \hline

                 \\
               {\bf Beamline/Polarimeter Commissioning:}   &                &                       \\
               \\
                Beam line and Compton              & 14 days        & 14 days                      \\
                                 \\
               {\bf Experiment Commissioning:}   &                &                       \\
               \\
                High Power Cryotarget                   & 7 days         &    7 days                   \\
                Main Detectors, Lumi's                                  & 4 days         &    4 days                \\
                Transverse Pol. measurement        & 2 days         &   2 days                 \\
                Initial $Q^2$ measurement                   & 4 days         &   4 days                    \\
                 QTOR Magnet                                & 2 days         &   2 days                     \\
                 Neutral Axis Studies                       & 5 days        &   5 days                   \\
                 0.1-5 nA beam and diagnostics,             & 4 days         &  4 days                    \\
                 Regions I, II, III tracking                & 7 days         &  7 days                     \\
                 Background Studies                 & 7 days         &  7 days                    \\
      \\
                 Commissioning subtotal & 56 days & 56 days
                 \\
                 \\
                {\bf Production: $e+p$ elastic on $LH_2$}             &  106 days      &  127 days       \\
                                        &               &                       \\
                {\bf Overhead:}
                \\                          &              &                       \\
                Configuration changes                   & 4 days          & 4 days \\
                Al window background                & 3 days        &  4 days                     \\
                Inelastic background                & 3 days         &   4 days                     \\
                Soft background measurements            & 3 days         &    4 days                  \\
                Polarimetry measurements                    & 4 days         & 4 days      \\
                $Q^2$ measurements                      & 12 days         &  12 days                      \\
                Systematics: $I_{beam}$ dependence, etc.     & 7 days       &   8 days        \\
                \\
                Overhead subtotal      & 36 days        &        40 days                \\
                                         &               &                       \\
          &              \\ \hline
                                        &               &                       \\
                {\bf Total:}      &  {\bf 198 days}      &  {\bf 223 days}           \\

                &               &       \\ \hline

\end{tabular}
\end{table}

\newpage
\section{Collaboration and Management Issues}
\label{collaboration}

The \Qweak{} collaboration presently consists of 86 individuals from 25 institutions. The collaboration list is
kept at the experiment's web page, at \verb+http://www.jlab.org/qweak+. A document server provides access to the
collaborations technical archive. The archive includes key documents such as the original 2001 proposal, the
2003 Technical Design Report  including the review committees findings, the 2003 project management plan
including all quarterly progress reports to the DOE, the 2004 jeopardy proposal, and this document the 2007
jeopardy update.

The \Qweak  experiment operates as a managed project. A Project Management Plan dated June 28, 2004 is in place
and defines our interaction with the DOE. In addition, the management plan describes the management
organization, the cost, schedule, and performance requirements and controls, contingency plans, and reporting.
The individual Work Packages of the experiment are described there along with their detailed cost and schedule
breakdowns.

The major capital construction funding was provided by the US DOE through Jefferson Lab (\$1.91M), the US NSF
through a MRI (\$590k) which has University matching funds (\$452k) associated with it, and the Canadian NSERC
($\sim$\$315k). Including a small additional NSF grant (\$50k), the total budget for the experiment is \$3.316M.
The experiment aims to begin installation in Hall C at Jefferson Lab in the fall of 2009.

Besides the Spokespersons, construction project Work Package Leaders, and Operations Team Leaders, the
collaboration has a Principal Investigator, a Project Manager, and an Institutional Council. The Institutional
Council consists of representatives from each of the major ``stakeholder" institutions.

The capital construction work within the formal project has been broken down according to a Work Breakdown
Structure described in the Project Management Plan. Each major WBS line item has a Work Package Leader
associated with it. These major activities,  Hall C infrastructure upgrades (such as the Compton polarimeter),
several smaller ``post" project management plan submission sub-systems, and other ``operations" related
activities are summarized in Table~\ref{table-wbs}. Activities listed under the category of ``operations " will
expand as we approach installation time to include: Hardware installation tasks, readiness reviews, sub-system
commissioning/calibration and tasks associated with production running.

\begin{table*}[ht]
\centering
\begin{tabular}{|l|l|l|l|}      \hline
Category & Title &
 Leader &
Institute  \\ \hline \hline
Management & Principal Investigator  & R. Carlini & JLab \\ \hline
Management & Spokespersons & R. Carlini, M. Finn& JLab, W\&M,  \\
&  &  S. Kowalski, S. Page &  MIT, UManitoba \\ \hline
Management & Project Manager & G. Smith & JLab \\ \hline
WP1 & Detector System & D. Mack & JLab \\ \hline
WP1.1 & Detector Design & D. Mack &JLab \\ \hline
WP1.2 & Detector Bars & D. Mack & JLab \\ \hline
WP1.3 & Detector Electronics & Larry Lee, Des Ramsay & TRIUMF \& UManitoba \\ \hline
WP1.4 & Detector Support & A. Opper & GWU \\ \hline
WP2 & Target System & G. Smith & JLab \\ \hline
WP3 & Experiment Simulation & N. Simicevic & LaTech \\ \hline
WP4 & Magnet & S. Kowalski & MIT \\ \hline
WP5 & Tracking System & D. Armstrong & W\&M \\ \hline
WP5.1 & WC1--GEMs & S. Wells/T. Forest & LaTech/Idaho State \\ \hline
WP5.2 & WC2--HDCs & M. Pitt & VPI \\ \hline
WP5.3 & WC3--VDCs & J. M. Finn & W\&M \\ \hline
WP5.4 & Trigger Counters & A. Opper & GWU \\ \hline
WP6 & Infrastructure & R. Carlini & JLab \\ \hline
WP7 & Magnet Fabrication & Wim van Oers & TRIUMF \& UManitoba \\ \hline
WP8 & Luminosity Monitor & M. Pitt & VPI \\ \hline
Operations& Magnet Mapping &  L. Lee & TRIUMF \& UManitoba\\ \hline
Operations& Profile Scanner &  J. Martin & UWinnipeg \\ \hline
Operations& Compton \& New Beamline  &  D. Gaskell & JLab \\ \hline
Operations&Compton e- Detectors &  J. Martin/D. Dutta & UWinnipeg/UMiss \\
 &  & H. Mkrtchyan & Yerevan \\ \hline
Operations & Polarized Beam Properties& Matt Poelker/K. Paschke& JLab/UVA \\ \hline
Operations & GEANT Simulations & K. Grimm/A. Opper & LaTech/GWU \\ \hline
Operations & Data Acquisition & P. King & Ohio University \\ \hline
\end{tabular}
\caption[beamspecs]{{\em Stakeholder and Management Structure of the \qw\ experiment. }}
\label{table-wbs}
\end{table*}

\newpage
\section{Manpower }

The data-taking for the \Qweak experiment will demand significant investment of time from the collaboration.
Assuming about 54 weeks calendar weeks of beam time, including both commissioning and production running, and
with three shifts/day, each staffed with three collaborators, we will need to staff 3400 shifts. In addition, we
will need 27 different 2-week long run `coordinatorships', another significant investment of time. Finally, we
will require on-site presence of ``experts'' for each of the subsystems, especially during the commissioning
phase.

Three individuals per shift will certainly be needed during the demanding commissioning phase. We might be able
to reduce that to two people during the presumably ``routine'' production running. Experience with the G0
experiment showed that this was possible, even for a demanding parity-violation experiment, during the later
parts of the production data-taking. However, along with the mandatory cryogenic target operator and the shift
leader, the continuous operation of a Compton polarimeter may require a 3rd person on shift; experience during
the HAPPEX experiments with the Hall A Compton showed that it was not until significant experience had been
obtained with the polarimeter that it could be usefully run without fairly constant attention.

The collaboration has grown significantly since the last update to the PAC, with 86 collaborators at present,
and we continue to welcome new collaborators and institutions. Fortunately, we have 7 faculty members who plan
to take a sabbatical leave at JLab during the installation and running of the experiment (D. Armstrong, J.
Birchall, P. King, A. Opper, S. Page, M. Pitt, J. Roche), and we hope that the PAC will encourage the Lab  to
support these sabbaticals. We presently have a total of 7 graduate thesis students already identified for the
experiment, with the expectation of more joining in the near future.

Using the present collaboration size, then, the average shift load will be 44 shifts per collaborator, spread out over two calendar
years, or 22 shifts/person/year, with about 1/3 of the collaborators also taking on a two-week long duty as run coordinator.  This
represents a significant but not unreasonable investment of time. Thesis students and postdocs, of course, will typically take a larger
share than faculty with teaching responsibilities.
         %

\bibliography{update}
\bibliographystyle{unsrt}

\end{document}